\documentclass[final,3p,times,nopreprintline,sort&compress]{elsarticle}
\usepackage{amsmath,amssymb,amsfonts,dsfont}
\usepackage{mathtools}
\usepackage[utf8]{inputenc}
\usepackage{graphicx}
\usepackage{slashed}
\usepackage{bm}
\usepackage{booktabs}
\usepackage{tabulary}
\usepackage{placeins}
\usepackage{ucs}
\usepackage[dvipsnames]{xcolor}
\usepackage[caption=false]{subfig}
\usepackage[
  pdftitle={Towards a theory of hadron resonances},
  pdfauthor={Maxim Mai, Ulf-G. Meissner, Carsten Urbach},
  bookmarks,
  colorlinks,
  linkcolor=myblue,
  citecolor=mymagenta,
  menucolor=black,
  urlcolor=myblue,
  plainpages=false,
  pdfpagelabels,
  hypertexnames=false]{hyperref}
\usepackage{cleveref}

\crefformat{figure}{Fig.~#2#1#3}
\crefformat{table}{Tab.~#2#1#3}
\crefformat{equation}{Eq.~(#2#1#3)}
\crefformat{section}{Sect.~#2#1#3}

\newcommand{\beq}{\begin{equation}}
\newcommand{\eeq}{\end{equation}}

\renewcommand{\Re}{\text{Re}\,}
\renewcommand{\Im}{\text{Im}\,}
\newcommand{\re}{\mathrm{Re}\,}
\newcommand{\tr}{\mathrm{Tr}\,}
\newcommand{\csw}{c_\mathrm{SW}}

\DeclarePairedDelimiterX\set[1]\lbrace\rbrace{#1}
\DeclarePairedDelimiter\abs{\lvert}{\rvert}%

\newcommand{\pcm}{\bm{p}_\mathrm{cm}}

\def\XXint#1#2#3{{\setbox0=\hbox{$#1{#2#3}{\int}$}
     \vcenter{\hbox{$#2#3$}}\kern-0.5\wd0}}

\allowdisplaybreaks[1]

\begin{document}

\renewcommand{\theequation}{\arabic{equation}}
\numberwithin{equation}{section}
\setcounter{tocdepth}{3}

\begin{frontmatter}

\title{Towards a theory of hadron resonances}

\author[Bonn,GWU]{Maxim Mai}
\author[Bonn,Juelich,tiflis]{Ulf-G.\ Meißner}
\author[Bonn]{Carsten Urbach}

\address[Bonn]{Helmholtz-Institut f\"ur Strahlen- und Kernphysik (Theorie) and
   Bethe Center for Theoretical Physics, Universit\"at Bonn, D--53115 Bonn, Germany}
\address[Juelich]{Institut f\"ur Kernphysik, Institute for Advanced Simulation and JARA-HPC,\\  Forschungszentrum J\"ulich,
  D--52425  J\"ulich, Germany}
\address[GWU]{The George Washington University, Washington, DC 20052, USA}
\address[tiflis]{Tbilisi State University,  0186 Tbilisi, Georgia}

\begin{abstract}
In this review, we present the current state of the art of our understanding of the spectrum of excited strongly interacting particles and discuss methods that allow for a systematic and model-independent calculation of the hadron spectrum. These are lattice QCD and effective field theories. Synergies between both approaches can be exploited allowing for deeper understanding of the hadron spectrum. Results
based on effective field theories and hadron-hadron scattering
in lattice QCD or combinations thereof are presented and discussed.
We also show that the often used Breit-Wigner parameterization is at
odds with chiral symmetry and should not be used in case of strongly coupled channels.
\end{abstract}


\end{frontmatter}
\tableofcontents

\section{Introduction: QCD and excited states}
\label{sec:intro}
For a mathematical concept to become a dogma in describing Nature it has to be confronted with observations. In the realm of the microscopic building blocks of matter ($\sim10^{-10}~\rm m$) such concepts are often far from the intuitive human experience, which is evolutionary engraved into logical concepts by dealing with everyday objects ($\sim10^0~\rm m$). For example, the  originally non-intuitive language of quantum mechanics\footnote{See, e.g., reflections of W.~Heisenberg's \emph{``Physics and Beyond: Encounters and Conversations''}~\cite{HeisenbergBOOK} on the early disputes on nature of quantum mechanics} is now an indisputable concept of any model of nuclear and particle physics. In that, one of the major breakthroughs of quantum mechanics is associated with the correct description of the pattern of excited states of atomic spectra. Going five orders of magnitude deeper ($\sim10^{-15}~\rm m$) along this path, it has been discovered then that strongly interacting particles also build a complex spectrum of ground (such as proton and neutron) and excited states, so-called hadrons. It is, therefore, believed that resolving the general pattern and microscopic structure of this spectrum holds the key to the correct understanding of the strong interactions. This is the topic of \emph{hadron spectroscopy} that is at the heart of the present review. Currently, the most universal language in addressing this area of research is the language of gauge field theories, which unifies the principles of special relativity and quantum mechanics incorporating local (gauge) invariance. Quantum chromodynamics (QCD) is a prime example of such a gauge theory describing the strong interaction.

Quantum chromodynamics is a truly remarkable theory passing all tests when compared to experiment for nearly five decades. Its Lagrangian can be written in a single line
\begin{equation}
{\cal L}_{\rm QCD} = -\frac{1}{4} \, G_{\mu\nu}^a G^{\mu\nu, a} +
\sum\limits_{f} \bar q_f (\mathrm{i} D\!\!\!/ - {\cal M}_f) q_f + \ldots\,,
\label{eq:L}
\end{equation}
where the ellipses denote the gauge-fixing and the $\theta$-term, which will not
be considered in what follows. Here,  $D_\mu = \partial_\mu - \mathrm{i}g A_\mu^a \lambda^a/2$ is the gauge-covariant
derivative, $A_\mu^a$ ($a=1,\ldots,8$) the gluon field, $G_{\mu\nu}^a =\partial_\mu A_\nu^a
- \partial_\nu A_\mu^a - \mathrm{i}g [A_\mu^b, A_\nu^c]$ the gluon field strength tensor,  $g$ is the
SU(3) color gauge coupling, $q_f$ a quark spinor of flavor $f$ ($f=u,d,s,c,b,t$)  and ${\cal M}_f$ is the diagonal quark matrix. The quarks come in two types, the
light ($q=u,d,s$) and heavy ($Q=c,b,t$) quark flavors, where light and heavy refers to the
QCD scale $\Lambda_{\rm QCD} \simeq 210$~MeV (for  $N_f = 5, \overline{\mathrm{MS}}, \mu = 2\, $GeV).
Note that the top quark decays too quickly to participate in the strong interactions,
so effectively $Q=(c,b)$. In the absence of the quark masses,
$\Lambda_{\rm QCD}$ is the only dimensionful parameter in QCD that is generated by dimensional
transmutation through the running of the strong coupling $\alpha_s = g^2/4\pi$.
The fundamental fields of QCD, the quarks and gluons, have never been observed in
isolation, which is called color confinement. They appear as constituents of the strongly interacting particles, \emph{the hadrons}. This particular feature makes QCD highly non-trivial but also very interesting.

The Lagrangian of QCD  allows us to define two special limits, in which the theory can
be analyzed in terms of appropriately formulated effective field theories (EFTs).
In the light quark ($f=u,d,s$) sector, the effective Lagrangian can be written
in terms of left- ($q_L$) and right-handed ($q_R$) quark fields, such that
\begin{equation}
{\cal L}_{\rm QCD} = \bar q_L \, \mathrm{i} D\!\!\!/ \, q_L  + \bar q_R \, \mathrm{i} D\!\!\!/ \, q_R 
+ {\cal O}(m_f / \Lambda_{\rm QCD})\,.
\end{equation}
As can be seen, left-  and right-handed  quarks decouple, which is reflected in the 
chiral symmetry.  It is explicitly broken by the finite but small quark masses $m_f$. 
Furthermore, chiral  symmetry is spontaneously broken, leading to the
eight pseudo-Goldstone bosons, the pions, the kaons and the eta. These are indeed the
lightest hadrons, with their squared masses proportional to $m_f$.
The pertinent EFT is chiral perturbation theory (CHPT). 

Matters are very different for the heavy $c$ and $b$ quarks, where the leading order
Lagrangian takes the form
\begin{equation}
{\cal L}_{\rm QCD} = \bar Q_f \,  \mathrm{i} v \cdot D \, Q_f + {\cal O}(\Lambda_{\rm QCD} / m_f)\,,
\end{equation}
with $v$ the four-velocity of the heavy quark and $Q_f$ denotes a quark spinor of
flavor $f$ ($f=c,b$). Note that to leading order, this Lagrangian
is independent of quark spin and flavor, which leads to  SU(2) spin and  SU(2) flavor symmetries, 
called HQSS and HQFS, respectively. The pertinent EFT to analyze the consequences is 
heavy quark effective field theory (HQEFT), which comes in different manifestations. Finally, in  heavy-light systems, where heavy quarks act as matter fields coupled to the light pions,
one can combine CHPT and HQEFT.

By construction, EFTs are limited to certain energy regions. A more general non-perturbative
method is lattice QCD, where the Euclidean version of the theory is put on a four-dimensional
space-time grid, characterized by a  given lattice spacing $a$ and a finite volume,
$V = L^3\times L_t$, with $L$ the spatial length ($L=Na$) and $L_t$ the extension in Euclidean time.
Observables can be calculated by Monte Carlo simulations on the lattice. To make contact
with Nature, one must consider the continuum limit $a\to 0$, the thermodynamic limit $V\to\infty$ and often has to extrapolate in the (light) quark masses down to the
physical values. All this induces some systematic uncertainties, but also comes with
additional value. On the one hand, varying the quark masses allows one to pin down
low-energy constants of pertinent EFTs that can often only be determined approximately (or not at all)
from continuum investigations and on the other hand, the volume dependence of the
measured energy levels encodes information about excited states, as discussed in more
detail below.

There are various reasons to consider \emph{excited states}, which are the topic of
this review. First, the spectrum of  QCD is arguably its least understood feature. The
hadron spectrum has for a long time been a playground of the constituent quark model, but we
know now that this only captures certain symmetry properties of QCD, but not its full dynamics.
This is evident from  the questions with respect to the nature of the XYZ and other
``exotic'' states, where the word exotic appears between quotation marks, because this
usually refers to states that can not be described within the (conventional) quark model.
It is very obvious that the quark model is much too simple, as it is, e.g., it does not account for a whole
class of important players in the hadron spectrum, the so-called hadronic molecules.
Since the beginning of this millennium, there has been a waste activity both experimentally
and theoretically to pin down the properties of these unusual (``exotic'') states,
see the recent reviews~\cite{Lebed:2016hpi,Esposito:2016noz, Hosaka:2016pey,Chen:2016qju,Chen:2016spr, Ali:2017jda,Guo:2017jvc, Olsen:2017bmm,Karliner:2017qhf,Brambilla:2019esw, Liu:2019zoy, Ali:2019roi,Yang:2020atz,JPAC:2021rxu,Chen:2022asf,Brambilla:2022ura}.
Generally, QCD permits for a whole set of bound states, which can roughly be categorized
as compact states of quarks and antiquarks, states dynamically generated from hadron-hadron
(or three-hadron) interactions, hybrid states made from quarks and gluons as well as
glueballs, which are arguably the most exotic states QCD offers. Note, however, that in the
limit of many colors $N_c \to \infty$, the glue sector completely decouples from the quark sector.
It is also important to note that high-precision data for spectrum studies
have been and will be produced with ELSA at Bonn, MAMI at Mainz, CEBAF at Jefferson Lab, the 
LHCb experiment at CERN, the BES\-III experiment at the BEPCII, Belle-II at KEK,
GlueX at Jefferson Lab and in  the future with PANDA at FAIR and other labs worldwide.
These data clearly pose a challenge for any theoretical approach.

In what follows, we discuss theoretical approaches that will eventually unravel the physics
behind the QCD spectrum. 
More precisely, we only consider methods that are
1) {\em largely model-independent}\footnote{The meaning of ``largely'' will become clear in what follows.}, 2) {\em can be systematically improved}, and 3) {\em
allow for uncertainty estimates}. If one of these conditions is not
fulfilled, a given method will not be considered further, such as the Schwinger-Dyson
approach, which is genuinely non-perturbative but lacks any power counting scheme.
Also, we eschew models here.
So that leaves us with lattice QCD (LQCD) and EFTs or
combinations thereof. LQCD can get ground-states  and some excited
states at (almost) physical pion masses, but the most  distinctive feature of excited states are
{\sl decays} (or the fact that the excited states have a width). Consequently, we will
only consider LQCD studies of hadron-hadron scattering, that allow to extract the mass
and the width of a given resonance (this also means that we will not discuss any
spectrum calculations based on two-point functions). Furthermore, we will also show how
the use of suitably tailored finite-volume EFTs can help in this daring endeavour.

This review is organized as follows: In Sect.~\ref{sec:reso}, we define what is meant by a resonance.
In Sect.~\ref{sec:LQCD}, we discuss the basics of lattice QCD, with an emphasis on the formalisms
to extract resonances. Sect.~\ref{sec:EFT} consider EFTs for resonances, either as explicit fields
or via some unitarization method. Sect.~\ref{sec:well} contains the results on well separated
resonances, which are only a handful. Results for the more general case of resonances with
multiple decay channels (coupled channels) are collected and discussed in Sect.~\ref{sec:cc}, where
we also discuss the new phenomenon of the two-pole structure first observed in case of the
enigmatic $\Lambda(1405)$. In \cref{sec:well} and \cref{sec:cc} we consider publications until end of May 2022.
We summarize our results in Sect.~\ref{sec:summary}. As it will become clear,
we still far away from a precise understanding of the hadron spectrum using the methods
discussed here, but it is also remarkable to see the progress that has been made in the last decade.


\section{What is a resonance?}
\label{sec:reso}
\begin{figure}[t]
\centering
\includegraphics[height=6.9cm]{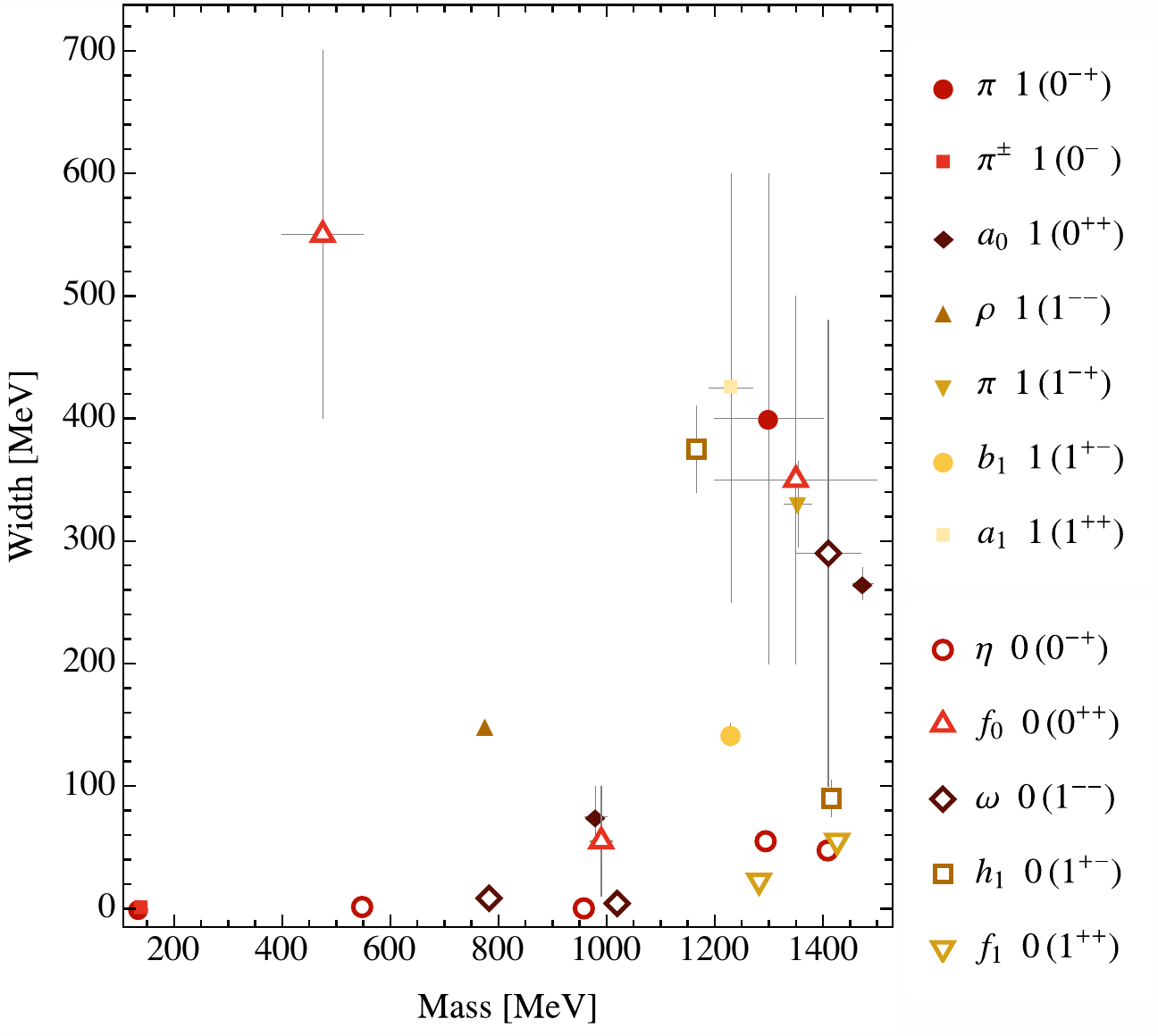}
\includegraphics[height=6.9cm]{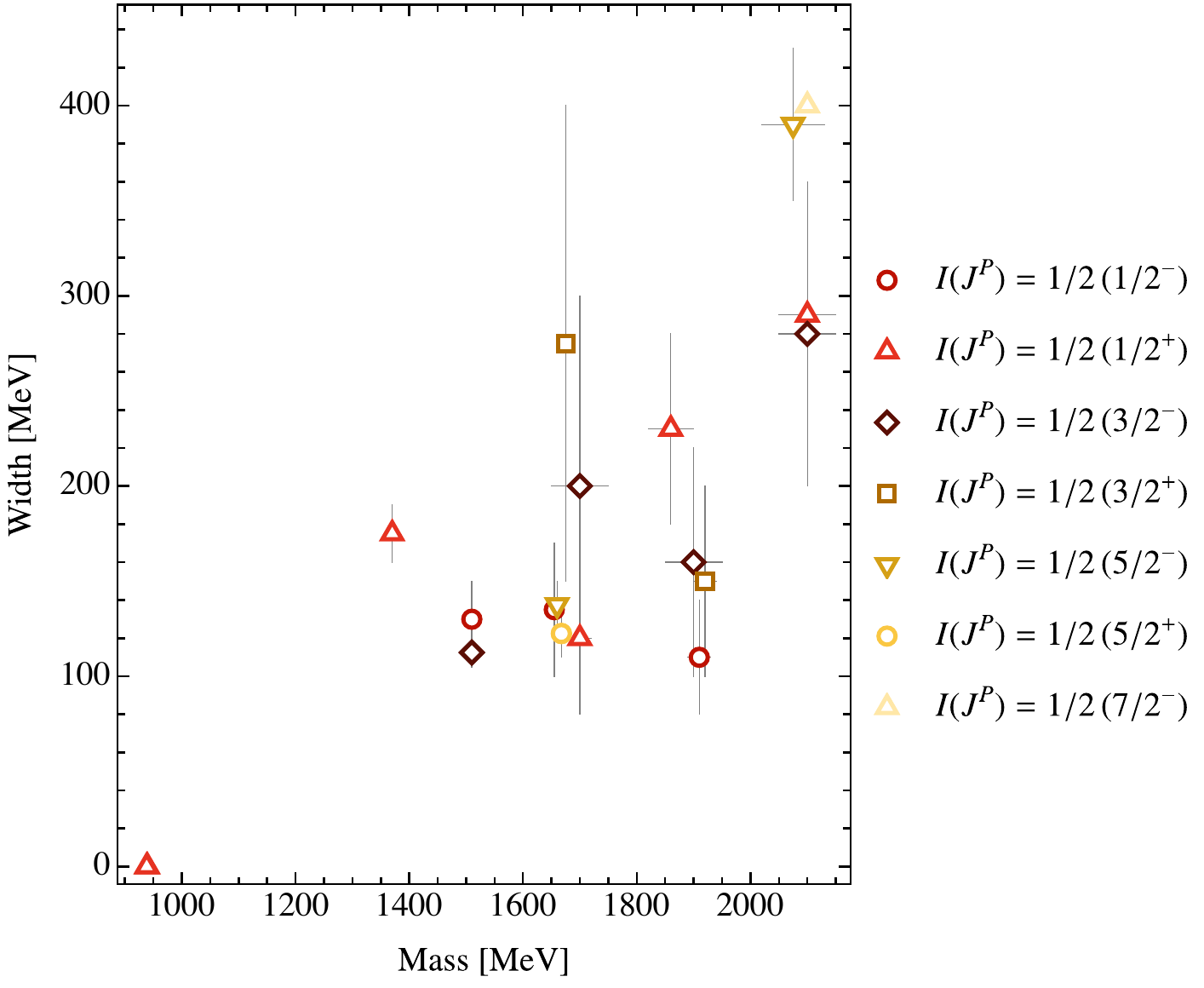}
\caption{Spectrum of unflavored mesons (left) and light baryons (right) categorized by isospin $I$ and total angular momentum and parity $J^P$. Values and uncertainty regions are taken from the current review of the Particle Data Group~\cite{ParticleDataGroup:2020ssz}.
\label{fig:spectrum}}
\end{figure}

The Particle Data Group (PDG)~\cite{ParticleDataGroup:2020ssz} lists around 100 excited mesonic
and around 50 confirmed (4 star) baryonic states. The basic parameters of these states, the hadrons, are their mass and decay width shown for the unflavored mesons and the lightest baryons in Fig.~\ref{fig:spectrum}. Most of these states are actually not stable, they are, broadly speaking, resonances. From the classical mechanics of a periodically driven oscillator, a resonance is characterized
by two properties at the resonance frequency: a peak in the amplitude of the oscillator and
a phase difference of $90^\circ$ between the driving force and the oscillator's response.

This concept can be transferred to quantum mechanics, where the peak is observed in the cross
section and the phase-shift between asymptotic incoming and outgoing states.
For the scattering of a plane wave (wave vector $\bm{k}\in\mathds{R}^3,~k:=|\bm{k}|$) with a angular momentum $\ell$ the total cross section $\sigma$ and phase-shift $\delta$ can be parameterized by the Breit-Wigner formula~\cite{Wigner:1946zz}
(note that we will encounter situations where this parameterization is inadequate, see e.g. the discussion in Sect.~\ref{sec:BDphiphi})
\begin{align}
\sigma(E)\propto\frac{4\pi}{k^2}(2\ell+1)\frac{\Gamma^2_{\ell}/4}{(E-E_R)^2+\Gamma_\ell^2/4}
\quad\text{and}\quad
\tan\delta\propto -\frac{\Gamma_\ell}{2(E-E_R)}\,.
\end{align}
This explains the name 'width' for $\Gamma$ representing twice the distance from the peak (at $E_R$) to the half of the maximal value of the total cross section. The enhancement of the cross section at $E=E_R$ is the very origin of the term 'resonance', see, e.g., Ref.~\cite{Wigner:1946zz}. Closely related to this is the so-called Flatt\'e{} parametrization~\cite{Flatte:1976xu} which allows to address coupled-channel systems, see, e.g., Refs.~\cite{Lesniak:1996qx,Kerbikov:2004uz,Baru:2004xg} for more details and  applications.

\begin{figure}[t]
\centering
\begin{minipage}{0.48\linewidth}
\includegraphics[width=\linewidth,trim=0 0 0 0,clip]{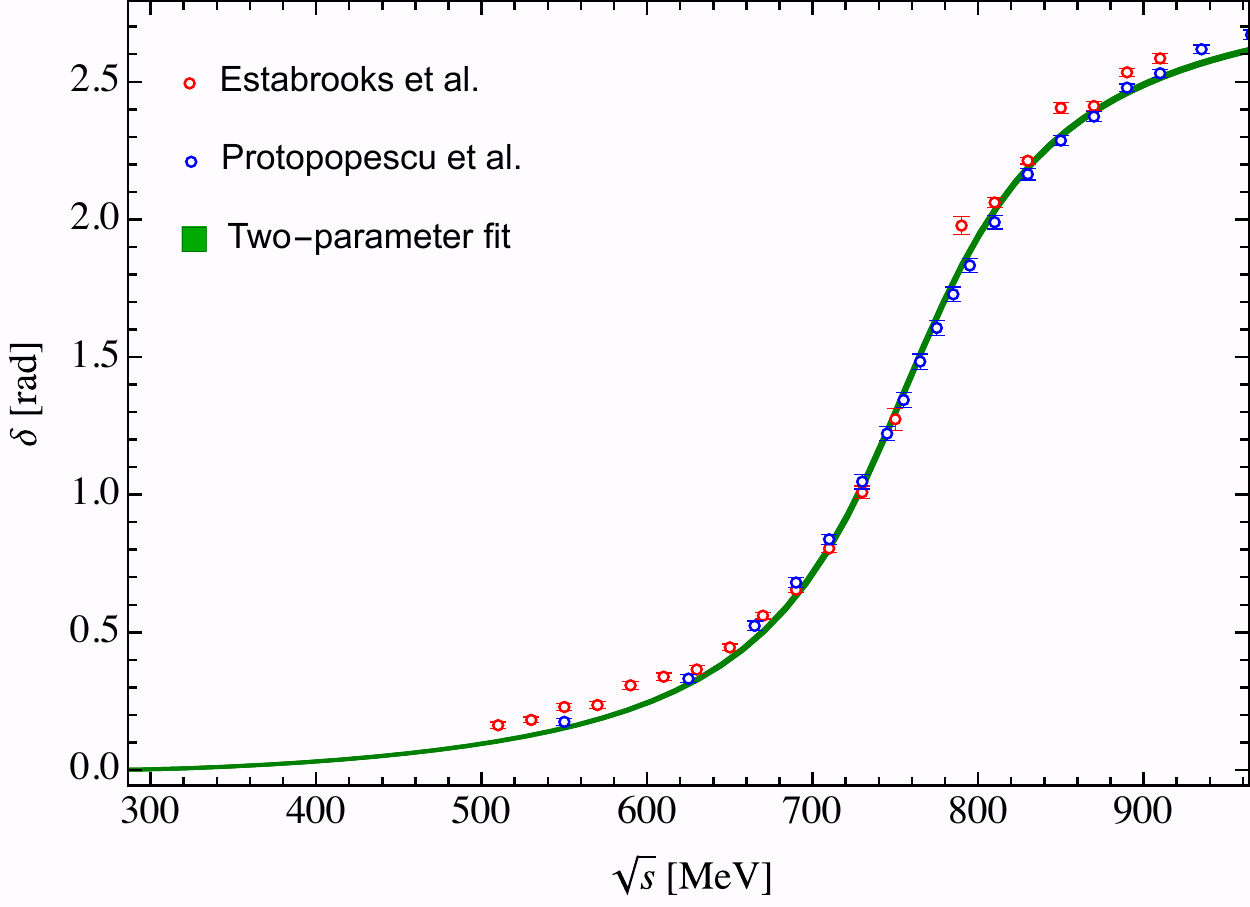}
\end{minipage}
~~~
\begin{minipage}{0.48\linewidth}
\includegraphics[width=\linewidth,trim=0 11cm 10cm 0,clip]{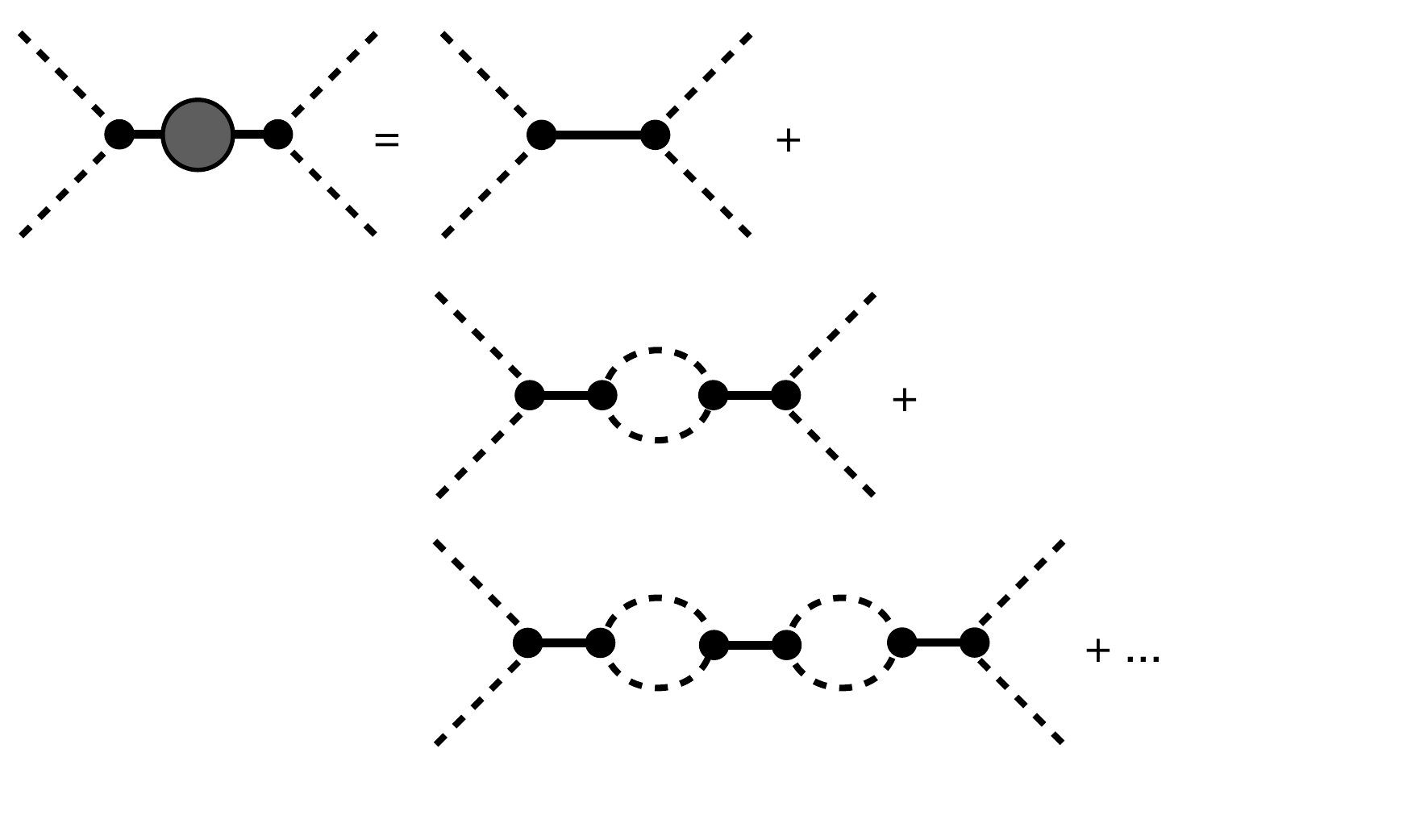}
\end{minipage}
\caption{ Left: A fit to experimental data from Refs.~\cite{Estabrooks:1974vu,Protopopescu:1973sh} using twice subtracted self-energy of the $\rho$ field, see Eq.~\eqref{eq:rho-self-energy}. Green band represents 1-sigma statistical uncertainty region of the two-parameter fit, respectively. Right: Diagrammatic representation of the $\rho$ resonance in $\pi\pi$ scattering via a bare state (full line) and self-energy due to coupling to asymptotically stable pion fields (dashed line).
\label{fig:rho-pipi}}
\end{figure}

There are exemplary cases, like for instance the $\rho$-resonance, where it is to very good
approximation sufficient to consider the P-wave ($\ell=1$) only. There, this quantum mechanical
picture of a resonance can be carried over to quantum field theory more or less directly.
However, in general the concept of resonances needs to be generalized to quantum field theories:
the enhancement of the cross section can be seen as a manifestation of a new quantum (resonance) field, which couples to asymptotically stable fields and acquires, thus, a finite width through self-energy contributions
Note in passing that often, an enhancement in the
cross section is not seen due to the strong background or coupled channel
effects. This is most clearly seen in the $P_{11}$ phase shift of pion-nucleon scattering, which is rather small in the vicinity of the
Roper resonance.
A prime As an example, consider a theory with two types of fields, an asymptotically stable pion (pseudo)scalar field and an unstable $\rho$ vector field. Given the coupling of the latter to two pions ($g_{\rho\pi\pi}$), the $\pi\pi$ scattering amplitude in the P-wave (see below for a formal definitions) reads
\begin{align}
T_P(s)\propto \frac{g_{\rho\pi\pi}^2}{s-M_\rho^2-\Sigma(s)}
\quad \text{and} \quad
\Sigma(s)=\int\frac{dk\, k^4}{(2\pi)^3}\frac{g_{\rho\pi\pi}^2}{2E_k(s-4E_k^2+i\epsilon)}\,,
\label{eq:rho-self-energy}
\end{align}
where $s$ denotes the total energy squared and $E_k=\sqrt{M_\pi^2+k^2}$. Note that the self-energy contribution ($\Sigma$) is divergent so that the quantities $g_{\rho\pi\pi}$ and $M_\rho$ need to be divergent as well since the full amplitude has to be finite. Over-subtracting the self-energy integral and introducing new subtraction constants allows then to express the scattering amplitude in terms of finite quantities only. These can be determined by fitting to experimental data as shown in Fig.~\ref{fig:rho-pipi} for the case of two subtractions.
We note further that often an enhancement in the
cross section is not seen due to the strong background or coupled-channel
effects. This is most clearly seen in the $P_{11}$ phase shift of pion-nucleon scattering, which is rather small in the vicinity of the
Roper resonance. Or as it is often stated: ``Not every bump is a resonance and not every resonance is a bump'', see, e.g., the discussion in
Ref.~\cite{Zichichi}

The introduction of auxiliary (resonance) fields is a useful tool in many applications of hadron physics, see, e.g., Refs.~\cite{Sadasivan:2020syi, Mai:2021vsw, Severt:2020jzc} for recent applications to multihadron systems. However, there are many examples, such as $f_0(500)$ or $\Lambda(1405)$, where such a representation is not sufficient. The most universal and modern approach to resonances deals directly with scattering amplitudes. In that, the so-called $S$-matrix -- originally introduced by Heisenberg~\cite{Heisenberg:1943zz} -- relates the asymptotic in- (three-momenta $\bm{p}_1,...,\bm{p}_m$) and outgoing (three-momenta $\bm{p}'_1,...,\bm{p}'_n$) states as
\begin{align}
\langle\bm{p}'_1,...,\bm{p}'_n|S|\bm{p}_1,...,\bm{p}_m\rangle
=
\langle\bm{p}'_1,...,\bm{p}'_n|(\mathds{1}+\mathrm{i}T)|\bm{p}_1,...,\bm{p}_m\rangle\,,
\end{align}
which defines the so-called $T$-matrix\footnote{Note that the definition of the $T$-matrix varies in the literature. In particular, the definition includes at times a different sign or momentum prefactor, which may have some technical advantages in some specific cases.}. The $S$-matrix obeys \emph{crossing symmetry}, i.e., the $S$-matrix element for a $n\to m$ transition can be converted analytically to the element describing $n+1\to m-1$ transitions etc.. Furthermore, and crucial for the matter of the present review is the principle of \emph{analyticity}. It is rooted in the requirement of causality~\cite{vanKampen:1953rya,Minerbo:1971gg} for physical processes and states that physical $S$-matrix elements are boundary values to an analytic functions in all inner products of all involved momenta ($\bm{p}^{(\prime)}_i\bm{p}^{(\prime)}_j$) or in closely related generalized Mandelstam variables, promoted  to their complex values. In that, the choice of variables is not unique but the number of independent ones is fixed.
First, since all in-/outgoing states are on the respective mass shell $\left(p^{(\prime)}_i{}^2=m^{(\prime)2}_i\right)$ only $3(m+n)$ combinations can be independent. Furthermore, energy-momentum conservation and choice of the reference frame constrain the number of independent invariants to $3(m+n)-4-6$, see Ref.~\cite{Eden:1966dnq} for more details. Finally, following the latter reference the $S$-matrix can be expressed as
\begin{align}
S=\sum_i {}_{\rm in}|i\rangle\langle i|_{\rm out}
\quad\text{for}\quad
{}_{\rm out}\langle i|j\rangle_{\rm in}=\delta_{ij}
\quad\text{and}\quad
\sum_i {}_{\rm in}|i\rangle\langle i|_{\rm in}=\sum_j {}_{\rm out}|j\rangle\langle j|_{\rm out}=\mathds{1}\,,
\end{align}
i.e., using complete and orthonormal set of physical states. This obviously leads to the \emph{unitarity} of the $S$-matrix, $SS^\dagger=S^\dagger S=\mathds{1}$, which physically ensures  probability conservation~\cite{Heisenberg:1943zz}. It is notable that the number of physical states depends on the energy of the system. For a generic two-body systems with identical particles of mass $\mu$ (equivalently for other cases) and total energy squared $s$ this yields schematically that
\begin{equation}
\vcenter{\hbox{\includegraphics[width=6cm,trim=0 18cm 16cm 0, clip]{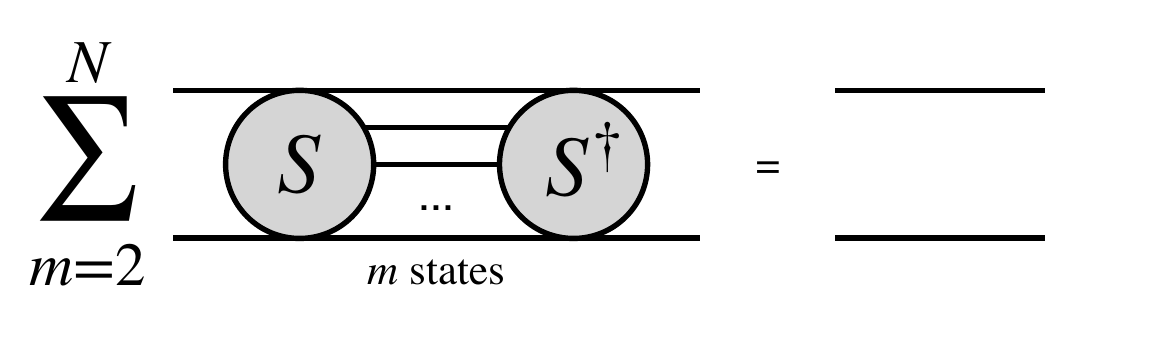}}}~
\begin{aligned}
(N)^2\le s/\mu^2 < (N+1)^2\,.
\label{eq:resonances-S-unitarity}
\end{aligned}
\end{equation}
Overall, unitarity, analyticity and crossing symmetry are the main principles of $S$-matrix theory going beyond expansion in Feynman diagrams. The pertinent matrix elements are expected to encode all information about the dynamics of the system including the properties of the resonances as  will be discussed below.

Unitarity of the $S$-matrix implies for the $T$-matrix that $(T-T^\dagger)=\mathrm{i}T^\dagger T$, leading for a general $n\to n$ transition to the following condition on the matrix elements and invariant matrix element
\begin{equation}
    {\mathcal{M}(\{p\}\to \{p'\})=(2\pi)^4\delta^{(4)}\left(\sum_{i=1}^n p_i-p_i'\right)\langle \bm{p}_1'\bm{p}_2'|T|\bm{p}_1\bm{p}_2\rangle}
\end{equation}
with
\begin{align}
\label{T-matrix Unitarity}
\langle \bm{p}'_1,...,\bm{p}'_n|(T-T^\dagger)| \bm{p}_1,...,\bm{p}_n\rangle&=
\mathrm{i}\int\prod_{i=1}^N\frac{\mathrm{d}^4k_i}{(2\pi)^{4}}\,(2\pi)\delta^+(k_i^2-m^2)
\langle \bm{p}'_1,...,\bm{p}'_n| T^\dagger|\{\bm{k}\}\rangle\,,
\langle\{\bm{k}\}|T| \bm{p}_1,...,\bm{p}_n\rangle \,,\\
\mathcal{M}(\{p\}\to\{p'\})-\mathcal{M}^*(\{p'\}\to\{p\})&=
\mathrm{i}\int\prod_{i=1}^N\frac{\mathrm{d}^4k_i}{(2\pi)^{4}}\,(2\pi)\delta^+(k_i^2-m^2)
\,(2\pi)^4\delta^{(4)}\left(P-\sum_{i=1}^N\,k_i\right)
\mathcal{M^*}(\{p'\}\to\{k\})\mathcal{M}(\{p\}\to\{k\})\,,\nonumber
\end{align}
where $\delta^+(k_i^2-m^2)$ selects the positive energy solution,
$P$ denotes the total four momentum of the system, and the  size of the complete set of intermediate states ($N$) depends on the total energy of the system, see Eq.~\eqref{eq:resonances-S-unitarity}.  The integration over intermediate momenta on the right-hand-side of the latter equation leads to another crucial and beautiful implication for the structure of the $S$-matrix.

\begin{figure}[t]
\centering
\includegraphics[width=\linewidth,trim=0 5cm 0 0,clip]{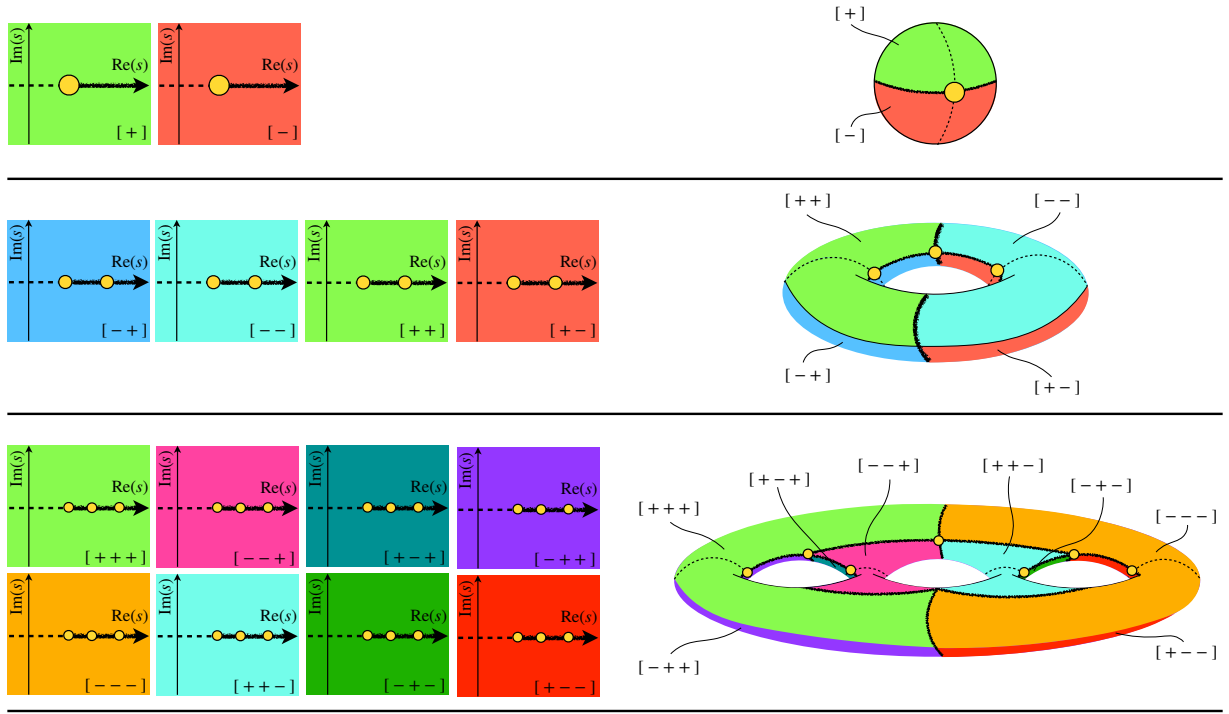}
\caption{Representation of the Riemann surface for one (top row) and two (middle row) and three (bottom row) open two-body thresholds. The value of $|s|=\infty$ is put for simplicity on the poles of the sphere and rim of the tori, respectively. Riemann Sheets are denoted by a sequence $[\pm\pm...]$ as discussed in the main text. Yellow dots depict the position of the branch points.
\label{fig:RiemannSheets}}
\end{figure}

As an example, consider a simple case of $2\to2$ scattering of identical particles, where by limitations due to total energy of the system or some quantum numbers only two-body intermediate states are allowed. Then, for the invariant matrix element $\mathcal{M}(p_1,p_2\to p_1',p_2')=(2\pi)^4\delta^{(4)}(p_1+p_2-p_1'-p_2')\langle \bm{p}_1'\bm{p}_2'|T|\bm{p}_1\bm{p}_2\rangle$ and the corresponding partial wave expansion
$\mathcal{M}_\ell(s)=1/(64\pi)\int_{-1}^{+1}dz \mathcal{M}(s,z)P_\ell(z)$ the unitarity relation simply implies that
\begin{align}
\label{eq:unitarity}
\Im \mathcal{M}_\ell^{-1}(s) = - \frac{2q(s)}{\sqrt{s}}\theta\left(\sqrt{s}-2m\right)
\qquad\text{with}\quad
q(s)=\sqrt{\frac{s}{4}-m^2}\,,
\end{align}
where $m$ denotes the mass of the involved field. This implies that a two-body system is described by a set of partial wave amplitudes of the form
\begin{align}
\label{eq:K-matrix}
\mathcal{M}_\ell(s)=\frac{\sqrt{s}}{2}\frac{1}{K_\ell^{-1}(s)-\mathrm{i} q}\,,
\end{align}
for some real function $K_\ell(s)$ referred to as the K-matrix and related to observable two-particle phase-shifts as $K^{-1}_\ell(s)=q(s)\cot\delta_\ell$. The key point is that due to multi-valuedness of the square root function, also the partial wave amplitude is in general multivalued. The energy at which the multi-valuedness first occurs (lowest energy at which participating states first go on-shell) is referred to as the branch point.
Thus, the domain of the function $T_\ell(s)$ is extended from $s\in\mathds{C}$ to $s\in\mathds{C}\times\mathds{C}$, representing a complex manifold referred to as the \emph{Riemann surface} consisting of two Riemann sheets, each covering $\mathds{C}$.
Generalizing this, a partial wave amplitude covering $m$ possible intermediate two-body channels is a complex-valued function on a $2^m$ sheeted Riemann surface. Three examples of a one-, two- and three-channel problem are shown in Fig.~\ref{fig:RiemannSheets}. There, individual sheets are shown together with a diffeomorphic mapping of those to a three-dimensional manifolds of gender zero, one and three, respectively. The latter mapping demonstrates clearly how all $2^m$ Riemann sheets are connected to one-another. Any measured (experimentally or as a result of a numerical lattice calculation) values are located on the real energy-axis of the so-called physical sheet, see the green-shaded sheet denoted by $[+\ldots+]$ referring to ${\rm Sgn}(\Im(q(s)))$ in each two-body channel. Generalizing this notation, we denote all sheets by the same type of sequence, see, e.g., Refs.~\cite{Cieply:2011fy, Cieply:2016jby}. Specifically, the unphysical sheet connected to the physical one between the first and second threshold is denoted by $[-+\ldots+]$, the one connected to the physical sheet between the second and third threshold by $[--+\ldots+]$ etc.\,.

\begin{figure}[t]
\centering
\includegraphics[width=0.9\linewidth,trim=0 5cm 3cm 3cm,clip]{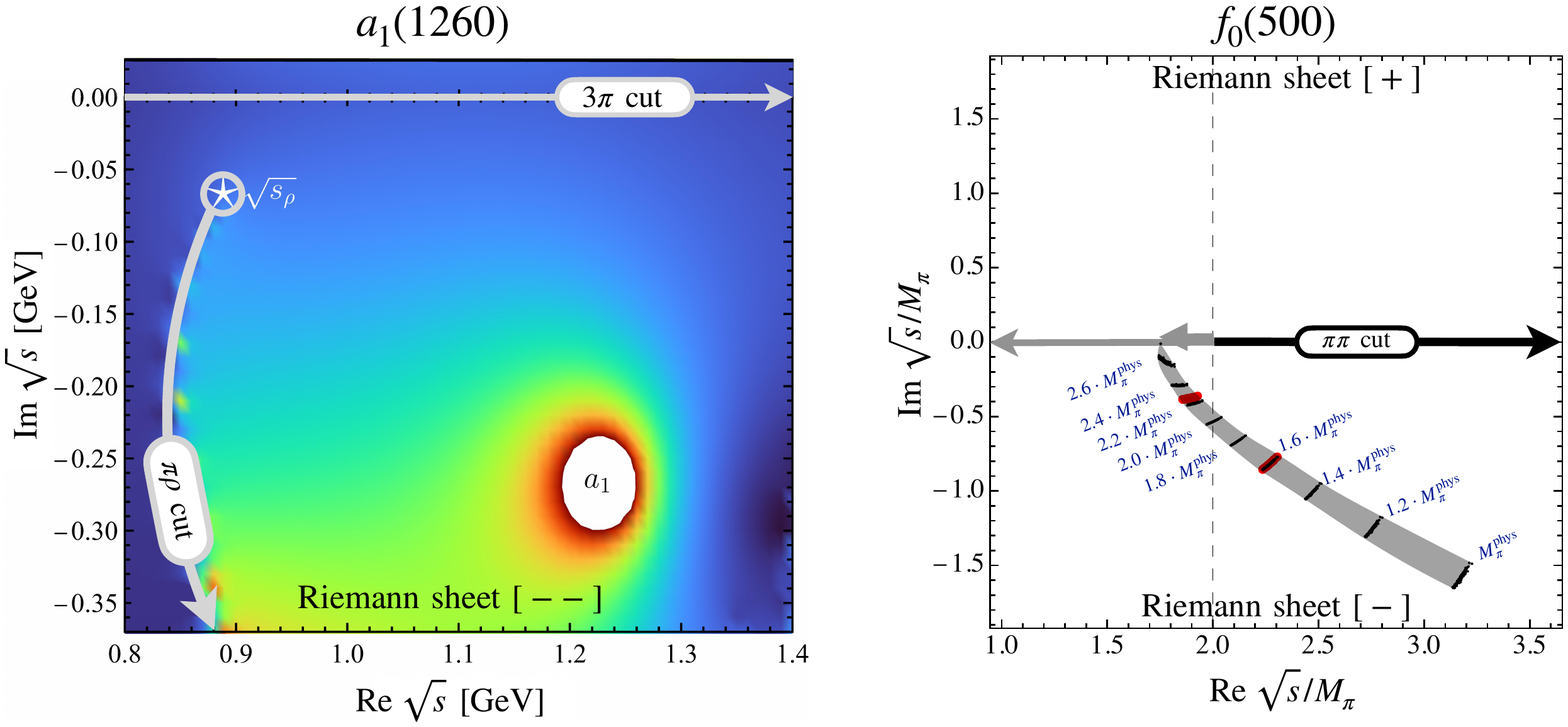}
\caption{Left: Unphysical $[--]$ Riemann sheet of the $3\pi$ system including resonance pole of the $a_1(1260)$ state. Double cut structure arises from three-pions being on-shell as well as two pions forming a $\rho$-resonance.  Figure adopted from Ref.~\cite{Mai:2021nul}.
Right: Pole movement of the $f_0(500)$-resonance as a function of the pion mass in terms of the physical pion mass ($M_\pi^{\rm phys}$). Black dots and gray band represent pole position resampled with respect to the lattice QCD calculations at the unphysical pion masses~\cite{Guo:2016zos, Guo:2018zss, Culver:2019qtx, Mai:2019pqr}.
\label{fig:pirho}}
\end{figure}

Coming back to the matter of resonances, we recall that the $S$-matrix (and with it also the scattering amplitude $T$) is a holomorphic function  -- smooth in a neighborhood of any complex-valued point -- with respect to the discussed complex-valued  (e.g., generalized Mandelstam) variables. The only allowed exception to this is the presence of bound states on the real axis below the production threshold. Furthermore, and as we have seen with explicit fields, excited hadrons or resonances can be associated with poles in the complex plane. Thus, these poles can only be located on the unphysical sheets of the analytically continued $S$-matrix. Typically, only closest to the physical axis located sheets ($[-+\ldots+]$, $[--+\ldots+]$, etc.) are searched for the resonances, as it is assumed that their influence on the physical processes is anti-proportional to the direct (perpendicular to $\Re s$) distance to the real energy axis. Of course, in general more complex situations occur and poles on remote Riemann sheets can be of importance as well, see, e.g., Refs.~\cite{Cieply:2016jby, Jido:2003cb}. So far we have discussed examples with two-body states only.
However, when more particles are present either in the in/out or intermediate states, the picture remains with the exception that if, e.g., two particles subsystems interact resonantly additional branch cuts occur in the complex $s$-plane~\cite{Doring:2009yv}. An example, of such a case is the $3\pi$ system with $I^G(J^{PC})=1^-(1^{++})$ quantum numbers, where two of three pions form a resonant subsystem corresponding to the $\rho$-meson, see, e.g., Refs.~\cite{Mai:2021nul,Sadasivan:2021emk}. This is depicted in the left panel of Fig.~\ref{fig:pirho}.

In summary, the modern approach to resonances utilizes the concept of the $S$-matrix. Induced by the requirement of unitarity (probability conservation), analyticity (causality of physical processes) and crossing symmetry, it is fixed on the real energy-axis by experimental or more recently lattice QCD results. When fixed, it is believed to contain all information about the exited states of hadrons. Specifically, any state is described by a complex-valued position of its pole on a given Riemann sheet. Such parameters are universal, i.e., they do not depend on the choice of a particular transition. Coupling to specific initial and final states is encoded in the residuum of the pole of a scattering amplitude to such states. This interpretation naturally unites the concept of bound states (poles on the real energy-axis on the physical Riemann sheet below production threshold), virtual states (poles on the real energy-axis on the unphysical Riemann sheets) and resonances (complex energy poles on the unphysical Riemann sheets). Interestingly, the transition between different regimes can be tested using lattice QCD calculations at different (unphysical) quark masses. The prominent example in this regard is the long-debated scalar isoscalar mesonic $f_0(500)$, see the recent review~\cite{Pelaez:2015qba}. The pion mass dependence of the $f_0(500)$ pole has been studied extensively with methods based on unitarity and chiral perturbation theory, see, e.g., Refs.~\cite{Dobado:1996ps, Hanhart:2008mx, Guo:2018zss}. The results of a recent study backed by lattice QCD at unphysical pion masses~\cite{Guo:2016zos, Guo:2018zss, Culver:2019qtx, Mai:2019pqr} are shown in the right panel of Fig.~\ref{fig:pirho}.


\section{Theoretical methods I: Lattice QCD}
\label{sec:LQCD}
In this section we introduce Lattice QCD, the formulation of QCD in Euclidean space-time regularized by confining the theory to a finite volume on a discrete hyper-cubic lattice. The main focus of this section is to familiarize the reader with the relevant concepts but also to enable the reader to understand the systematics of this method. More details can be found in textbooks on the topic of lattice field theory.

\subsection{QCD on a lattice}

We recall the renormalizable QCD Lagrangian that conserves parity and
is invariant under time reversal, Eq.~\eqref{eq:L}, for a single quark flavor $q$
\begin{equation}
  \mathcal{L} = -\frac{1}{4}(G_{\mu\nu}^a)^2 + \bar{q}(\mathrm{i}\gamma^\mu
  D_\mu - m_q)q\,.
\end{equation}
Here,
$\bar{q}=q^\dagger\gamma_0$ and
\begin{equation}
G_{\mu\nu}^a = \partial_\mu A_\nu^a - \partial_\nu A_\mu^a + g f^{abc}
A_\mu^b A_\nu^c
\end{equation}
is the field strength tensor. The structure constants of SU$(3)$ are denoted by $f^{abc}$, while $g$ is the bare gauge coupling constant and $m_q$ is the bare quark mass parameter. The the vector potential is denoted by $A$. This theory needs regularisation and renormalization.

For the lattice regularisation we start in Euclidean space-time,
i.e., after analytically continuing to purely imaginary times
$t\to -\mathrm{i} t$. The Euclidean action then reads
\begin{equation}
  S\ =\ \int \mathrm{d}^4x\left[
    \frac{1}{4}(G_{\mu\nu}^a)^2 + \bar{q}(\gamma_\mu D_\mu + m_q) q\right]\,,
\end{equation}
with Euclidean $\gamma$-matrices satisfying
\begin{equation}
  \{\gamma_\mu,\ \gamma_\nu\}\ =\ 2\,\delta_{\mu\nu}\,.
\end{equation}
The Euclidean partition function $\mathcal{Z}$ is then given by
\begin{equation}
  \mathcal{Z}\ =\ \int\mathcal{D}\bar{q}\ \mathcal{D} q\ \mathcal{D}A \mathrm{e}^{-S}\,,
\end{equation}
with the Euclidean action being real and bounded from below. This
allows one to interpret the exponential factor as a probability
weight and enables Monte Carlo integration methods to be applied.

In order to make such a Monte Carlo approach work in practice, the
system is confined into a finite volume with spatial extents
$a_sL_k\,,\ k=1,2,3$, and temporal extent $a_t L_4$. This
hypercube is then discretized with lattice spacings $a_s$ in spatial
and $a_t$ in temporal directions. Denoting here and in the following Euclidean four-vectors by
$\vec x$, the set of lattice sites can be
written as
\begin{equation}
  \Lambda\ =\ \{ \vec n:\quad n_\mu = 0, 1, ..., L_\mu-1\,; \mu=1,2,3,4\}\,.
\end{equation}
In doing so the theory is regularized both in the infrared by the
finite volume and the ultraviolet by the discretization. Typically
one chooses $L_k=L, k=1,2,3$ and $L_4=T$ in lattice simulations, a
notation that we will also adopt where possible.

Furthermore, the fermionic fields $\bar{q}, q$ are defined only on
the lattice sites $\bar{q}(\vec n)\,,\  q(\vec n)$ and the integral
over the volume becomes a sum
\begin{equation}
  \int\mathrm{d}^4 x\quad\to\quad \frac{1}{L_1 L_2 L_3 L_4}\sum_{\vec
    n\in\Lambda}\,,
\end{equation}
and the functional integrals for the Grassmann valued fermionic fields
read
\begin{equation}
  \mathcal{D}\bar{q}\ \mathcal{D} q\quad\to\quad \prod_{\vec n\in\Lambda}\,
  \mathrm{d}\bar{q}(\vec n)\ \mathrm{d} q(\vec n)\,.
\end{equation}
The lattice gauge field is represented by SU$(3)$-valued fields
$U_\mu(\vec n)\equiv U_{\vec n,\mu}$ in the fundamental
representation, which connect lattice sites $\vec n$ and $\vec n +
\hat\mu$, where $\hat\mu$ is the unit vector in direction $\mu$.
This allows one to also define the functional integral over the gauge
fields on the lattice by the transition
\begin{equation}
  \int\ \mathcal{D}A\quad \to\quad \int\prod_{\vec n\in\Lambda,\mu}
  \mathrm{d}U_{\vec n,\mu}\,,
\end{equation}
denoting the invariant Haar measure~\cite{Haar:1933} by $\mathrm{d}U$. The
parallel transports $U_{\vec n,\mu}$ are related to the gauge
potential $A_\mu$ via
\[
U_{\vec n,\mu}\ =\ \mathrm{e}^{\mathrm{i} g a A_\mu(\vec n + \hat\mu/2)}\,.
\]
For the gauge fields usually periodic boundary conditions are used,
though also open boundary conditions are used to mitigate topological
freezing~\cite{Luscher:2012av}. For the fermionic fields boundary
conditions are in general implemented by a phase factor
\begin{equation}
   q(\vec n + L_\mu\hat\mu) = \mathrm{e}^{\mathrm{i}\theta_\mu} q(\vec n)\,,
\end{equation}
with angles $\theta_\mu\in[0,2\pi[\,,\ \mu=1,2,3,4$.

\subsection{Lattice actions}


It remains to discretize the different elements in the Euclidean action, while maintaining gauge invariance. For convenience, we assume
$a_s=a_t=a$ here for the moment. The pure gauge part can be written in
terms of the $U$-fields as follows
\begin{equation}
  S_g\ =\ -\frac{\beta}{3}\sum_{\vec n}\left(b_0\sum_{1\leq\mu<\nu}
  \re\tr\left(U^{1\times 1}_{\vec n, \mu,\nu}\right) +
  b_1\sum_{\mu\neq\nu}\re\tr\left(U^{1\times2}_{\vec n, \mu\nu}\right)\right)\,.
\end{equation}
Here, the bare inverse coupling $\beta=6/g^2$, while $U^{1\times 1}_{\vec n,
  \mu,\nu}$ represents the so called plaquette loop
\begin{equation}
  U^{1\times 1}_{\vec n, \mu,\nu}\ =\ U_{\vec n,\mu}^{~} U_{\vec
    n+\hat\mu,\nu}^{~} U_{\vec n + \hat\nu,\mu}^\dagger U_{\vec n,\nu}^\dagger\,,
\end{equation}
which corresponds to the smallest closed loop in the $\mu-\nu$ plane on a
lattice. $U^{1\times2}_{\vec n, \mu\nu}$ is defined analogously as the
Wilson loop in the $\mu-\nu$ plane with edge lengths $1$ and $2$ in
direction $\mu$ and $\nu$, respectively.

The parameter choice $b_0=1, b_1=0$ corresponds to the original Wilson
gauge action~\cite{Wilson:1974sk}. Up to corrections of order $a^2$
this action reproduces the continuum gauge action. Improved choices
for $b_0, b_1$ have been introduced for instance in
Refs.~\cite{Weisz:1982zw,Iwasaki:1985we} on which most modern lattice
QCD simulations are based on.

For the discretization of the Dirac operator $D_\mu$ there are many
different choices possible, all differing by lattice artefacts and
symmetry properties. The most prominent ones are Wilson's
discretization~\cite{Wilson:1974sk}, the so-called staggered or
Kogut-Susskind discretization~\cite{Kogut:1974ag,Susskind:1976jm}, the
overlap~\cite{Neuberger:1997fp,Neuberger:1998wv} and domain-wall
discretizations~\cite{Kaplan:1992bt,Furman:1994ky}. The latter two
allow for exact chiral symmetry on the
lattice~\cite{Luscher:1998pqa}.

For $a_s\neq a_t$ the gauge action needs to be split into one term
involving only spatial loops and a second one involving space-time
loops, which requires the introduction of inverse couplings $\beta_s$
and $\beta_t$. $\beta$ (and $\beta_t$ and $\beta_s$) are related to
the lattice spacing value(s) via the QCD $\beta$-function. For details we refer to Refs.~\cite{Morningstar:1997ff}.

Most relevant for this review are discretizations based on the Wilson
discretization. The corresponding part of the action for a doublet of
mass-degenerate Wilson type quarks $ q$ -- which we interpret as a
vector in space, time, spin and colour -- reads
\begin{equation}
  \label{eq:Sq}
  S_q = \bar{ q}\left[ D_W(U) + m_0 + \mathrm{i} \mu_q \gamma_5 \tau^3 +
    \frac{\mathrm{i}}{4} \csw \sigma^{\mu\nu}
    \mathcal{F}^{\mu\nu}(U)  \right]  q\,,
\end{equation}
where $\mu_q$ is the bare twisted mass parameter and $\tau^3$ is the third Pauli matrix acting in flavor
space.
Furthermore, $D_\mathrm{W}$ represents the massless Wilson Dirac operator
\begin{equation}
  D_\mathrm{W}(U) = \frac{1}{2}\left[\gamma_\mu(\nabla_\mu^{~} +
    \nabla_\mu^\star) - ar\nabla_\mu^\star\nabla_\mu^{~}\right]\,,
\end{equation}
with $\nabla_\mu$ and $\nabla_\mu^\star$ the gauge covariant forward
and backward difference operators given by
\begin{equation}
\nabla_\mu q(\vec n)\ =\ U_{\vec n,\mu}^{~}\,  q(\vec n + \hat \mu) -
 q(\vec n)\,,\qquad \nabla_\mu^\star q(\vec n)\ =\  q(\vec n) -
U_{\vec n-\hat\mu, \mu}^\dagger\, q(\vec n-\hat\mu)\,,
\end{equation}
respectively.
The Wilson parameter is usually set to $r=1$ and $m_0$ is the bare Wilson mass
parameter. Finally, the so-called Sheikoleslami-Wohlert improvement coefficient \cite{Sheikholeslami:1985ij} is
denoted by $c_{\mathrm{sw}}$ with $\mathcal{F}^{\mu\nu}$ being a discretization of the field strength tensor,
see for instance Ref.~\cite{Jansen:1996yt}.

With $\mu_q=0$ and $\csw=0$ the action \cref{eq:Sq} corresponds to
Wilson's discretization. With only $\mu_q=0$ we have so-called clover
improved lattice fermions~\cite{Luscher:1996ug}, which are with
appropriate tuning of the value of $\csw$ $\mathcal{O}(a)$ improved\footnote{Note that this is on-shell improvement and additional, operator specific improvement coefficients might be required.},
i.e., lattice artefacts are coming at $a^2$, while the pure Wilson
discretization has artefacts proportional to $a$. With $\mu_q\neq 0$
the discretization is called Wilson clover twisted mass
fermions~\cite{Frezzotti:1999vv,Becirevic:2006ii}, which has the
property of automatic $\mathcal{O}(a)$
improvement~\cite{Frezzotti:2003ni} independent of the choice for
$\csw$ for particular choices of $m_0$ and $\mu_q$ (so-called maximal
twist).

The different Wilson type discretizations come with their own
advantages and disadvantages. This is to a large extend a technical
issue which for the scope of the present review merely affects the size of
lattice artefacts. However, also the symmetry properties of the
actions differ, most notably for twisted mass fermions where isopsin
and parity symmetries are broken at the level of $a^2$ lattice
artefacts~\cite{Frezzotti:2003ni}. Moreover, the renormalization patterns
are different: for instance can the pion decay constant $f_\pi$ be
determined without the need of multiplicative renormalization in the
twisted mass formulation, while with Wilson and Wilson clover fermions
this is not the case.

Further diversity in the lattice discretization comes by the usage of
gauge field smearing in the covariant derivative. Smearing is used to
reduce ultraviolet fluctuations in the gauge field and, thus, lattice
artefacts in observables. Typically so-called stout
smearing~\cite{Morningstar:2003gk} or HEX
smearing~\cite{Capitani:2006ni} is used.

Finally, lattice QCD simulations are performed with different numbers
of dynamical quark flavors: $N_f=2$, $N_f=2+1$ and $N_f=2+1+1$ for
simulations with mass degenerate up and down quarks plus strange or
strange and charm quarks. The not active quark flavors are in turn
assumed to be infinitely heavy and, therefore, decoupled. The number of
flavors is particularly important, in the context of this review, to
understand which decay channels are open in a lattice calculation. For example, if
strange quarks are not dynamically simulated, a sufficiently heavy
resonance made from up and down quarks cannot decay into two kaons.

Note that in order to prove positivity of the transfer matrix for
Wilson's discretization anti-periodic boundary conditions in
time for the fermionic fields are required~\cite{Luscher:1976ms}. This, in turn, is
a prerequisite for the so-called Osterwalder-Schrader reflection
positivity, which is mandatory for the analytical continuation from
Euclidean to Minkowski space~\cite{Osterwalder:1973dx,Osterwalder:1974tc}.

\subsubsection{Valence versus Sea Quarks}

Lattice simulations can involve more than the lattice action used for the Monte Carlo simulations: \emph{valence} quarks can be added using a different discretization than the one used for the \emph{sea} quarks. This is called a \emph{mixed action} approach. It is even possible to add valence quarks that are not present in the sea, in case of which one speaks about \emph{partially quenched} simulations.

Working in a mixed action situation can be useful if properties of the valence quarks are required or desirable, which are too resource demanding for the full Monte Carlo simulation. Such properties can be for instance additional symmetries or reduced lattice artefacts. It requires a matching procedure of sea and valence actions: typically one computes one or several hadron masses with only sea and with only valence quarks and tunes the valence parameters until sea and valence hadrons agree (within statistical uncertainties). Typical examples are valence quarks with exact chiral symmetry on top of a sea action with the Wilson or staggered fermion discretization.

A partially quenched configuration is typically used to include strange or charm quarks in the valence sector that are not present in a $N_f=2$ or $N_f=2+1$ sea action. This allows one for instance to study $K$-, $D$- and $D_s$-mesons in lattice QCD simulations with $N_f=2$ dynamical quarks. It is important to realise that valence quarks not present in the sea cannot annihilate. Still in some cases with saturated quantum numbers, such simplified calculations may reflect the physical system, e.g., the recent calculation of $K^-K^-K^-$ scattering~\cite{Alexandru:2020xqf}.

\subsubsection{Heavy Quarks}
\label{sec:heavyLat}

As discussed in the introduction, heavy and light quarks are usually treated differently, which is also the case on the lattice. The main reason is the following: every quark flavor comes with its quark mass parameter $m_q$. In a lattice simulations the relevant quantity is actually the quark mass in lattice units, i.e., $am_q$. In order for the discretization to be meaningful one would argue that $am_q\ll 1$ should be fulfilled. The lattice spacing is typically around $0.1\ \mathrm{fm}$. Thus, for the charm quark $am_c\approx 0.8$, but for the bottom quark $am_b>2$. Even with $a\sim 0.05\ \mathrm{fm}$, which is nowadays used in some simulations, $am_b$ is above the cutoff. On the other hand, dynamical effects of quarks become less and less important with increasing mass.

These are the main reasons why lattice simulations should be performed including at least up and down and strange quarks dynamically. Since $N_f=2$ simulations are easier to perform and require less resources, they were explored before simulating with $N_f=2+1$ dynamical flavors, and the corresponding simulations are still being analysed. For the charm quark it is not entirely clear whether or not sea effects can be neglected. The main reason for not including it in the dynamical simulation are potentially large lattice artefacts of $\mathcal{O}(a m_c)$. However, it is certain that the bottom quark will not be included as a dynamical degree of freedom in lattice simulations in the near future.

We remark here that simulations with Wilson twisted mass quarks at maximal twist can only be performed either in- or excluding a strange/charm doublet. However, so-called automatic $\mathcal{O}(a)$ improvement means that effects of $\mathcal{O}(am_c)$ are absent and the relevant lattice artefacts start at $\mathcal{O}((am_c)^2)$.

The question remains of how to treat charm and bottom quarks in the valence sector when studying heavy-light mesons, where a charm or bottom quark is combined with one of the three light quarks, or when one is interested in quarkonia. There are two strategies:
\begin{itemize}
    \item Treat heavy quarks relativistically. While this appears to become more or less the standard for the charm quark, the bottom requires either an improvement program, see Ref.~\cite{El-Khadra:1996wdx}. Or clever strategies need to be devised~\cite{ETM:2009sed}. In the latter case the idea is to construct suitable ratios with a well defined static limit. Computing these ratios in the charm quark mass region and smoothly interpolating to the static limit allows one to compute $B$-physics observables without being compromised by large systematic uncertainties stemming from $am_b>1$.

    \item Treat heavy quarks in an effective theory. The choices are heavy quark effective theory with expansion parameter $1/m_q$ or non-relativistic QCD with the expansion parameter $v$ denoting the modulus of the four velocity of the heavy quark. HQEFT is appropriate for heavy-light systems, while NRQCD is more appropriate for quarkonia.

\end{itemize}
In both cases, the zeroth-order Lagrangian reads in Euclidean space-time
\begin{equation}
\label{eq:LHQEFT}
    \mathcal{L}_0 = q_+^\dagger(-D_0 + m_q)q_+\,,
\end{equation}
with fields projected to quark (antiquark) fields $q_\pm=(1\pm\gamma_0) q /2$. In this static limit the propagator is just a straight Wilson line. The Lagrangian \cref{eq:LHQEFT} can be systematically improved to include higher order corrections in the respective expansion parameter. We remark here that in \cref{sec:cc}, where we compile recent results, the lattice results have been all obtained by treating the charm quark relativistically.

\subsection{Resonances in a finite volume}

In the previous subsection we discussed that Monte Carlo simulations
are enabled by working in Euclidean space-time. This has certain
consequences for the observables, which can be investigated, even if
Osterwalder-Schrader reflection positivity holds. Generally speaking,
observables connected to real time turn out to be problematic, such as
scattering amplitudes, as was pointed out long ago by Maiani and
Testa~\cite{Maiani:1990ca}.

\subsubsection{Lüscher's method}
\label{sec:Luescher}

A way out was found by Lüscher, who devised a method based on the
works~\cite{PhysRev.103.1565,PhysRev.105.767} now coined as
Lüscher method~\cite{Luscher:1986pf,Luscher:1990ux}. Generally
speaking, the method is based on the observation that the energy
spectrum of the lattice Hamiltonian depends on the volume. And this
dependence on the volume encodes information on the scattering
properties, because with decreasing volume the interaction probability
increases.

In practice the situation is more complicated because there are
different contributions to finite-volume induced energy shifts. In
particular, there are contributions which are for large enough $L$
exponentially suppressed and of the form
$\exp(-M_\pi\cdot L)$~\cite{Gasser:1986vb,Gasser:1987ah}. These
exponential finite-volume effects compete with those only power
suppressed in $1/L$, which are the ones related to infinite-volume
scattering properties.

Lüscher's method is by now well established and applicable for the
case that the exponential finite-volume effects are negligible
compared to the power suppressed ones. A
detailed summary of the two particle formalism can be found
in the recent review articles~\cite{Briceno:2017max,Morningstar:2017spu,Lee:2021kfn}. Most recently, the Lüscher method has been generalized also to the three-body
case~\cite{Polejaeva:2012ut, Mai:2017bge,Mai:2018djl, Mai:2019fba,Culver:2019vvu,Doring:2018xxx, Briceno:2018aml, Blanton:2019igq,Briceno:2019muc,Jackura:2019bmu,Brett:2021wyd,Alexandru:2020xqf,Romero-Lopez:2019qrt, Pang:2019dfe, Blanton:2020jnm, Muller:2020wjo, Muller:2021uur, Hansen:2021ofl, Blanton:2021mih, Blanton:2021eyf,Mai:2021nul,Guo:2020kph, Guo:2020spn, Guo:2020wbl, Guo:2020ikh}, see also recent reviews~\cite{Hansen:2019nir,Mai:2021lwb}.

Instead of writing here the most general formalism, we will focus on
the particular case of scattering of two equal mass mesons below
inelastic threshold. The observable
to be determined from lattice data are phase-shifts $\delta_\ell(k^2)$ for
partial wave $\ell$ as a function of the squared scattering
momentum $k^2$. Lüscher's method is formulated by means of the following
finite-volume quantisation condition
\begin{equation}
  \label{eq:luescher_formula}
  \det \left[\mathcal{M}_{\ell m,\ell'm'}(k^2, L, \pcm) - \delta_{\ell \ell'}
  \delta_{mm'}\,  \cot\left(\delta_\ell(k^2)\right) \right]\ =\ 0\,,
\end{equation}
where the determinant acts in angular momentum space.
The lattice input to this formula is encoded in the analytically known
matrix valued function $\mathcal{M}_{\ell m,\ell'm'}(k^2, L, \pcm)$, which
contains the famous Lüscher $\mathcal{Z}$-function and which is
in general not diagonal in angular momentum. $\mathcal{M}$ depends on
the squared scattering momentum $k^2$ (or equivalently the scattering
energy), the lattice extend $L$ and the total momentum of the two
particle system in the centre-of-mass frame $\pcm$, which is quantized
due to finite volume
\begin{equation}
  \pcm = \frac{2 \pi}{L} \cdot \bm{d} \,, \quad \bm{d} \in \mathbb{Z}^3 \,.
\end{equation}
Given the (lattice) energy level of the two body systems
$E_\mathrm{CM}$ in the centre-of-mass frame, the scattering momentum
is then given by
\begin{equation}
  \label{eq:scatteringk}
  k^2\ =\ \frac{E_\mathrm{CM}^2}{4} - M^2\,,
\end{equation}
where $M$ is the infinite-volume single meson mass value.
Momentum sectors are usually classified\footnote{Note that the first ambiguity in using this nomenclature arises first at $\abs{\bm{d}}^2=\sqrt{2^2+2^2+1}=\sqrt{3^2}$, which is usually too high for any practical lattice calculations.} by $\abs{\bm{d}}^2$. The set
of equivalent momenta is denoted as
\begin{equation}
  \set{\bm{d}} \equiv \set{\bm{z} \in \mathbb{Z}^3 \,, \quad {\bm{z}}^2 = \bm{d}^2}\,.
\end{equation}
It becomes apparent that for each set of values $\{k^2, L, \abs{\bm{d}}^2\}$ one
value of $\delta_l$ can be determined. Therefore, in order to determine
the phase-shift for a dense as possible set of scattering momenta,
significant effort was put into generalising Lüscher's formalism
for different moving frames~\cite{Rummukainen:1995vs,Moore:2005dw,Feng:2010es}. For two particle system the formalism is spelled
out for general spin in Ref.~\cite{Briceno:2014oea}.

In principle, the matrix $\mathcal{M}$ is dense, because angular
momentum is no longer a good quantum number even for the zero-momentum
case, because the continuum rotation groups is broken down to the
octrahedal or cubic group. All infinite-volume angular momenta fall, therefore,
in one of the ten irreducible representations $\Gamma$ of the octahedral
group. When considering also non-zero total momentum of the
multi-particle system, symmetries are further reduced to so-called
little groups (or stabilisers) with corresponding irreducible
representations. Every lattice irreducible representation contains an
infinite tower of continuum angular momenta, but not all.

Still, even this reduced amount of symmetry helps in simplifying
\cref{eq:luescher_formula}: when operators are projected into the
lattice irreducible representations $\Gamma$ (so-called subduction), the matrix
$\mathcal{M}$ becomes block diagonal significantly simplifying the
calculation. We will not go into full details here, since this is
rather technical, but refer to the original
literature~\cite{Bernard:2008ax,Gockeler:2012yj,Dudek:2012gj}.

\subsubsection{The Michael-McNeile method}

This method was developed in Ref.~\cite{McNeile:2002az} and it is
based on the assumption that it is sufficient to consider a two-state
transfer matrix $T$. Let us be specific and write the formalism for
the $\rho$-resonance~\cite{McNeile:2002fh}. Here, the two relevant
states would be a $\rho$ state and a two pion state. The method can easily be
generalized to the $\Delta$-resonance, for instance, see Ref.~\cite{Alexandrou:2013ata}. The two-state
transfer matrix reads then
\begin{equation}
  T = e^{-a\tilde E}
  \begin{pmatrix}
    e^{-a\Delta/2} & a x\\
    a x & e^{-a\Delta/2}\\
  \end{pmatrix}\,,
\end{equation}
with $x=\langle\rho|\pi\pi\rangle$ a transition amplitude. The $\rho$
state is assumed to have energy $\tilde E-\Delta/2$ and the $\pi\pi$ state
$\tilde E+\Delta/2$. The transfer matrix $T$ has eigenvalues
$\lambda = \exp(-aE)$ with
\begin{equation}
E \approx \tilde{E} \pm \sqrt{\frac{\Delta^2}{4} + x^2}\,.
\end{equation}
Alternatively, expressed this differently, the $\rho$ energy $E_\rho$ and the $\pi\pi$
energy $E_{\pi\pi}$ are related to $\tilde E$ and $\Delta$ via
\begin{equation}
  \tilde{E} = \frac{1}{2}(E_\rho + E_{\pi\pi})\,,\quad \Delta =
  E_{\pi\pi} - E_\rho\,.
\end{equation}
Using the expectation, that this $\rho$ state is predominantly created
in a lattice calculation by a quark-antiquark bilinear operator with
the correct quantum numbers, while the two pion state is predominantly
generated by an operator consisting of two bilinears. From these, the
energies $E_{\pi\pi}$ and $E_\rho$. can be measured.

Making the assumption that the energies of the two hadronic states,
here $\rho$ and $\pi\pi$, are close, then the transition amplitude can
be determined from the following ratio of Euclidean correlation
functions~\cite{McNeile:2000xx,McNeile:2002az}
\begin{equation}
  \frac{\langle\rho(0)\ \pi\pi(t)\rangle}{\sqrt{\langle\rho(0)\ \rho(t)\rangle\langle\pi\pi(0)\ \pi\pi(t)\rangle}}
  \approx x t + \textrm{const}\,.
\end{equation}
Then, using Fermi's Golden Rule one can relate $x$ to the width via
\begin{equation}
  \Gamma = 2\pi\langle x^2\rangle \rho(E)\,,
\end{equation}
with $\rho(E)$ the density of states and $\langle x^2\rangle$
indicates the average over spatial directions. $\rho(E)$ can be
estimated, see Ref.~\cite{McNeile:2002fh}.

\subsubsection{The HAL QCD method}

The HAL QCD method was developed in
Refs.~\cite{Ishii:2006ec,Aoki:2009ji,Aoki:2012tk,Aoki:2012bb,HALQCD:2018gyl}.
We will only describe the basic idea here, referring to the original works for more details. The HAL QCD method relies on the
Nambu-Bethe-Salpeter wave function $ q(\bm{r})$, which is used to
define a non-local and energy independent potential $U$ from
\begin{equation}
(E_k - H_0) q(\bm{r}) = \int \mathrm{d}\bm{r}^\prime U(\bm{r}, \bm{
r}^\prime)  q(\bm{r}^\prime)
\end{equation}
below inelastic threshold. Here $E_k = k^2/(2\mu)$ with $\mu$ the
reduced mass and $H_0 = -\nabla^2/(2\mu)$. Details on how to implement
this in the lattice QCD framework can be found in the aforementioned
references.
Note that it is well known from nuclear physics that the used $p/\mu$-expansion is not well converging.

\subsubsection{Optical potential methodology}

Yet another methodology aims in extracting global properties of scattering amplitudes from the finite-volume spectrum without mapping out scattering quantities ($\cot\delta$) for each individual energy eigenvalue. This relatively new approach relies the so-called ordered double limit~\cite{DeWitt:1956be} $(\lim_{{\rm Im}E\to 0+}\lim_{L\to \infty})$, with $E$ denoting the total energy of the system. Such an approach was first introduced in Ref.~\cite{Agadjanov:2016mao}. As demonstrated there on synthetic lattice data, it indeed allows to access scattering amplitudes without usual complication when dealing with multi-channel or multi-particle systems. For related works see Refs.~\cite{Hansen:2017mnd, Guo:2020ikh, Briceno:2020rar,Bulava:2019kbi}. Typically, the price to pay for the universality of such an approach is a much more dense finite-volume spectrum required as an input compared to that of traditional quantization condition methodology, see Sect.~\ref{sec:Luescher}.

\subsection{Lattice Energy Levels}
\label{sec:Elevels}

As pointed out above, the important input from a lattice calculation
are energy levels of single- and multi-particle hadron systems, like
for instance a single pion and two pions. Energy levels are determined
from Euclidean correlation functions $C(t, \bm{p})$, which are
constructed from expectation values
\begin{equation}
  C(t-t^\prime, \bm{p})\ =\ \langle\, \mathcal{O}(t, \bm{p})^\dagger
  \mathcal{O}^\prime(t^\prime, \bm{p})\,\rangle\,,
\end{equation}
where $\mathcal{O}, \mathcal{O}^\prime$ are (multi-)local operators
with certain quantum numbers. By using the time evolution operator,
invariance under translations in time and by
inserting a complete set of energy eigenstates, one can show that these
correlation functions exhibit the following dependency on Euclidean
time $t$
\begin{equation}
  C(t, \bm{p})\ \propto\ \sum_n\ \mathrm{e}^{-E_n(\bm{p})\cdot t}\,.
\end{equation}
This dependence must be modified in the presence of (e.g.,
periodic) boundary conditions, see below.
From the exponential decay of these correlation functions at large
enough Euclidean times the ground state can be extracted, if the
statistical precision allows. However, as mentioned before one needs
to determine as many energy levels as possible to estimate phase
shifts for as many as possible scattering momenta. Therefore, one
applies the so-called generalized eigenvalue
method~\cite{Michael:1982gb,Luscher:1990ck,Blossier:2009kd,Fischer:2020bgv} for which one needs to solve a generalized eigenvalue problem (GEVP).
It consists of defining a suitable list of independent operators
$\mathcal{O}^i(t,\bm{p})$ for $i=1, \ldots, n$ for given quantum
numbers. Using these operators, a correlator matrix
\begin{equation}
  \label{eq:Corr}
  \mathcal{C}_{ij}(t-t^\prime,\bm{p})\ =\ \langle
  \mathcal{O}^i(t,\bm{p})^\dagger\ \mathcal{O}^j(t^\prime, \bm{p})\rangle
\end{equation}
can be computed. Next, one solves the generalized eigenvalue problem
\begin{equation}
  \label{eq:GEVP}
  \mathcal{C}(t, \bm{p})\ \eta^{(k)}(t, t_0)\ =\ \lambda^{(k)}(t, t_0)\
  \mathcal{C}(t_0, \bm{p})\ \eta^{(k)}(t, t_0)
\end{equation}
for eigenvectors $\eta^{(k)}$ and eigenvalues
$\lambda^{(k)}\,,\ k=1,\ldots, n$. For the eigenvalues one can again
show that
\begin{equation}
  \label{eq:lambdat}
  \lambda^{(k)}(t, t_0)\ \propto\ \mathrm{e}^{-E_k(t-t_0)}\,.
\end{equation}
Apart from allowing one to determine more than the ground state (if
statistical precision permits), the generalized eigenvalue method
makes it possible to analytically estimate the residual systematic
effects introduced by using a correlator matrix of finite size while
infinitely many states contribute
theoretically~\cite{Luscher:1990ck,Blossier:2009kd,Fischer:2020bgv}.
\Cref{eq:lambdat} holds for large $t$. The corrections due to
$t\neq\infty$ are of the order $\exp(-\Delta E t)$, where $\Delta E$
depends on the choice of $t_0$. For $t_0>t/2$ one has
$\Delta E = E_{n+1}-E_k$, while otherwise
$\Delta E = \min_{l\neq k}|E_l -E_k|$. Clearly, the former is
favourable, but also often unfeasible. Note that matrix elements can also be computed using the generalized eigenvalue method using then both the eigenvalues $\lambda^{(k)}$ and
-vectors $\eta^{(k)}$.

The operators $\mathcal{O}^k$ are usually constructed by resembling
the quark content of hadronic states. For mesons a single-hadron state
is usually constructed from $h(x)\propto \bar{q} D \Gamma  q(x)$,
with $ q$ quark fields and
where $D$ and $\Gamma$ represent here generically one or several
covariant derivatives and $\gamma$-structures, respectively, which
need to be adjusted to match the desired quantum numbers. For baryonic
states the quark content needs to be adjusted accordingly.
Single-hadron operators with certain fixed momentum values $h(t,
\bm{p})$ can constructed from the $h(x)$ by Fourier transforming.
Multi-hadron operators can then be constructed from the single-hadron operators straightforwardly.

Estimating all the correlations needed for the construction of the
correlator matrix \cref{eq:Corr} requires significant computational
resources and different methods have been designed to reduce this
effort as much as possible. A widely used method is dubbed
\emph{distillation}~\cite{HadronSpectrum:2009krc,Morningstar:2011ka},
which makes it particularly easy to build a large operator basis.

\subsection{Scale Setting and Renormalisation}

Lattice QCD simulations are performed with very few relevant input
parameters: the bare quark masses and the inverse square coupling
$\beta$. In particular, the lattice spacing $a$ is not an input
parameter, but must be fixed in the so-called scale setting procedure:
for each quark mass parameter and the lattice spacing one physical
observable like a hadron mass, a decay constant or ratios thereof are
needed. In principle, arbitrary choices for scale setting quantities
are possible, but different choices will only affect lattice
artefacts. For more details and an overview we refer to
Ref.~\cite{Sommer:2014mea}.

However, it is important to point out that lattice results obtained at
finite lattice spacing by different groups are not readily comparable:
they might differ by lattice artefacts. This needs to be kept in mind
when lattice data is interpreted. In principle, only continuum
extrapolated results can be compared reliably.

The Lüscher method formulated above represents a finite-volume
method, but it does not include effects from discretising space-time.
Thus, the finite-volume lattice energy levels $E(L, a)$, which
are used as input to the Lüscher method, differ from their continuum
counter parts $E(L)$ by lattice artefacts
\[
E(L, a)\ =\ E(L) + \mathcal{O}(a^m)\,,
\]
of a certain order $m\geq 1$, usually $m=1$ or $m=2$ depending on the
lattice action used. In principle, one would need to perform
an extrapolation of the finite-volume energy levels to the continuum
limit $a\to 0$ first and then apply the Lüscher method. However, this
is impractical as it would require to take this limit at fixed value
of $aL$ and fixed values of the renormalized quark masses. Therefore,
for small enough lattice spacing values one assumes the following series
one can expand
\[
\mathcal{M}(E(L, a))\ =\ \mathcal{M}(E(L)) + \mathcal{O}(a^m)\,,
\]
which means in practice one extracts phase-shifts equal to their
continuum counterparts only up to lattice artefacts.

One more comment is in order here: since QCD is a quantum field theory,
observables may require multiplication by renormalisation constants
$Z(1/a)$ when the cutoff $1/a$ is removed in order to remove
divergencies. This is not relevant for the Lüscher method because
energy levels do not require renormalisation since they are
eigenvalues of the lattice Hamiltonian.

\subsection{Systematic Uncertainties}
\label{subsec"systematic_uncertainties}

Since lattice QCD simulations are based on Monte Carlo methods, all
lattice results come with statistical uncertainties. However,
when interpreting lattice QCD results one also should always keep the
possible systematic uncertainties in mind, most importantly those that
are not quantified by the authors. In general, a complete lattice
calculation should control the following effects
\begin{itemize}
\item \emph{lattice artefacts:}~the discretization needs to be
  removed by taking the limit $a\to0$. Depending on the lattice action
  used, leading lattice artefacts are of order $a$ or order $a^2$.

  If such an extrapolation cannot be performed, because there are only
  fewer than three lattice spacing values available, or the range in
  lattice spacing values is too small, one can still estimate the
  effects parametrically. With $a$ being a length scale, there are
  certain natural scales available in QCD which can be combined with
  $a$ to a dimensionless combination: firstly, there is
  $\Lambda_\mathrm{QCD}$ and correction of order
  $(a\Lambda_\mathrm{QCD})^m$ can be expected with $m=1$ or $m=2$. If   both, $a$ and $\Lambda_\mathrm{QCD}$ are known, the order of the
  expected effect can be computed. Other dimensionless combinations
  are $am_q$ with $m_q$ the light, strange or charm quark mass. In
  particular the combination with the charm quark mass $am_c$ can be
  quite sizable.

\item \emph{finite-volume effects:}~the dependence on the finite
  volume needs to be investigated and in principle an extrapolation to
  the thermodynamic limit, i.e., infinite volume, needs to be
  performed. finite-volume effects strongly depend on the quantity
  under investigation and often guidance from effective field theory is available~\cite{Gasser:1986vb,Gasser:1987ah}.

\item \emph{extrapolation to physical pion mass:}~often lattice QCD
  simulations are performed at unphysically large values of the pion
  mass. If this is the case an extrapolation to physical pion mass
  value needs to be performed. For this there is often guidance from
  chiral perturbation theory~\cite{Gasser:1983yg,Gasser:1984gg} available.

\end{itemize}
The Flavour Lattice Averaging Group (FLAG) has developed guidelines
-- so called quality criteria -- which can be found in
Ref.~\cite{Aoki:2021kgd}. They might help to judge the reliability of
a given lattice QCD calculation. Apart from these general systematic effects, there are systematic effects particular to the calculation of hadronic resonances from
lattice QCD, which we quickly discuss in the following:
\begin{itemize}
    \item As mentioned before exponential finite-volume effects must in principle be negligible compared to the power suppressed ones in order to apply Lüscher's method. This requires a
    delicate balance, because for larger $L$ also the desired effects
    become smaller quickly. Moreover, there are ways to reduce the
    influence of such exponential finite-volume
    effects~\cite{Albaladejo:2012jr,Romero-Lopez:2018rcb}.

    \item We have discussed above that for the generalized eigenvalue
    method a list of appropriate operators must be chosen. The actual
    choice is important for several reasons. First of all the operators
    need to have overlap with the desired states. It turned out to be
    important to include both single- and multi-hadron operators,
    discussed for instance in  Refs.~\cite{Wilson:2015dqa,Briceno:2017max}. Otherwise, energy
    levels might be missed leading to a wrong interpretation of the
    lattice results, as shown explicitly for $\pi N$ scattering in the Roper channel in Ref.~\cite{Padmanath:2017oya}.

    \item The second point related to the generalized eigenvalue method and
    the operator choice is the estimation of energies based on the
    Euclidean time  dependence of the eigenvalues \cref{eq:lambdat}. For
    too small $t$-values residual excited state contaminations due to
    the finite size of the correlator matrix \cref{eq:Corr} make a reliable
    determination of the energy level impossible. On the other hand, at
    too large Euclidean times statistical noise is generally
    exponentially increasing. Moreover, with periodic boundary
    conditions in time there are so-called thermal pollution effects
    distorting the signal at large Euclidean time~\cite{Feng:2010es}.

\end{itemize}

\subsection{Example: The \texorpdfstring{$\rho$}{}-resonance}

We will close this section with the example of the
$\rho$-resonance. More precisely, we consider the $\rho^0$ decaying
into $\pi^+\pi^-$ pair. We will discuss the simplest possible case with
only P-wave contributions included and all higher partial waves
neglected. For the $\rho$-resonance this appears to be a relatively
good approximation. Moreover, we concentrate on the elastic region
with energy levels between $2M_\pi$ and $4M_\pi$.

The most basic set of fermionic interpolating operators can be constructed
from two types of operators. First, a single $\rho^0$ interpolator
\begin{equation}
  \mathcal{O}_{\Gamma}(t, \bm{x}) = \frac{1}{\sqrt{2}} (\bar{u}(t,
  \bm{x})\, \Gamma\, u(t, \bm{x}) -
  \bar{d}(t, \bm{x})\, \Gamma_{\alpha\beta}\, d(t, \bm{x}))\,,
  \label{eq:RhoBilinear}
\end{equation}
with $\Gamma\in\{\mathrm{i}\gamma_i, \gamma_0\gamma_i\}$. This operator
projects to an isospin $|1, 0\rangle$ state with quantum numbers
$J^{PC}=1^{--}$. Second, a two pion operator projecting to isospin
$I=1$
\begin{equation}
  \mathcal{O}_{\pi\pi}(t, \bm{x}_1, \bm{x}_2)
  = \frac{1}{\sqrt{2}} \left[ \mathcal{O}_{\pi^+}(t, \bm{x}_1)\, \mathcal{O}_{\pi^-}(t, \bm{x}_2)
  - \mathcal{O}_{\pi^-}(t, \bm{x}_1)\, \mathcal{O}_{\pi^+}(t, \bm{x}_2)
  \right]\,.\\
  \label{eq:operator_PgpPgp}
\end{equation}
Here, we have used
\begin{equation}
  \mathcal{O}_{\pi^+}(t, \bm{x}) = \bar{d}(t, \bm{x})\, \gamma_5 u(t, \bm{x}) \,,\qquad
  \mathcal{O}_{\pi^-}(t, \bm{x}) = \bar{u}(t, \bm{x})
  \gamma_5 d(t, \bm{x}) \,.
  \label{eq:operator_Pgp}
\end{equation}
Each of these operators can be projected to definite momentum via a
Fourier transformation
$\mathcal{O}(t, \bm{p}) = \sum_{\bm{x}} \mathcal{O}(t,\bm{x})\,\exp(\mathrm{i}\bm{x}\bm{p})$.

These operators are actually sufficient to study the case of zero
total momentum $\bm{p}_\mathrm{cm} = 0$. Since we are interested in
the P-wave case, the two pions then need to have opposite equal non-zero momentum. One then builds a correlator matrix for
instance from the operators
$\mathcal{O}^1=\mathcal{O}_{\mathrm{i}\gamma_1}(t,\bm{p}=0)$,
$\mathcal{O}^2=\mathcal{O}_{\gamma_0\gamma_1}(t,\bm{p}=0)$, and
$\mathcal{O}^3 = \mathcal{O}_{\pi\pi}(t, \bm{p}_1,
\bm{p}_2=-\bm{p}_1)$ with, e.g., $\bm{p}_1=2\pi/L(1,0,0)$. More
operators can be included with larger modulus of $\bm{p}_1$ to increase
the correlator matrix. Moreover, one can include operators with
$\gamma_2$ and $\gamma_3$ for the single particle operator.

This correlator matrix is used to solve the
generalized eigenvalue problem \cref{eq:GEVP} and determines the
interacting energy levels $E_\mathrm{cm}$ of the two pion system.
At this point one needs to take care of so-called thermal pollution
when working with periodic boundary conditions. In general, the
leading contribution from thermal states to this two pion system reads
\begin{equation}
    \varepsilon_t(t, \bm{p}_1, \bm{p}_2)\ \propto\
    e^{-E_{\pi}(\bm{p}_1) T}\, e^{-(E_{\pi}(\bm{p}_2) -E_{\pi}(\bm{p}_1)) t}\
    +\ e^{-E_{\pi}(\bm{p}_2) T} e^{-(E_{\pi}(\bm{p}_1) -E_{\pi}(\bm{p}_2)) t}\,.
\end{equation}
Here, $E_\pi(\bm{p})$ corresponds to the single pion energy at momentum $\bm{p}$.
This term comes about because in the two particle system, one of the two pions
can propagate via the boundary. For the special case we are discussing here,
we have $\bm{p_1}=-\bm{p_2}$, and thus $\varepsilon_t\propto \exp(-E_\pi(\bm{p_1}T))$
is independent of $t$. However, depending on the value of $T$, the constant will
distort the form of the correlator for $t$-values around $T/2$. The constant can
be removed by considering the discrete derivative in Euclidean time of the correlator
matrix instead of the correlator matrix itself
\[
\tilde{\mathcal{C}}(t) = \mathcal{C}(t+1) - \mathcal{C}(t)\,.
\]
The element of $\tilde{\mathcal{C}}$ will have a $\sinh$ ($\cosh$) form in
time, if it was $\cosh$ ($\sinh$) in $\mathcal{C}$. But the constant will be
absent in $\tilde{\mathcal{C}}$. In addition, taking this difference can
reduce correlation of different time-slices significantly, see, e.g., Ref.~\cite{Ottnad:2017bjt,Dimopoulos:2018xkm}.

In general, the thermal pollution contributions is time-dependent, calling for more sophisticated measures, which can be
found in the literature. Most importantly, one may use \emph{weighting and shifting}~\cite{Dudek:2012gj}, where one first divides by the (known)
time-dependent part in the pollution (weighting), then applies the
derivative from above (shifting). Other possibilities include the usage
of appropriate ratios~\cite{Helmes:2018nug}, which are, however, often not
compatible with the GEVP.

The pion mass $M_\pi$ can be determined directly from the Euclidean
correlation functions of the operators
$\mathcal{O}_{\pi^\pm}(t,\bm{p})$ without solving a GEVP.

We have restricted ourselves to very few operators for this example. Of
course, there are many more operators which can be included in the
list. These become typically more difficult to construct, because they
contain for instance derivatives. However, they will improve on the
one hand the accuracy to which the energy levels can be estimated. And
on the other hand more energy levels become accessible.

\begin{figure}
  \centering
  \includegraphics[width=0.7\linewidth,trim=0 4.5cm 7.cm 4.2cm, clip]{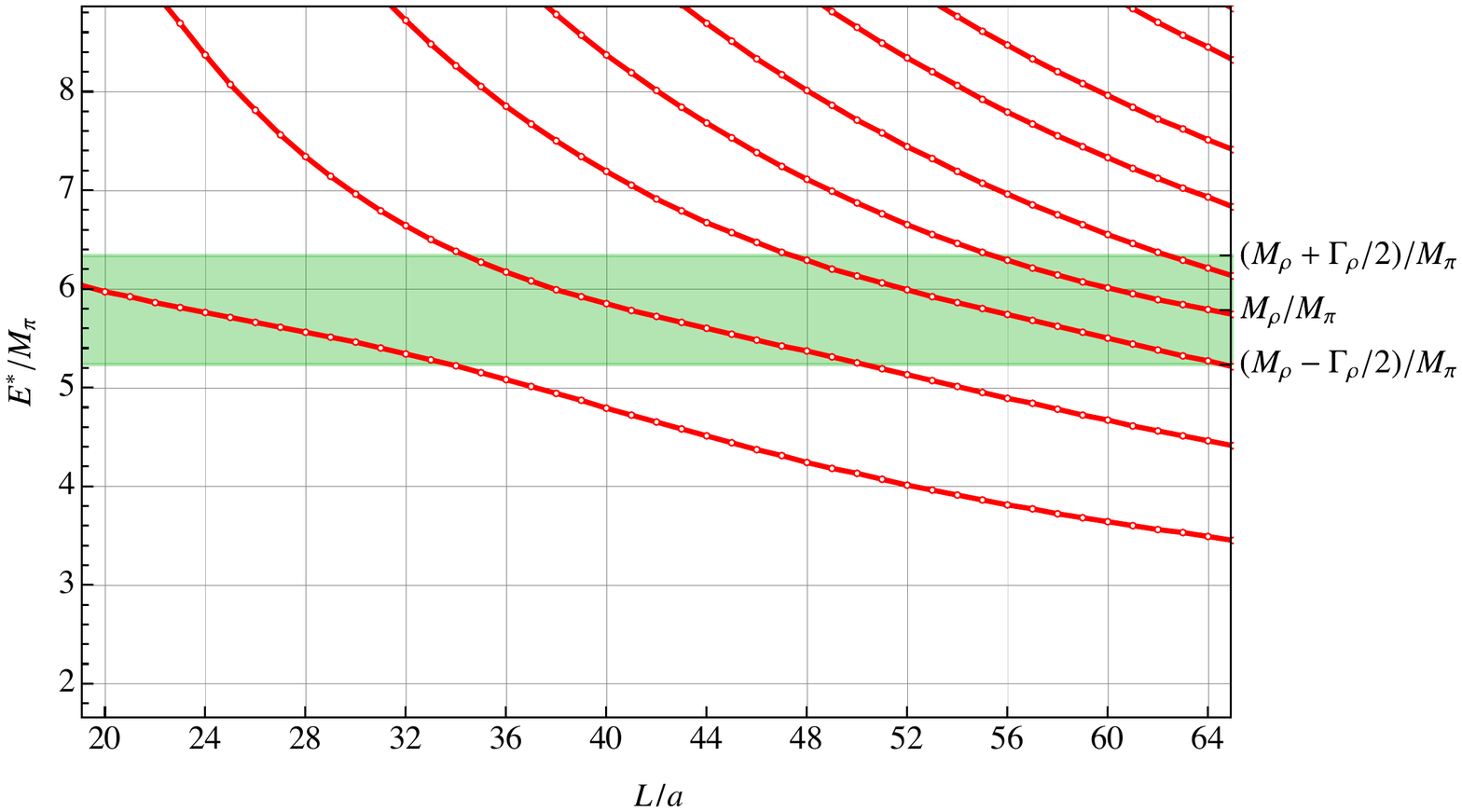}
  \caption{Energy spectrum of a $I=1$, $\ell=1$, $\pi\pi$ system at physical pion mass in a finite cubic volume of side length $aL$ ($a=0.1~{\rm fm}$). The spectrum is obtained from Eq.~\eqref{eq:luescher_formula} using Breit-Wigner parametrization of the phase-shifts with PDG parameters~\cite{ParticleDataGroup:2020ssz}. Periodic boundary conditions are applied and projection to the $T_{1g}$ irrep are performed.}
  \label{fig:ALCrho}
\end{figure}

For the zero total momentum case considered here the Eq.~\eqref{eq:luescher_formula} reduces to
\begin{equation}
  \label{eq:rhoLuescher}
  \cot(\delta_1) = \frac{\mathcal{Z}(1; q^2)}{\pi^{3/2} q}\,,\qquad
  q = k \frac{L}{2\pi}\,,
\end{equation}
with $k(E_\mathrm{cm})$ defined in Eq.~\eqref{eq:scatteringk} with $M\equiv
M_\pi$ and the Lüscher zeta function
\begin{equation}
  \mathcal{Z}(1; q^2) = \frac{1}{\sqrt{4\pi}}\sum_{\vec d\in\mathbb{Z}^3}
  \left(|\vec d|^2 - q^2\right)^{-s}\,.
\end{equation}
Now, one is left to solve Eq.~\eqref{eq:rhoLuescher} numerically for the
phase-shift $\delta_1$ with $E_\mathrm{cm}$ and $M_\pi$ as input. Obviously, the procedure can be inverted, predicting the finite-volume spectrum, assuming a specific form of the $\pi\pi$ interaction determining the left-hand-side of Eq.~\eqref{eq:rhoLuescher}. For the simple case of Breit-Wigner parametrization discussed in the beginning of this review, the predicted spectrum is depicted in Fig.~\ref{fig:ALCrho}. The distinctive feature of the so-called avoided level crossing is evident there~\cite{Wiese:1988qy}.


\section{Theoretical methods II: EFTs for resonances}
\label{sec:EFT}
Chiral perturbation theory (CHPT) is the low-energy effective field theory (EFT) of QCD
\cite{Weinberg:1978kz,Gasser:1983yg,Gasser:1984gg}. First and foremost, it is
the theory of the Goldstone bosons, the pions in the two-flavor ($u,d$) case and
the pions, the kaons and the eta in the three-flavor ($u,d,s$)
sector. It has enjoyed considerable successes, see
Refs.~\cite{Meissner:1993ah,Ecker:1994gg,Pich:1995bw,Bernard:2006gx,Bijnens:2006zp,Holstein:2008zz,Hermansson-Truedsson:2020rtj}
for reviews. The Goldstone bosons also couple to matter fields, in particular
the nucleons (protons and neutrons) in the SU(2) case or the low-lying baryon octet
($N,\Lambda,\Sigma,\Xi$) for three flavors, see, e.g.,
Refs.~\cite{Gasser:1987rb,Jenkins:1990jv,Bernard:1992qa} for early works and
Refs.~\cite{Bernard:1995dp,Bernard:1995dp,Scherer:2009bt,Geng:2013xn} for reviews.
A cornerstone of CHPT is the power counting, which we briefly discuss here.
Symbolically, any matrix element admits an expansion in small momenta/energies/masses $Q$
over the hard scale $\Lambda$ of the form
\begin{equation}
{\cal M} = \sum_\nu \left(\frac{Q}{\Lambda}\right)^\nu f_\nu(Q/\mu,g_i)~,
\end{equation}
where $\mu$ is a regularization scale (often the scale of dimensional regularization),
the $g_i$ are low-energy constants (LECs), the $f_\nu$ are functions of order one (``naturalness'')
and the index $\nu$ is bounded from below, which leads to a systematic and
controlled expansion. In case of pure Goldstone boson interactions, $\nu=2$ is the
smallest possible value due to the derivative nature of the interactions. This expansion
can be mapped onto a well-defined quantum field theory with tree and loop graphs,
where the infinite part of the LECs absorb the UV infinities generated by the
loop diagrams at a given order.

Since CHPT is an EFT, its range of applicability is limited to  momenta and
energies below some hard scale. This scale of chiral symmetry breaking $\Lambda_\chi$ is
often identified as $\Lambda_\chi = 4\pi F_\pi$, with $F_\pi \simeq 92\,$MeV the pion
decay constant, thus, $\Lambda_\chi \simeq 1.2\,$GeV~\cite{Manohar:1983md}. However,
the true limitation  to CHPT sets in earlier and is channel-dependent, related to the
appearance of certain  {\em resonances} with appropriate quantum numbers. Prominent
examples are the broad $f_0(500)$ for pion-pion interactions in the channel with
$\ell=I=0$, with $\ell$ and $I$ denoting the total angular momentum and isospin of the
two-pion system, respectively,
the much narrower $\rho(770)$ for $\ell=I=1$ or the lowest-lying resonance in
pion-nucleon scattering, the $\Delta(1232)$. Of course, the appearance of such
resonances is not restricted to the light quark sector, but leads to similar limitations
in EFTs involving the heavy $c,b$ quarks, as will be discussed later.

Therefore, we need to extend CHPT to cope with resonances. This can be done in two
ways. First, one can construct EFTs with explicit resonance fields, see \cref{sec:explicit}.
The main obstacle here is the fact that resonances decay and it is not trivial to write
down a consistent  power counting that accounts for the different momentum/energy
scales that necessarily appear. Second, one can use unitarization methods, which amount
to a resummation of the chiral expansion similar to the geometric series, that is the
expansion of the form $1+x$ is substitute by $1/(1-x)$, which allows for the generation
of resonances. This was first done in Refs.~\cite{Truong:1988zp,Dobado:1989qm}
and critically re-examined in~\cite{Gasser:1990bv}. The upshot is that
such a unitarization procedure induces some model-dependence,
as discussed in \cref{sec:uni}. In a few selected cases, one can combine the
chiral expansion with dispersion relations to extract resonance properties, which
we will not consider in detail here but rather refer to
\cite{Caprini:2005zr,Pelaez:2015qba,Pelaez:2021dak} (and references therein).

\subsection{EFTs with explicit resonance fields}
\label{sec:explicit}

\begin{figure}[t!]
\centering
\includegraphics[width=0.5\linewidth]{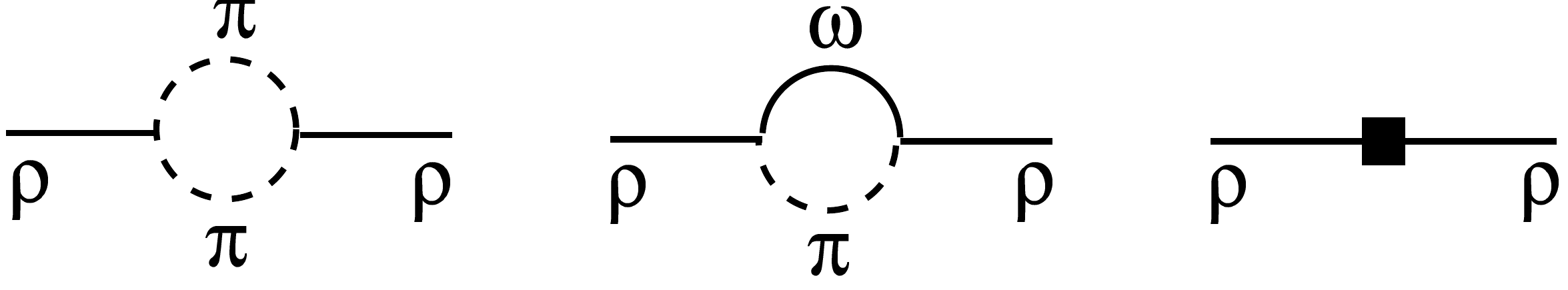}
\caption{Leading one-loop contributions to the $\rho$-meson self-energy.
         The filled square denotes a counterterm. In the complex mass scheme,
         the latter is a complex-valued quantity.
  \label{fig:rhoSE}}
\end{figure}

Before addressing the ways of explicitly including vector mesons, let us
discuss the problems with the power counting that arises for unstable particles.
As our prime example, we take the $\rho\to \pi\pi$ decay and follow the
arguments given in~\cite{Bruns:2004tj}.
Consider the leading one-loop correction to the vector meson mass given by the
leftmost diagram in Fig.~\ref{fig:rhoSE}. Denote the integral corresponding to
this self-energy contribution by $I$. It can not straightforwardly be calculated because
the large vector meson mass $M_V$ obscures the power counting as in the case of the
nucleon~\cite{Gasser:1987rb}.  However, the integral can be split into a ``soft'' and
a ``hard'' part~\cite{Ellis:1997kc}, such that $I= I_{\rm soft}+I_{\rm hard}$, where the
soft part is entirely generated by small momenta $Q\ll M_V$ and, thus, is consistent
with the power counting, whereas the hard part with momenta $Q\simeq M_V$ leads to the
breaking of the power counting and needs to be treated separately. In a more
formal language, this corresponds to  the method of dimensional counting \cite{Gegelia:1994zz}
or the strategy of regions \cite{Beneke:1997zp}. In fact, the soft parts of any diagram
satisfy the expected power counting. Working out  $I_{\rm soft}$, one finds that the result
corresponds to a series of tadpole graphs, involving only one Goldstone boson propagator,
scaling as ${\cal O}(Q^3)$. This can of course not be the whole story, because the amplitude of
Fig.~\ref{fig:rhoSE} (leftmost diagram) has an imaginary part due to the production of two
Goldstone bosons in the intermediate state, while the tadpole sum does not have such an
imaginary part. In order to take only $I_{\rm soft}$ as the regularized amplitude, one would
have to write complex coefficients in the effective Lagrangian, which in general, one
does not want to do (for an exception, see \cref{sec:CMS}).
A direct calculation of the full scalar loop integral shows that the imaginary part indeed does
not satisfy the power counting mentioned above, i.e, it does not scale $\sim Q^3$ .
This is related to the fact that for large external four-momenta squared, $P^2$, of
the heavy external particle, the Goldstone bosons produced in the decay of this particle are
not to be considered as soft. Below the threshold, we have $P^2 < 4M_\pi^2$,
so $P^2$ can not be considered as being very large compared to the scale $M_\pi^2$ in that
region, and we would have to take the full integral $I$ as the soft part, and not just
$I_{\rm soft}$. This phenomenon of the ``missing imaginary part'' was also pointed out in the
framework of heavy meson effective theory in Ref.~\cite{Bijnens:1997rv}. For other approaches
to include vector mesons in chiral EFTs, see, e.g.,
\cite{Jenkins:1995vb,Bijnens:1996nq,Bijnens:1997ni,Bijnens:1997ni,Ruiz-Femenia:2003jdx,Rosell:2004mn,Lutz:2008km,Bruns:2008ub,Terschlusen:2012xw}. The case of baryon resonances will be discussed in \cref{sec:EFTbaryons}.

\subsection{The complex mass scheme}
\label{sec:CMS}
The complex-mass renormalization scheme  is a method that was
originally introduced for precision $W,\,Z$-physics, see, e.g.,~\cite{Stuart:1991xk,Denner:1999gp}
and later transported to chiral EFT~\cite{Djukanovic:2009zn}.
Let us first give a brief outline of the complex-mass scheme (CMS), following
Ref.~\cite{Denner:2006ic}. Consider first an instable particle at tree level. The CMS
amounts to treating the mass of this particle
consistently  as a complex quantity, defined as the location of the pole in
the complex $k^2$-plane of the corresponding propagator with momentum $k$. It can be shown that
this scheme is symmetry-preserving and leaves the corresponding Ward identities intact.
Extending  this to one loop, one splits the real bare masses into complex renormalized
masses and complex counter\-terms. This is important, as only renormalized masses are observable.
The corresponding Lagrangian yields Feynman rules with complex masses and counterterms, which allows for standard perturbative calculations. This is essentially a rearrangement of contributions that is not affected by double counting. The imaginary part of the
particle mass appears in the propagator and is resummed in the Dyson series. In contrast to this, the imaginary part of the counterterm is not resummed. One can show that in such a case
gauge invariance remains valid, and unitarity cancellations are respected order by order
in the perturbative expansion. This also requires  integrals with complex internal masses, as
worked out in Ref.~\cite{Beenakker:1988jr}. For further discussions\
of the method, the reader is referred to Refs.~\cite{Denner:2006ic,Denner:2014zga} (and references therein).
In case of a chiral EFT, the perturbative expansion proceeds as usual in terms of small
momenta and quark masses,  with a proper treatment of the heavy particle mass in loop
diagrams (like the heavy-baryon scheme~\cite{Jenkins:1990jv,Bernard:1992qa}
or the so-called infrared-regularization~\cite{Becher:1999he}
or the extended-on-mass-scheme discussed below~\cite{Fuchs:2003qc}).

Let us now calculate the mass and the width of the $\rho$ within the CMS to leading
one-loop order ${\cal O}(Q^3)$, following Ref.~\cite{Djukanovic:2009zn}.
The pertinent Lagrangian is given by \cite{Gell-Mann:1962hpq,Meissner:1987ge,Bando:1987br,Ecker:1989yg}
\begin{eqnarray}
{\cal L} &=& {\cal L}_{\pi}+{\cal L}_{\rho\pi}+{\cal L}_\omega +{\cal L}_{\omega\rho\pi}+\ldots~,\nonumber\\
{\cal L}_\pi + {\cal L}_{\rho\pi} &=& \frac{F^2}{4}\, \langle\partial_\mu U (\partial^\mu U)^\dagger\rangle
+\frac{F^2\,M^2}{4}\,\langle U^\dagger+U\rangle
-\frac{1}{2}\,\langle\rho_{\mu\nu}\rho^{\mu\nu}\rangle + \left[ M_{\rho}^2 +
\frac{c_{x}\,M^2}{4}\,\langle U^\dagger+U\rangle \right]\, \langle \hat{\rho}_\mu
\hat{\rho}^{\mu}\rangle~,\nonumber\\
{\cal L}_\omega + {\cal L}_{\omega\rho\pi}  &=& -\frac{1}{4} \omega_{\mu\nu}\omega^{\mu\nu} +\frac{1}{2}M_{\omega}^2\,
\omega_{\mu}\omega^\mu + \frac{1}{2}\,
g_{\omega\rho\pi}\,\epsilon_{\mu\nu\alpha\beta}\, \omega^{\nu}\, \langle\rho^{\alpha\beta} u^\mu\rangle~,
\label{eq:Lrpw}
\end{eqnarray}
where
\begin{eqnarray}
U&=&u^2={\rm exp}\left(\frac{\mathrm{i}\vec{\tau}\cdot\vec{\pi}}{F}\right)~,~~
\Gamma_\mu = \frac{1}{2}\,\bigl[ u^\dagger\partial_\mu u+u\partial_\mu u^\dagger\bigr]~,~~
u_\mu=  \mathrm{i} \left[ u^\dagger \partial_\mu u-u \partial_\mu u^\dagger \right]~, \nonumber\\
\rho^\mu & = & \frac{\vec\tau\cdot\vec\rho\,^\mu}{2}~,~~ \rho^{\mu\nu}  =
\partial^\mu\rho^\nu-\partial^\nu\rho^\mu - \mathrm{i} g\left[\rho^\mu,\rho^\nu\right]~,~~
\hat{\rho}_\mu = \rho_\mu -\frac{\mathrm{i}}{g}\Gamma_\mu~,~~ \omega^{\mu\nu} = \partial^\mu\omega^\nu-\partial^\nu\omega^\mu~.
\end{eqnarray}
Here, $F$ is the pion decay constant in the chiral limt, $M^2$ is the leading term in the quark mass expansion
of the pion mass squared, $\langle ... \rangle$ denotes a trace in flavor space, $M_\rho$ and $M_\omega$ refer to the
bare $\rho$ and $\omega$ masses, $g$ is the $\rho$-coupling subject to the constraint $M_\rho^2 = 2\,g^2 F^2$,
the so-called KSFR relation \cite{Kawarabayashi:1966kd,Riazuddin:1966sw}, $g_{\omega\rho\pi}$ parameterizes the
strength of the $\omega\rho\pi$ vertex and $c_x$ is a LEC related to the quark mass expansion of the $\rho$ mass,
which also affects the $\rho\to\pi\pi$ vertex.  Next, one performs standard renormalization, i.e., the bare
parameters (as indicated by a subscript 0) are expressed in terms of the normalized ones and a number
of counterterms, leading to (we only display the ones contributing at leading loop order)
\begin{equation}
\rho^\mu_0 = \sqrt{Z_\rho}\,\rho^\mu~,~~ M_{\rho,0} = M_R + \delta M_R~,~~ c_{x,0}  =  c_x+\delta c_x ~.
\end{equation}
Now, one applies CMS and chooses
\begin{equation}
  M_R^2=(M_\chi - i\, \Gamma_\chi/2)^2~,
\end{equation}
as the pole of the $\rho$-meson propagator in the chiral limit, where the pole mass and the width of the
$\rho$ meson in the chiral limit are denoted by $M_\chi$ and $\Gamma_\chi$, respectively. These are input parameters.
In this scheme, one  includes $M_R$ in the propagator and the counterterms, which are
complex-valued quantities now, are treated perturbatively.
As noted before, the mass of the $\rho$ is not a small quantity, thus, one has to specify a power counting
that accounts for that.  Let $Q$ collectively denote a small quantity.
The pion propagator counts as ${\cal O}(Q^{-2})$ if it does not
carry large external momenta and as ${\cal O}(Q^{0})$ if it does.
The vector-meson propagator counts as ${\cal O}(Q^{0})$
if it does not carry large external momenta and as ${\cal O}(Q^{-1})$ if it does.
The pion mass counts as ${\cal O}(Q^{1})$, the vector-meson mass as ${\cal O}(Q^{0})$,
and the width as ${\cal O}(Q^{1})$.  Vertices generated by the effective Lagrangian of Goldstone
bosons ${\cal L}_\pi^{(n)}$ count as ${\cal O}(Q^n)$.  Derivatives acting on heavy vector mesons, which
cannot be eliminated by field redefinitions, count as ${\cal O}(Q^0)$. The contributions of
vector meson loops can be absorbed systematically in the renormalization of the parameters of the
effective Lagrangian. Therefore, such loop diagrams need not be included for energies
much lower than twice the vector-meson mass. Note also that the smallest order resulting from the various
assignments is defined as the chiral order of the given diagram.

Now we are in the position to evaluate the two-point function (2PF). The mass and width of the $\rho$ meson are
extracted from the complex pole of the 2PF. The 2PF, that is the sum of all one-particle irreducible diagrams,
is parameterized as
\begin{equation}
\Pi^{ab}_{\mu\nu}(p)= \delta^{ab}\,\left[ g_{\mu\nu}\Pi_{1} (p^2)+p_\mu p_\nu\,\Pi_2 (p^2)\right]\,.
\label{eq:VSEpar}
\end{equation}
The dressed propagator, expressed in terms of the self-energy, has the form
\begin{equation}
S^{a b}_{\mu\nu}(p) = -\delta^{ab}\,\frac{g_{\mu\nu}-p_\mu p_\nu \left(1+ \Pi_2(p^2)\right)\left(M_R^2+\Pi_1(p^2)
 +p^2 \Pi_2(p^2)\right)^{-1}} {p^2 -  M_R^2-\Pi_1(p^2)+\mathrm{i}\,\eta}\,,
\label{eq:dressedprop}
\end{equation}
with $\eta \to 0^+$ and the pole of the propagator is found as the (complex) solution to
the equation:
\begin{equation}
z- M_R^2-\Pi_1(z)=0\,.
\label{eq:poleeq}
\end{equation}
In the vicinity of the pole $z$, the dressed propagator takes the form
\begin{equation}
S^{a b}_{\mu\nu}(p) = -\delta^{ab}\left[\frac{Z^r_\rho \left(g_{\mu\nu}-
    {p_\mu p_\nu}/{z}\right)}{p^2-z+\mathrm{i}\,\eta}+R\right]\,,
\label{eq:proppole}
\end{equation}
where $Z^r_\rho={1}/({1- \Pi_1'(z)})$ and $R$ denotes the non-pole part (the remainder).
The counterterms $\delta M_R$ and $\delta Z_\rho$ are fixed by requiring
that in the chiral limit $M_R^2$ is the pole of the dressed
propagator and that the residue $Z_\rho^r$ is equal to one. The solution to Eq.~\eqref{eq:poleeq}
admits a perturbative expansion of the form
\begin{equation}
z=z^{(0)}+ z^{(1)}+z^{(2)}+\ldots\,,
\label{eq:zexp}
\end{equation}
where the superscripts $(i)$ denote the $i^{\rm th}$ loop order. All of these terms further admit a chiral
expansion. For example, the tree level result to third order in the chiral expansion reads $z^{(0)}= M_R^2+c_x M^2$,
which is consistent with the general result for the quark mass expansion of the  vector meson
mass~\cite{Bruns:2004tj}. More interesting is the result at leading one-loop order. The corresponding
contributions to the self-energy are depicted in Fig.~\ref{fig:rhoSE}. The contributions of the pion loop
and the $\pi\omega$ loop to $\Pi_1$ are given by
\begin{eqnarray}
D_{\pi\pi} & = & -\frac{g^2 \mu ^{4-n}}{d-1} \,  \left[2 \,I_{M }-\left(P^2-4
M^2\right) \,I_{MM}\right]\,, \nonumber\label{eq:pipiloop} \\
D_{\pi\omega} & = & \frac{(d-2)\,g_{\omega \rho \pi}^2\,\mu^{4-d}}{4\,
(d-1)}\,\left[ M^4\,I_{MM_{\omega}} -\left(2\,I_{MM_{\omega}}
M_\omega^2+I_M-I_{M_\omega}+2\,I_{MM_\omega} P^{2}\right) M^2\right.\nonumber\\
&& \left.+I_{MM_{\omega}} P^{2 }+M_\omega^2 \left(I_{MM_{\omega}} M_\omega^2+I_M-I_{M_\omega}\right)
-\left(2 I_{MM_{\omega}} M_\omega^2+I_M+I_{M_\omega}\right) P^{2}\right]\,,
\label{eq:piomloop}
\end{eqnarray}
in terms of the loop integrals
\begin{eqnarray}
I_{m_1m_2} & = & \frac{\mathrm{i}}{(2 \pi)^{d}}\,\int
\frac{d^dk}{\left[k^2-m_1^2+\mathrm{i}\,0^+\right]\left[(P+k)^2-m_2^2+\mathrm{i}\,0^+\right]}\,,\nonumber\\
I_m & = & \frac{\mathrm{i}}{(2 \pi)^{d}}\,\int \frac{d^dk}{k^2-m^2+\mathrm{i}\,0^+}\,,
\label{oneandtwoPF}
\end{eqnarray}
with $d$ the number of  space-time dimensions, $\mu$ the scale of dimensional regularization and $P$ the four-momentum
of the vector meson. Note, however, that the $\pi\pi$-loop only starts to contribute at ${\cal O}(Q^4)$
to $\Pi_1$, as the large component is $\propto P^\mu$. Note further that this diagram contains a power-counting
violating imaginary part, which is cancelled by the imaginary part of the complex counterterm, see Fig.~\ref{fig:rhoSE}.
This is the major advantage of the CMS scheme. In the calculation of the $\pi\omega$ loop, one uses $M_\omega = M_\rho$,
which is good approximation ($\rho$-$\omega$ mixing in chiral EFT is discussed in
Refs.~\cite{Urech:1995ry,Gokalp:2003dr,Kucukarslan:2006wk,Chen:2017jcw}, see also the review~\cite{OConnell:1995nse}).
Next, the counterterm contributions are adjusted such that the pole in the chiral limit stays at $M_R$, leading to
\begin{eqnarray}
\delta M_R & = & -\frac{1}{3}\, g^2 M_R \,\lambda + \frac{g^2}{288\,\pi ^2}
M_R \left(-3 \ln(M_R^2/\mu^2) +3\mathrm{i} \pi +5\right) + \frac{1}{3} g_{\omega\rho\pi}^2 M_R^3
\lambda+\frac{g_{\omega\rho\pi}^2}{288\pi^2} M_R^3 \left(3 \ln(M_R^2/\mu^2) +1\right)\,,\nonumber\\
\delta c_x & = & 4g^2 \lambda - \frac{g^2}{8\pi^2} \left( 1 - \ln(M_R^2/\mu^2) + \mathrm{i}\pi \right) +
g_{\omega\rho\pi}^2 M_R^2 \lambda - \frac{g_{\omega\rho\pi}^2}{32\pi^2} M_R^2 \left(1- \ln({M_R^2}/{\mu^2})\right)\,,
\nonumber\\
\lambda &=& \frac{1}{16\,\pi^2}\left\{\frac{1}{d-4}-\frac{1}{2}\,\left[\ln(4\pi)+\Gamma'(1)+1\right]\right\}\,,
\label{eq:deltaZ}
\end{eqnarray}
with $\Gamma'(1) = 0.5772$ the Euler-Mascheroni constant. Note that these terms all involve powers of
the large mass $M_R$ and are, thus, power-counting violating. However, they can all be absorbed in the
complex-valued counterterms. Using now the renormalized version of the
KSFR relation, one can eliminate the coupling $g$ and obtains for the pole mass and the width of the $\rho$
meson by expanding the contributions to ${\cal O}(Q^4)$
\begin{eqnarray}
\label{eq:rhomass}
M_\rho^2  & = & M_\chi^2 +c_x M^2_\pi -\frac{g_{\omega\rho\pi}^2 M^3_\pi M_\chi}{24 \pi }+ \frac{M^4_\pi}{32\pi^2 F^2_\pi}
\left(3 -2\, \ln(M^2_\pi/M_\chi^2)\right) -\frac{g_{\omega\rho\pi}^2}{32 \pi ^2} M^4_\pi \left(\ln(M^2_\pi/M_\chi^2) -1\right)
\,, \\
\Gamma & = & \Gamma_\chi +\frac{\Gamma_\chi ^3}{8M_\chi^2}-\frac{c_x \Gamma_\chi  M^2_\pi}{2 M_\chi^2}
-\frac{g_{\omega\rho\pi}^2 M^3_\pi \Gamma_\chi}{48 \pi \,M_\chi} +\frac{M^4_\pi}{16\,\pi \,F^2_\pi M_\chi}\,,
\label{eq:rhowidth}
\end{eqnarray}
where we have identified the leading terms in the quark mass expansion of the pion mass and the
pion decay constant with their physical values. A few more remarks are in order. Since the
power-counting breaking terms are all absorbed, we end up
with a well behaved chiral expansion featuring terms $\sim\! M^2_\pi, M^3_\pi$ and $M^4_\pi$. Note further
that there will be finite contributions from the neglected diagrams of ${\cal O}(Q^4)$ but
no new non-analytic terms. This agrees with the general structure of the chiral expansion worked out
in~\cite{Bruns:2004tj}. The non-analytic terms displayed here agree with the calculation
of Ref.~\cite{Leinweber:2001ac}. To get an idea about the size of the corrections, one plugs in
$F_\pi=0.092\,{\rm GeV}, M_\pi=0.139\,{\rm GeV}\,, g_{\omega\rho\pi} = 16 \,{\rm GeV^{-1}}, M_\chi\approx M_\rho=0.78\,{\rm GeV}$
and obtains $M_\rho^2 =  M_\chi^2+ 0.019c_x - 0.0044$ (in GeV$^2$) and $\Gamma = \Gamma_\chi +
0.21\Gamma_\chi^3-0.016c_x\Gamma_\chi -0.0058\Gamma_\chi + 0.0011$ (in GeV). One notices that the
corrections to the chiral limit mass are rather small, as also found in Ref.~\cite{Bruns:2004tj}.
In \cref{sec:well}, we will use Eqs.~(\ref{eq:rhomass},\ref{eq:rhowidth}) to analyze lattice QCD data
from the ETMC collaboration. For further work on vector meson properties using the CMS, see
Refs.~\cite{Djukanovic:2005ag,Djukanovic:2014rua}.

\subsection{The complex mass scheme at two loops}
\label{sec:CMS2}

It is important to consider the CMS beyond the one-loop level, as discussed below on the examples of
the decay $\omega \to 3\pi$, $\Delta \to N\pi$ or the Roper decays $N^*(1440)\to N\pi, N\pi\pi$.
We will first show on the example of the first decay that the CMS can indeed by extended
to two-loop order and then use this knowledge to derive so far unknown constraints on the
mentioned baryon decay modes.

\begin{figure}[t!]
\centering
\includegraphics[width=0.5\linewidth]{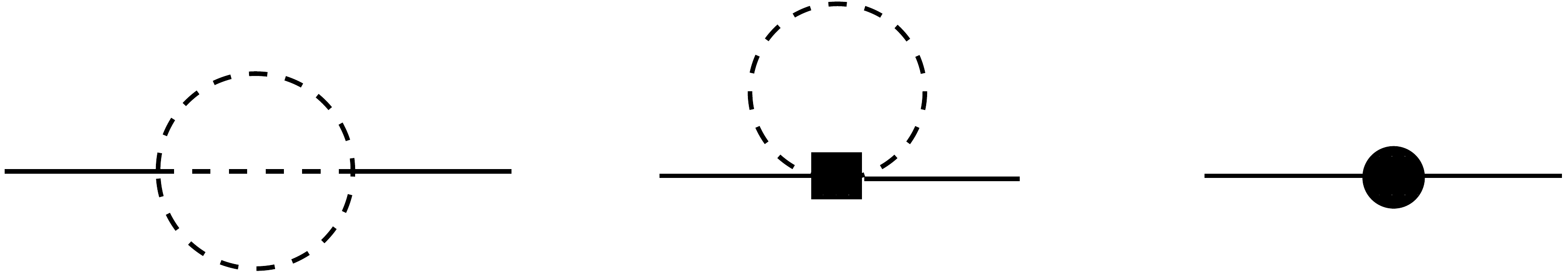}
\caption{Self-energy at two loops for a heavy particle (solid lines) coupling to light particles (dashed lines).
         The filled square and the filled circle denote a one-loop and a two-loop counterterm, respectively.
  \label{fig:SE-2loop}}
\end{figure}

Consider first the $\omega$ meson. As its main decay is $\omega\to 3\pi$, the self-energy has its first
non-trivial contribution at two-loop order, see the left diagram in Fig.~\ref{fig:SE-2loop}, where the $\omega$
is represented by the solid line and the dashed lines denote pions. The leading interaction
Lagrangian  reads \cite{Meissner:1987ge}:
\begin{equation}
{\cal L}_{V\Phi\Phi\Phi}^{(1)}=\frac{\mathrm{i}\,h}{4 F_\pi^3}\,\epsilon^{\mu\nu\alpha\beta}
\,\langle V_\mu \partial_\nu\Phi \partial_\alpha\Phi\partial_\beta\Phi\rangle~,
\end{equation}
with $V_\mu$ the vector field corresponding to the $\omega$ meson, the pions are
given by $\Phi= \pi^a \tau^a$, as we are considering two flavors here, and $h$ is
a coupling constant. The self-energy corresponding to this two-loop diagram takes the form:
\begin{eqnarray}
\Sigma^{\mu\nu} &=& \frac{72 \pi^2 h^2}{F_\pi^6}\,\left(p^\mu p^\nu-p^2 g^{\mu\nu}\right) I_{d+2}~,\nonumber\\
I_{d+2}&=&\frac{1}{(2\pi)^{2 (d+2)}}\int\frac{d^{d+2}k_1 d^{d+2}k_2}{\left[k_1^2-M^2+\mathrm{i}\eta\right]
\left[k_2^2-M^2+i\eta\right]\left[(p+k_1+k_2)^2-M^2+\mathrm{i}\eta\right]}~,
\label{eq:omSE}
\end{eqnarray}
with $\eta\to 0^+$ and $d$ is the number of space-time dimension. We see that the study of the
$\omega$ self-energy reduces to a scalar integral in six dimension. To further analyze this
self-energy, while avoiding any complications due to the spin and  chiral symmetry, the authors of Ref.~\cite{Djukanovic:2015gna} investigated a field theoretical model Lagrangian of interacting
scalar fields in six space-time dimensions
\begin{equation}
\label{eq:lagrange}
\mathcal{L}=\frac{1}{2}(\partial_{\mu}\pi\partial^{\mu}\pi-M^2\pi^2)+
\frac{1}{2}(\partial_{\mu}\Psi\partial^{\mu}\Psi-m^2\Psi^2)-
\frac{g}{3!}\,\pi^3\Psi +\mathcal{L}_1~,
\end{equation}
where the masses of the light and heavy scalar fields $\pi$ and $\Psi$, respectively, satisfy the condition
$M\ll m$, since  $\Psi$ represents an unstable particle. This Lagrangian $\mathcal{L}_1$ contains all
possible terms which are consistent with  Lorentz symmetry and with the invariance under the simultaneous
transformations $\pi\to -\pi$ and $\Psi \to -\Psi$. The power counting is build on small quantities $Q$
like the mass $M$, small external four-momenta of the $\pi$ or small external three-momenta of the $\Psi$.
The two-loop diagram shown in Fig.~~\ref{fig:SE-2loop} should, thus, scale as $Q^{(2d-4)}$, which at first
sight is messed up due to the complications generated by the large scale $m$. However,
using the ``dimensional counting analysis'' of Ref.~\cite{Gegelia:1994zz} or, equivalently, the ``strategy
of regions''~\cite{Beneke:1997zp}, after some lengthy algebra, one can identify and subtract the power
counting breaking terms generated from the heavy scale $m$. These are canceled exactly by the complex counterterms
that appear in the one-loop and tree-level diagrams at this order, see Fig.~\ref{fig:SE-2loop}. More precisely,
the one-loop counterterms originate from the one-loop diagrams contributing to elastic $\pi\Psi$ scattering,
whereas the two-loop counterterm arises from a subset of the terms in the Lagrangian ${\mathcal L}_1$.
Thus, the CMS scheme is also applicable at two-loop order, and it leads to a consistent power counting,
if and only if the subtraction of one-loop sub-diagrams in the renormalization of the two-loop diagrams is properly
done, for details see~\cite{Djukanovic:2015gna}. Calculations using the chiral effective Lagrangians are more involved due to the complicated structure
of the interactions, but the general features of the renormalization program do not change.

\subsection{The width of the lightest baryon resonances from EFT}
\label{sec:EFTbaryons}

We now consider the width of the two lowest baryon resonances, the $\Delta(1232)$ and the Roper $N(1440)$
at two-loop order in the CMS. Ultimately, these widths should be calculated on the lattice, see Sect.~\ref{sec:LQCD}, but the
detailed two-loop studies reveal in the case of the $\Delta(1232)$ some intriguing correlations
bewteen LECs, that will also be useful in a presicion extraction from lattice QCD data, and in case
of the Roper give insights into the important decay into a nucleon and two pions. Other resonances
like, e.g.,  the $\Lambda(1405)$ have also been considered as explicit fields in chiral Lagrangians, see, e.g.,  Refs.~\cite{Savage:1994pf,Lee:1995ku},
but in such cases that involve coupled-channels, unitarization methods are superior. These are
discussed in \cref{sec:uni}.

Consider first the $\Delta(1232)$-resonance. Here, the new small scale $\Delta \equiv m_\Delta - m_N \simeq 300\,$MeV arises, that must be dealt
with consistently and also, it is known that the $\Delta$ couples strongly to the $N\pi$ system, making
it an important ingredient in nuclear physics~\cite{Ericson:1988gk}. Extensions of CHPT including the
Delta (or the decuplet baryons in the three-flavor case) have been pioneered in
Refs.~\cite{Jenkins:1991es,Jenkins:1991bs,Jenkins:1991ne,Hemmert:1996xg,Hemmert:1997ye}, where in particular
in the last two references, the so-called small scale expansion (SSE) has been introduced. In the
SSE, the set of small parameters is extended to include the Delta-nucleon mass splitting,
\begin{equation}
Q \in \left\{ \frac{p}{\Lambda}, \frac{M_\pi}{\Lambda}, \frac{\Delta}{\Lambda}\right\}~,
\end{equation}
and one often uses $\epsilon$ instead of $Q$ to distinguish from the standard case with $\Delta =0$.
For other approaches to include these spin-3/2 fields, see, e.g.,
Refs.~\cite{Lutz:2001yb,Pascalutsa:2002pi,Bernard:2003xf,Kolomeitsev:2003kt,Pascalutsa:2005vq,Pascalutsa:2005ts,Hacker:2005fh,Bernard:2005fy,Wies:2006rv,Djukanovic:2007bw,Semke:2011ez,Lutz:2018cqo}.
It is important to note that the splitting $ m_\Delta - m_N$ does not vanish in the chiral limit,
which has important consequences for the corresponding EFT \cite{Gasser:1979hf,Bernard:1998gv,Meissner:2005mb}.
The decoupling of the Delta in the chiral limit enforces that contributions involving the spin-3/2 field
must vanish as $\Delta \to \infty$. Explicit examples are worked out in~\cite{Bernard:1998gv}. However,
in the limit of colors, $N_c$, going to infinity~\cite{tHooft:1973alw,Witten:1979kh} the situation
is opposite, namely the Delta becomes degenerate with the nucleon and, thus, can not be integrated out.
The consequences of this scenario have been first worked out in Refs.~\cite{Dashen:1993as,Jenkins:1993af,Dashen:1994qi},
for an early  review see \cite{Jenkins:1998wy} and further work, e.g.,   in
Refs.~\cite{Goity:1996hk,Carlson:1998gw,Flores-Mendieta:2000ljq,Schat:2001xr,Ahuatzin:2010ef,Goity:2007zs,Ahuatzin:2010ef,Flores-Mendieta:2012fxp,Flores-Mendieta:2021yzz}.

\subsubsection{The width of the \texorpdfstring{$\Delta(1232)$}{} }

Consider first the width of the $\Delta$ at two-loop order~\cite{Gegelia:2016pjm}.
The  pertinent effective Lagrangian contains, besides many
other terms, the leading $\pi \Delta$ and $\pi N\Delta$ couplings,
parametrized in terms of the LECs $g_1$ and $h$, respectively,
\begin{eqnarray}
{\cal L}^{(1)}_{\pi\Delta} &=& -\bar{\Psi}_{\mu}^i\xi^{\frac{3}{2}}_{ij}\Bigl\{\left(\mathrm{i}\slashed{D}^{jk}
-m_{\Delta}\delta^{jk}\right)g^{\mu\nu}
-\mathrm{i}\left(\gamma^\mu D^{\nu,jk}+\gamma^\nu D^{\mu,jk}\right) +\mathrm{i} \gamma^\mu\slashed{D}^{jk}\gamma^\nu
+m_{\Delta}\delta^{jk} \gamma^{\mu}\gamma^\nu\nonumber\\
&&+ {g_1}\frac{1}{2}\slashed{u}^{jk}\gamma_5g^{\mu\nu}
 +g_2\frac{1}{2} (\gamma^\mu u^{\nu,jk}
+u^{\nu,jk}\gamma^\mu)\gamma_5
+ g_3\frac{1}{2}\gamma^\mu\slashed{u}^{jk}\gamma_5\gamma^\nu \Bigr\}\xi^{\frac{3}{2}}_{kl}
{\Psi}_\nu^l\,,\nonumber\\
{\cal L}^{(1)}_{\pi N\Delta} &=& h\,\bar{\Psi}_{\mu}^i\xi_{ij}^{\frac{3}{2}}
\Theta^{\mu\alpha}(z_1)\ \omega_{\alpha}^j\Psi_N+ {\rm h.c.}\,,\\
{\cal L}^{(2)}_{\pi N\Delta}&=&\bar{\Psi}_{\mu}^i\xi_{ij}^{\frac{3}{2}}\Theta^{\mu\alpha}(z_2)
\, \left[\mathrm{i}\,b_3\omega_{\alpha\beta}^j\gamma^\beta+\mathrm{i}\,\frac{b_8}{m}\omega_{\alpha\beta}^ji\,D^\beta\right]
\Psi_N+{\rm h.c.}\,,\nonumber\\
{\cal L}^{(3)}_{\pi N\Delta}&=&\bar{\Psi}_{\mu}^i\xi_{ij}^{\frac{3}{2}}\Theta^{\mu\nu}(z_3)\biggl[
  \frac{f_1}{m}[D_\nu,\omega_{\alpha\beta}^j]\gamma^\alpha \mathrm{i}\,D^\beta
-\frac{f_2}{2m^2}[D_\nu,\omega_{\alpha\beta}^j]
\{D^\alpha,D^\beta\}
+ f_4\omega_\nu^j\langle\chi_+\rangle+f_5[D_\nu,\mathrm{i}\chi_-^j]\biggr]\Psi_N+ {\rm h.c.}\,,\nonumber
\label{eq:delta}
\end{eqnarray}
where $\Psi_N$ and $\Psi_\nu$ are the isospin doublet field of the nucleon
and the vector-spinor isovector-isospinor
Rarita-Schwinger field  of  the $\Delta$-resonance
with bare masses $m$ and $m_{\Delta 0}$, respectively.
$\xi^{\frac{3}{2}}$ is the isospin-$3/2$ projector,
$\omega_\alpha^i=\frac{1}{2}\,\langle\tau^i u_\alpha \rangle$ and $\Theta^{\mu\alpha}(z)=g^{\mu\alpha}
+z\gamma^\mu\gamma^\alpha$. Using field redefinitions the off-shell parameters $z$  can be absorbed in
LECs of other terms of the effective Lagrangian and, therefore, they can be chosen arbitrarily
\cite{Tang:1996sq,Krebs:2009bf}. We fix the off-shell structure
of the interactions with the Delta by adopting $g_2=g_3=0$ and $z_1=z_2=z_3=0$. Thus, $g_1$ parameterizes
the leading $\pi\Delta$ vertex. For vanishing external sources, the covariant derivatives are given by
\begin{eqnarray}
D_\mu \Psi_N & = & \left( \partial_\mu + \Gamma_\mu 
\right) \Psi_{N}\,, ~~~
\Gamma_\mu =   \frac{1}{2}\,\left[u^\dagger \partial_\mu u +u
\partial_\mu u^\dagger 
\right]=\tau_k\Gamma_{\mu,k}\,, \nonumber\\
\left(D_\mu\Psi\right)_{\nu,i} & = &
\partial_\mu\Psi_{\nu,i}-2\,\mathrm{i}\,\epsilon_{ijk}\Gamma_{\mu,k} \Psi_{\nu,j}+\Gamma_\mu\Psi_{\nu,i}
\,. \label{cders}
\end{eqnarray}

\begin{figure}[t]
\begin{minipage}[c]{0.48\textwidth}
\includegraphics[width=1\textwidth]{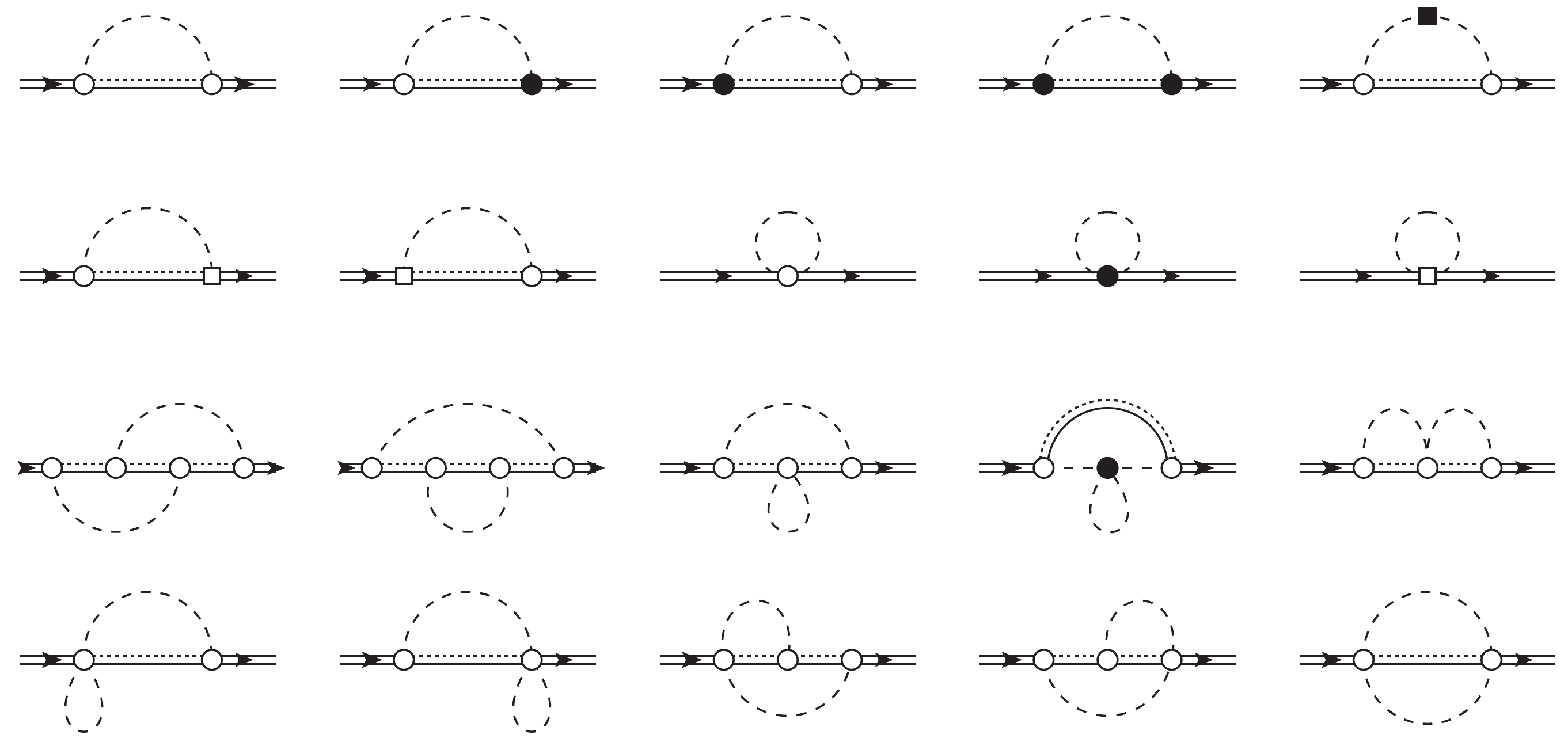}
\end{minipage}
\begin{minipage}[c]{0.48\textwidth}
\includegraphics[width=1\textwidth]{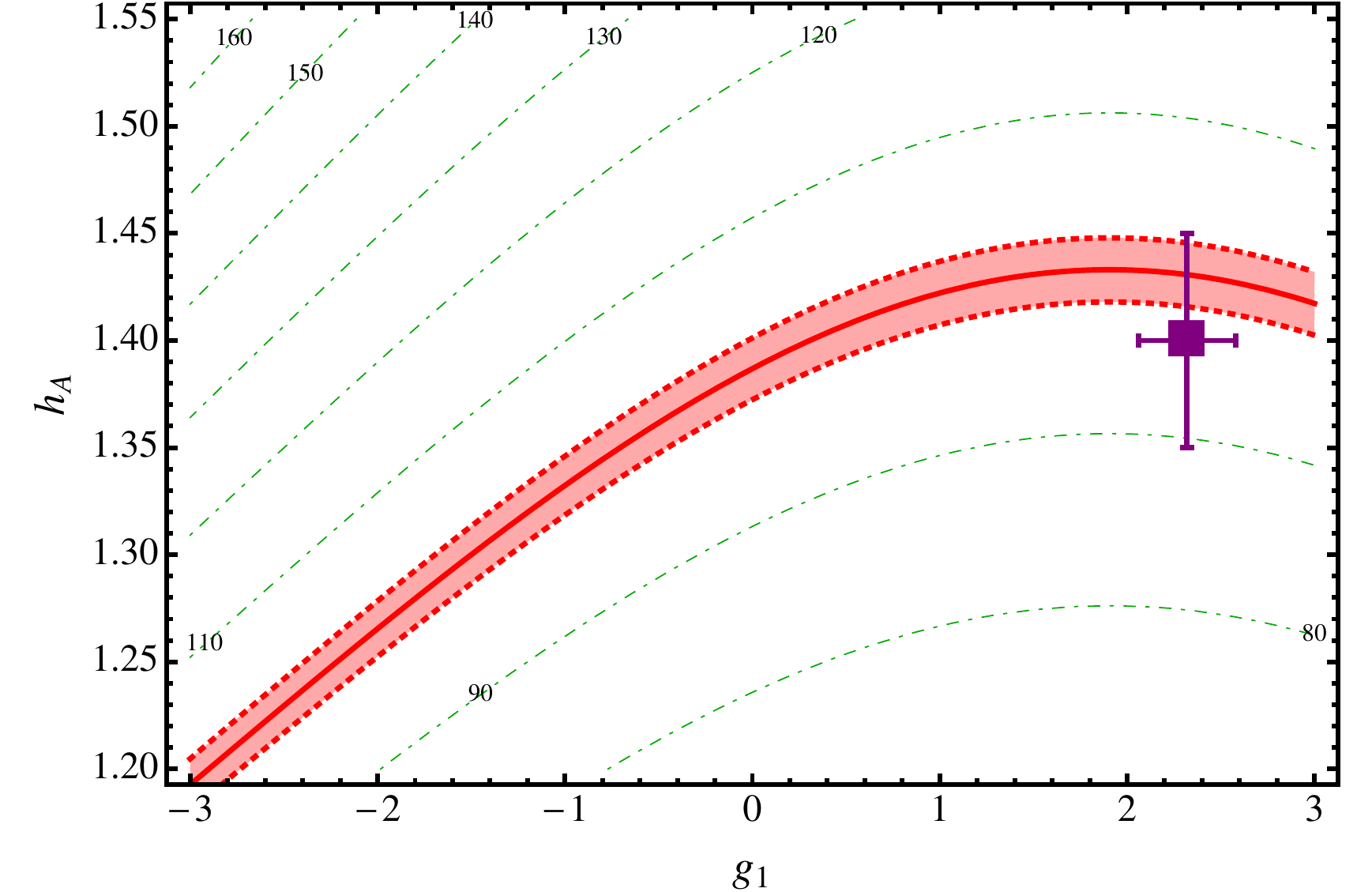}
\end{minipage}
\caption{Left panel:  One and two-loop self-energy diagrams contributing to the width
of the Delta resonance up-to-and-including
fifth order according to the standard power counting. The dashed and double solid lines
represent the pions and the Delta resonances, respectively.
The double (solid-dotted) lines in the loops correspond to either nucleons or Deltas.
Open and filled circles refer to LO and NLO vertices from the meson-baryon Lagrangian,
where as the filled box represents an NLO mesonic term.
Right panel:
Correlation between the leading $\pi \Delta$ and $\pi N \Delta$ couplings.
The central line corresponds to $\Gamma_\Delta = 100$~MeV  while the band is obtained by varying
$\Gamma_\Delta$ in the range of 98 to 102~MeV. The dot-dashed lines show the correlation for other values of the width of the Delta.
The box with the error bars are the results from the analysis of Ref.~\cite{Yao:2016vbz}.
Figure adopted from~\cite{Gegelia:2016pjm}.
}
\label{fig:Delta}
\vspace{-4mm}
\end{figure}

The power counting relies on the fact that $m_\Delta - m_N$ is a small quantity.
More precisely, the small parameters are the external momenta, the pion mass
and the nucleon-Delta mass splitting, collectively denoted as $\varepsilon$.
However, there are many LECs in Eq.~(\ref{eq:delta}), so how can one
one possibly make any prediction?  Let us evaluate the $\Delta$ self-energy at the complex pole,
\begin{equation}
  z - m_{\Delta}^0 - \Sigma(z) =0 \quad {\rm with} \quad
  z=m_\Delta-\mathrm{i}\,\displaystyle\frac{\Gamma_\Delta}{2}~.
\end{equation}
The corresponding diagrams for the one- and two-loop self-energy contributing to the
width of the Delta resonance up to order $\varepsilon^5$ are  displayed in Fig.~\ref{fig:Delta} (left panel),
where the counterterm diagrams are not shown. The one-loop diagrams are easily worked out.
For the calculation of the two-loop graphs one uses the Cutkosky rules for unstable
particles, that relate the width to the pion-nucleon scattering amplitude,
$\Gamma_\Delta \sim |A(\Delta\to N\pi)|^2$~\cite{Veltman:1963th}.
One finds a remarkable  reduction of parameters that is reflected in the relation
\begin{eqnarray}
h_A &=& h - \left(b_3\Delta_{23}+b_8\,\Delta_{123}\right)-\left(f_1\Delta_{23}+ f_2\,\Delta_{123}\right)\Delta_{123}
+2(2f_4-f_5)M_\pi^2~,\nonumber\\
\Delta_{23} &=& -\Delta = m_N-m_\Delta~,~~~~
\Delta_{123}= \frac{M_\pi^2+m_N^2-m_\Delta^2}{2m_N}~,
\end{eqnarray}
which means that all of the LECs appearing in the $\pi N\Delta$ interaction at second
and third order, the $b_i \,(i=3,8)$ and $f_i \, (i=1,2,4,5)$, respectively, merely lead to a renormalization
of the LO $\pi N\Delta$ coupling $h$,
\begin{equation}
h_A = h - \left(b_3\Delta_{23}+b_8\,\Delta_{123}\right)
-\left(f_1\Delta_{23}+f_2\,\Delta_{123}\right)\Delta_{123}+2(2f_4-f_5)M_\pi^2~,
\end{equation}
and, consequently, one finds a very simple formula for the decay width $\Delta \to N\pi$,
\begin{equation}
\Gamma(\Delta\to N \pi) =  \left(53.9\,{h}_A^2+0.9g_1^2 {h}_A^2-3.3g_1^{} {h}_A^2
-1.0\,{h}_A^4\right)~{\rm MeV}~.
\end{equation}
This leads to a novel correlation that is independent of the number of colors, as $N_c$ was not
used as a parameter in the calculation. This correlation between $h_A$ and $g_1$ is depicted in the right panel of
Fig.~\ref{fig:Delta}. It is obviously fulfilled by the analysis of  Ref.~\cite{Siemens:2016jwj}, that showed that
the inclusion of the $\Delta$ alleviates the tension between the threshold and subthreshold regions in the
description of $\pi N$ scattering found in baryon CHPT, see also~\cite{Becher:2001hv}.

\subsubsection{The width of the Roper resonance}
Next, consider the calculation of the width of the Roper-resonance, the $N^*(1440)$, at two-loop
order~\cite{Gegelia:2016xcw}, improving  the one-loop results from Ref.~\cite{Djukanovic:2009zn}.
A remarkable feature of the Roper is the fact that its decay width into a nucleon and a pion
is similar to the width  into a nucleon and two pions, $\Gamma (R\to N\pi) \simeq
\Gamma (R\to N\pi\pi)$. Any model that is supposed to describe the Roper must account for
this fact. In CHPT, consider the  effective chiral Lagrangian of pions, nucleons and
Deltas coupled to the Roper~\cite{Borasoy:2006fk,Djukanovic:2009gt,Long:2011rt},
\begin{eqnarray}
{\cal L}_{\rm eff}&=&{\cal L}_{\pi\pi}+{\cal L}_{\pi N}+{\cal L}_{\pi \Delta}+{\cal L}_{\pi R}
+ {\cal L}_{\pi N\Delta}+{\cal L}_{\pi NR}+{\cal L}_{\pi\Delta R}~,
\end{eqnarray}
with
\begin{eqnarray}
{\cal L}_{\pi R}^{(1)}&=&\bar{\Psi}_R\left\{i\slashed{D}-m_R+\frac{1}{2}{g_R}\slashed{u}
\gamma^5\right\}\Psi_R~,\nonumber\\
{\cal L}_{\pi R}^{(2)} &=& \bar{\Psi}_R\left\{c_1^R\langle\chi_+\rangle\right\}\Psi_R + \ldots~,\nonumber\\
{\cal L}_{\pi NR}^{(1)} &=& \bar{\Psi}_R\left\{\frac{1}{2}{g_{\pi NR}}\gamma^\mu\gamma_5 u_\mu\right\}\Psi_N
+ {\rm h.c.}~,\nonumber\\
{\cal L}^{(1)}_{\pi \Delta R} &=& {h_R}\,\bar{\Psi}_{\mu}^i\xi_{ij}^{\frac{3}{2}}\Theta^{\mu\alpha}(\tilde{z})\
\omega_{\alpha}^j\Psi_R+ {\rm h.c.}~,
\end{eqnarray}
where  $g_R$, $g_{\pi NR}$ and $h_R$, respectively, are the leading Roper-pion, Roper-nucleon-pion and Delta-Roper-pion couplings. Here, ${\Psi}_R$ denotes the Roper isospin doublet field and all other notations are as in the preceding subsection and in Ref.~\cite{Gegelia:2016xcw}.

\begin{figure}[t]
\centering
\includegraphics[width=0.48\textwidth]{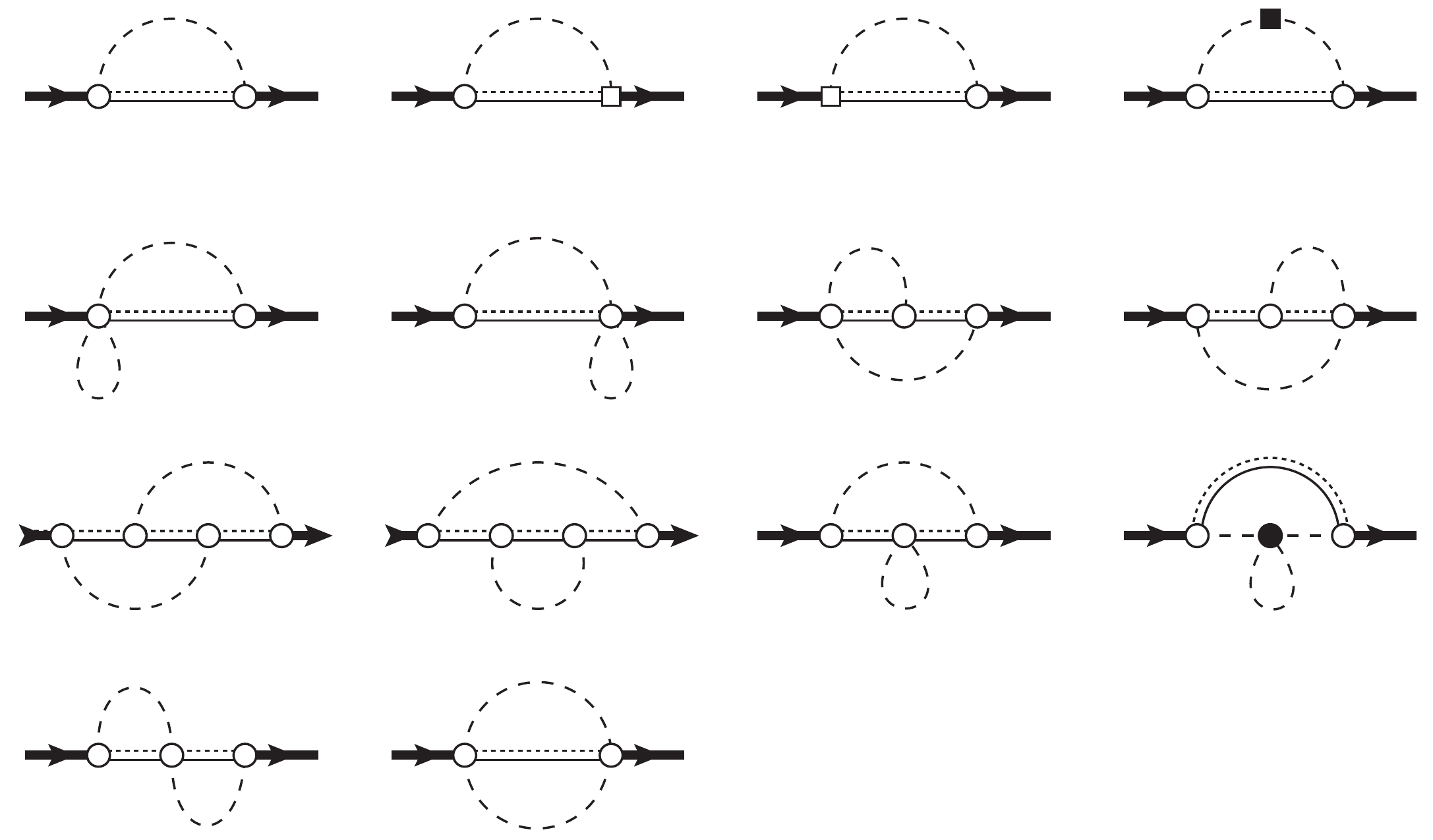}~~~~~
\includegraphics[width=0.48\textwidth]{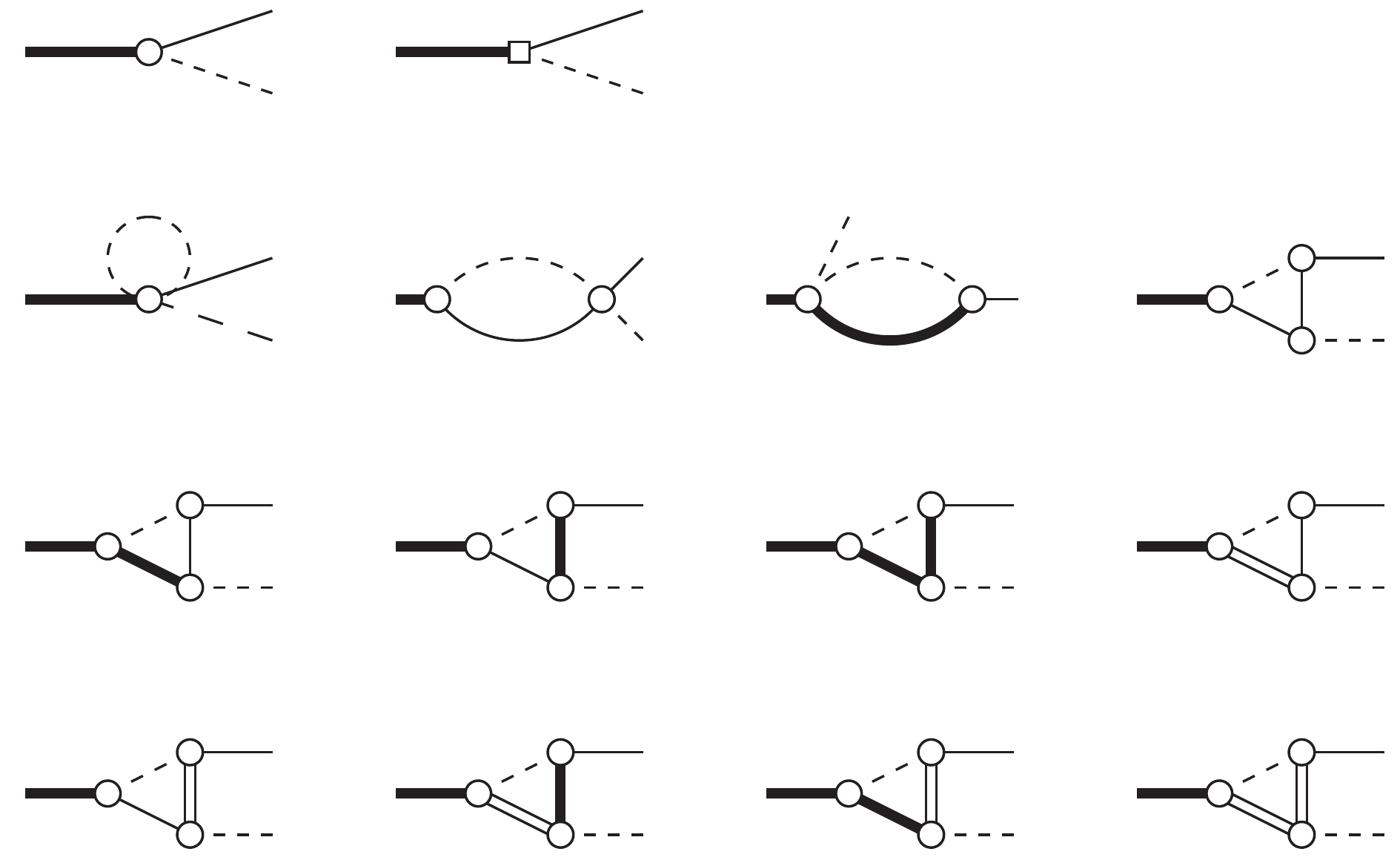}
\caption{Left panel: One and two-loop self-energy diagrams of the Roper resonance up-to-and-including  fifth order according to the standard power counting. The dashed and thick solid lines represent the pions and the Roper resonances, respectively. The thin solid lines in the loops stand for either nucleons, Roper or Delta resonances. For further notations, see Fig.~\ref{fig:Delta}.
Right panel: Feynman diagrams contribution to the decay $R\to N\pi$ up to
leading one-loop order. Dashed, solid, double and thick solid lines
correspond to pions, nucleons, Deltas and Roper resonances, respectively.}
\label{fig:Roper}
\vspace{-4mm}
\end{figure}

In this case, the power counting is more complicated, but can be set
up around the complex pole of the Roper resonance as (for more details, see~\cite{Gegelia:2016xcw}), assigning
the following counting rules:
\begin{equation}
m_R-m_N \sim \varepsilon~,~~ m_R-m_\Delta \sim \varepsilon^2~,~~ m_\Delta-m_N \sim \varepsilon^2~,~~
M_\pi \sim \varepsilon^2~,
\end{equation}
where $\varepsilon$ denotes a small parameter. Again, let us calculate the self-energy to two loops
at the complex pole $z_R = m_R - \mathrm{i}\Gamma_R/2$. The pertinent diagrams are shown in the left panel
of Fig.~\ref{fig:Roper}. By applying the cutting rules to these self-energy diagrams,
one obtains the graphs contributing to the decay amplitudes of the Roper resonance into the $\pi N$
and $\pi\pi N$ systems, leading to the total width
\begin{equation}
\Gamma_R = \Gamma_{R\to N\pi} +  \Gamma_{R\to N\pi\pi}~.
\end{equation}
A somewhat lengthy calculation  of the diagrams in the right panel of Fig.~\ref{fig:Roper} leads to:
\begin{equation}
\Gamma(R\to N \pi) = 550(58) \, g_{\pi NR}^2\ {\rm MeV}~,
\end{equation}
while the two-pion decay is given at this order by tree diagrams with intermediate nucleons and Deltas,
\begin{eqnarray}
\Gamma(R\to N\pi\pi) &=&\Bigl(1.5(0.6)\,g_A^2 \,g_{\pi NR}^2 -2.8(1.0)\, g_A^{} \, g_{\pi NR}^2\,g_R^{}
+1.5(0.6)\,g_{\pi NR}^2\, g_R^2 \nonumber\\
&& ~+ 3.0(1.0)\,g_A^{}\, g_{\pi NR}^{} \,h_A^{} h_R^{}
-3.8(1.4)\,g_{\pi NR}^{}\,g_R^{} \,h_A^{} h_R^{} +9.9(5.5)\,h_A^2h_R^2\Bigr)~{\rm MeV}\,.
\end{eqnarray}
The total width, thus, depends on five LECs. The uncertainties in the round brackets are generated by the
uncertainties in the LECs. We use $g_A =1.27$ and $h_A=1.42\pm 0.02$. The latter value is the real
part of this coupling taken  from Ref.~\cite{Yao:2016vbz}.
As for the other unknown parameters,  the authors of~\cite{Gegelia:2016xcw} fixed $g_{\pi NR}$ so as to
reproduce the width  $\Gamma_{R\to \pi N}=(123.5\pm 19.0)$~MeV from the PDG.
This yields $g_{\pi NR}=\pm (0.47\pm 0.04)$. In what follows, let us take the
positive sign for our central value and use the negative one as part of the
error budget. Further,  assume $g_R=g_A$ and $h_{R}=h_A$, the so-called maximal mixing assumption~\cite{Beane:2002ud}.
Then, one can make a prediction for the two-pion decay width of the Roper,
\begin{equation}
\Gamma (R\to N\pi\pi) = (41 \pm 22_{\rm LECs} \pm 17_{\rm h.o.})~{\rm MeV}~,
\end{equation}
which is consistent with the PDG value of  ($67\pm 10$)~MeV. The error due to the neglect of the higher orders (h.o.)
is simply estimated by multiplying the $\varepsilon^5$ result (central value) with $\varepsilon = (m_R-m_N)/m_N
\simeq 0.43$. Clearly, to make further progress, we need an improved determination of the LECs $g_R$ and $h_R$.
This could be addressed within LQCD. Note also  that this scheme has
been used to consider the electromagnetic transition form factors of the Roper~\cite{Gelenava:2017mmk}.

\subsection{Unitarization methods}
\label{sec:uni}

\subsubsection{General discussion}

In the last section we have seen that perturbative calculations based on the effective field
theories indeed can allow one to access properties of certain resonances. Such a methodology
is advantageous insofar as it allows to separate and identify in a systematic way the dominant
(long-range) effects
from short-range physics. However, it can only be applied to well separated, low-lying resonances as otherwise
no consistent power counting scheme can be set up. In the present section we discuss another class of genuinely
non-perturbative approaches. These allow one to deal with the effects of coupled channels and higher-lying
resonances, but of course one has to pay a certain prize, as discussed below.

Lattice QCD is a tool to access QCD Green's functions and transition amplitudes in a
non-perturbative and systematical way. However, since the energies and momenta are inherently real,
the extraction of resonance parameters requires again an additional step, the
analytical continuation to complex energies. This again requires knowledge of analytical
expression of the transition amplitudes as functions of, e.g.,  Mandelstam variables,
see \cref{sec:reso}. As discussed there, $S$-matrix unitarity plays a crucial role
constraining the form of such amplitudes, leading in, e.g., the two-body case to the
famous K-matrix parametrization~\cref{eq:K-matrix}. In that, the typical workflow includes
defining a general parametrization of the real-valued K-matrix (using, e.g.,~Pad\'e,
Chew-Mandelstam forms), fixing the parameters of such expressions from fits to either
experimental or lattice results and finally the extraction of poles for
complex-valued energies. In this sense also modern statistical and machine learning
techniques can indeed be utilized to reduce the parameter space as shown for example
in Ref.~\cite{Landay:2018wgf}. More details on such data driven techniques can be found in
recent reviews~\cite{Briceno:2017max, JPAC:2021rxu}. Dealing with QCD at low energies, chiral symmetry can further lead to additional constraints on transition amplitudes. In regard of lattice QCD results, this is enormously useful allowing one to trace out the quark mass dependence of amplitudes and ultimately resonance parameters. This methodology
runs under the name of chiral extrapolations, see \cref{subsec"systematic_uncertainties} and \cref{fig:pirho} for an explicit example. Such constraints can even be wrapped up in model-independent conditions of functional form of, e.g., resonant $\pi\pi$ amplitudes~\cite{Bruns:2017gix}. Reversing this logic, one can also use
low-energy effective theories (such as CHPT) to identify dominant interaction extending the region of applicability by the so-called unitarization procedure.

\begin{figure}[t]
\centering
\includegraphics[width=0.47\linewidth]{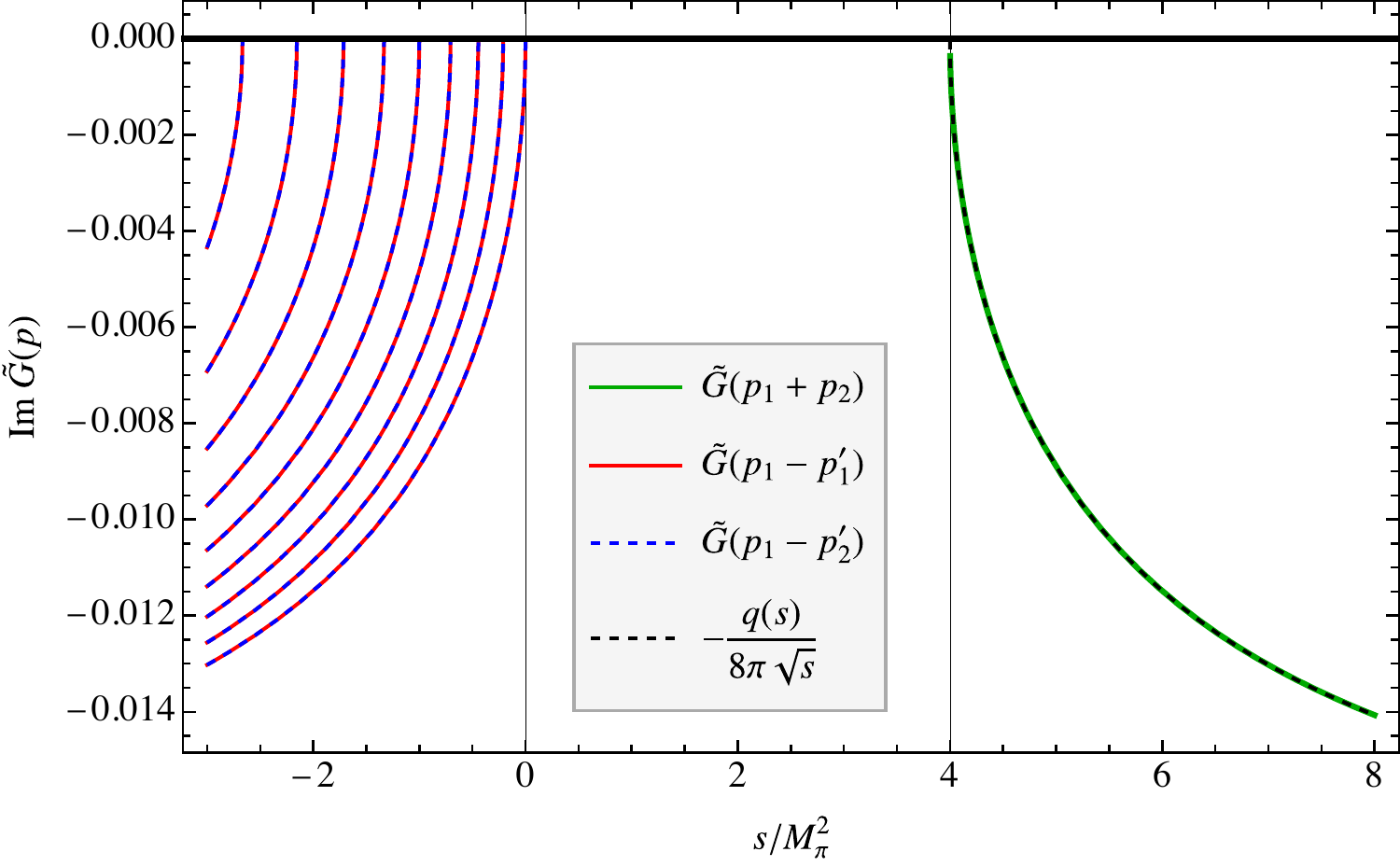}~~~
\includegraphics[width=0.47\linewidth,trim=5cm 5cm 5cm 5cm,clip]{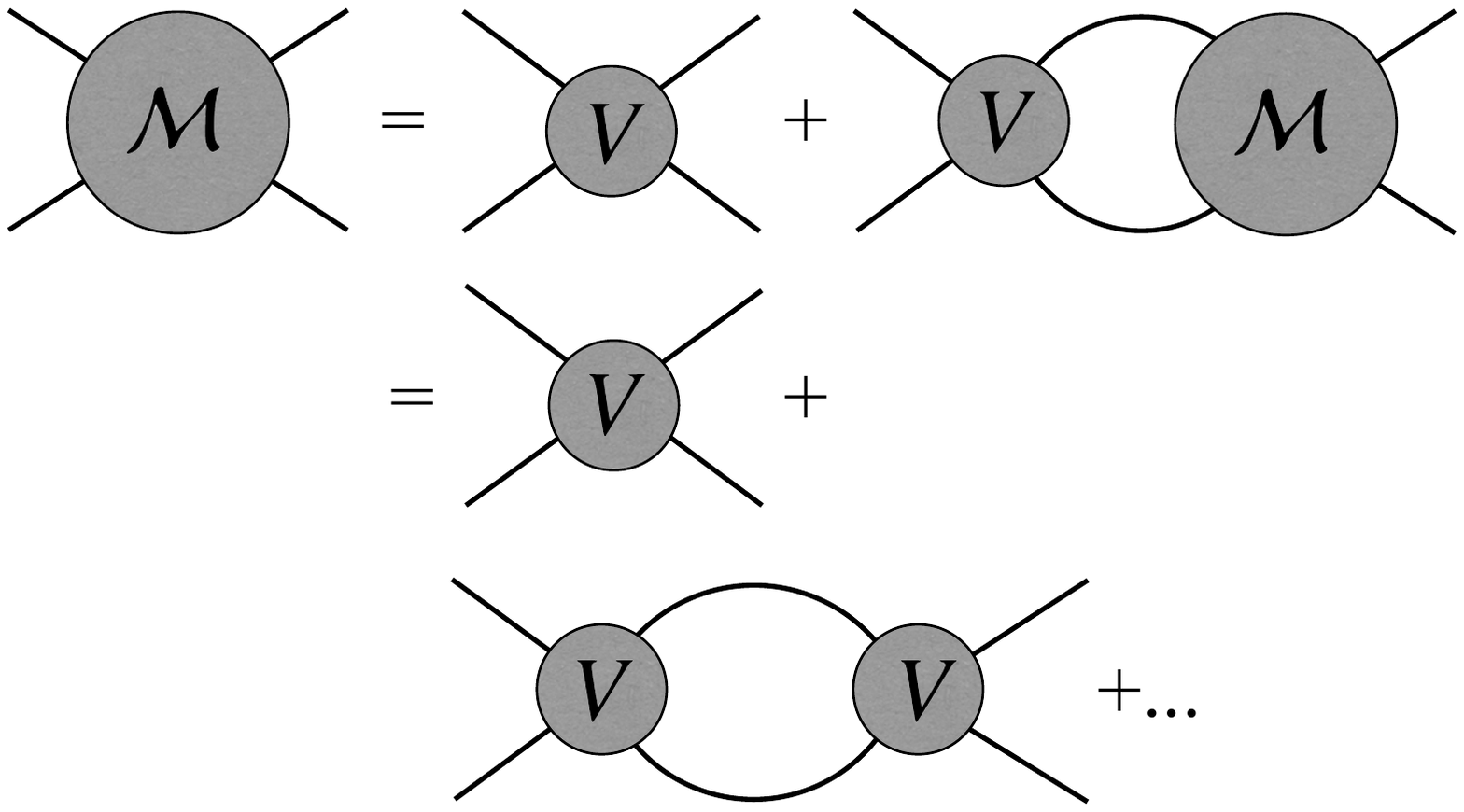}
\caption{Left: Imaginary part of one loop diagrams appearing in the next-to-leading
scattering amplitude in the $\phi^4$ theory. Notably, the $t$ and $u$ channel contributions
(different lines correspond to variation of the angle between in- and outgoing momenta) lead
to an imaginary part solely in the unphysical region.
Right: diagrammatic representation of a unitary scattering amplitude~\eqref{eq:BSE}.
\label{fig:2b-unitarity}}
\end{figure}

For pedagogical reasons we begin with the scalar $\phi^4$-theory which allows for a simpler treatment. Later we will show how this applies to  chiral Lagrangians, leading to the so-called \emph{Chiral Unitary approaches}(UCHPT) or {\em unitarization schemes}.
The Lagrangian of the $\phi^4$-theory reads
\begin{align}
\label{eq:phi4-Lagrangian}
\mathcal{L}=\frac{1}{2}\left(\partial^\mu\phi \partial_\mu\phi-M^2\phi^2\right)
-\frac{\lambda}{4!}\phi^4\,,
\end{align}
with $M$ the particle mass and $\lambda$  a coupling constant.
This simple form of the interaction leads to the fact that the scattering amplitude for the process
$\phi(p_1)\phi(p_2)\to
\phi(p_1')\phi(p_2')$ separates into an infinite series of Feynman diagrams ordered in
powers of $\lambda$. Specifically, the first two terms of this series read
\begin{align}
\label{eq:phi4-expansion}
\mathcal{M}_1(p_1',p_1;p)&=-\lambda\,,\nonumber\\
\mathcal{M}_2(p_1',p_1;p)&=-\lambda^2\left(\tilde G(p_1+p_2)+\tilde G(p_1-p_1')
+\tilde G(p_1-p_2')\right)\,,
\end{align}
abbreviating $\mathcal{M}(p_1',p_1;p)$ as $\mathcal{M}(p_1,p_2\to p_1',p_2')$ with
$p=p_1^{(\prime)}+p_2^{(\prime)}$. Here, $\tilde G(p)$ is the one-loop function in $d$
dimensions defined by
\begin{align}
\tilde G(p)=\int\frac{d^dk}{(2\pi)^d}\frac{\mathrm{i}}{k^2-M^2+\mathrm{i}\epsilon}\frac{1}{(k-p)^2-M^2+\mathrm{i}\epsilon}\,,
\end{align}
which is logarithmically divergent for $d=4$. Such divergences are removed in the usual
sense of perturbative renormalization, i.e., absorbing them in the renormalized particle mass $M$,
the coupling $\lambda$ and the field renormalization factor $Z$ at any given order. The exact
prescription does not matter for the present discussion and can be found in standard textbooks.
Picking one, we note that in dimensional regularization and the $\overline{\rm MS}$ subtraction
scheme the finite part of the loop integral becomes
\begin{align}
\tilde G(p)=\frac{1}{16\pi^2}\left(-1+2\ln\left( \frac{M}{\mu}\right)-\frac{4q(p)}
{\sqrt{p^2}}{\rm tanh}^{-1}\left(\frac{2q(p)\sqrt{p^2}}{4M^2-p^2}\right)\right)
\quad\text{with}\quad
q(p)=\sqrt{p^2/4-M^2}
\,,
\label{eq:dimregloop}
\end{align}
where $\mu$ denotes the regularization scale. One notes that the expression depends only on
$p^2$ and, furthermore, that for general values of $p^2\in\mathds{R}$ the above expression is
complex-valued with the imaginary part existing only for $p^2>4M^2$. In the center-of-mass
system and for $z=\bm{p}_1\cdot\bm{p}_1'/(|\bm{p}_1||\bm{p}_1'|)$, the two-body Mandelstam
variables read
\begin{align}
s=(p_1+p_2)^2\,,~~
t=(p_1-p_1')^2=-\frac{s-4M^2}{2}\left(1-z\right)\,,~~
u=4M^2-s-t\,,~~
\end{align}
which means that in the physical region ($s>4M^2$) only the first term of $\mathcal{M}_2$ in
\cref{eq:phi4-expansion} can develop an imaginary part. The other two terms become
complex-valued only for $s<s_0(z)$ with $s_0(z)\le 0$ depending on the value of $z\in[-1,1]$.
This is depicted in the left panel of \cref{fig:2b-unitarity}.
The opening of the imaginary part for negative values of $s$ is what leads to the so-called
left-hand cut of the complex $s$-plane. Taking a closer look on the expansion in
\cref{eq:phi4-expansion} one realizes that
\begin{align}
\mathcal{M}_2-\mathcal{M}_2^*=
-|\mathcal{M}_1|^2 \underbrace{(\tilde G(s)-\tilde G^*(s))}_{=-({2q(s)})/({8\pi\sqrt{s}})}\,,
\end{align}
while unitarity, Eq.~\eqref{T-matrix Unitarity}, for the unprojected matrix elements yields
${\rm Im}~\mathcal{M}=\frac{q(s)}{8\pi\sqrt{s}}|\mathcal{M}|^2$, meaning that the unitarity
condition is fulfilled perturbatively only. In general, restoration of unitarity to all orders is what is referred to as \emph{unitarization}. Various ways exist in this regard, some of most popular ones will be discussed below.

A large class of approaches starts from the following general ansatz
\begin{align}
\mathcal{M}(p_1',p_1;p)=V(p_1',p_1;p)-\int\frac{d^dk}{(2\pi)^d}V(p_1',k;p)G(k;p)
\mathcal{M}(k,p_1;p)\,,
\label{eq:BSE}
\end{align}
in terms of a two-body propagator $G$ and an interaction kernel $V$. Of course, regularization
of the integral equation is understood. Indeed, the form of the two-body propagator $G$ is
what defines the specific ansatz, which for the \emph{Bethe-Salpeter equation} (BSE) and
scalar particles reads $G(k;p)= \mathrm{i}/(k^2-M^2+\mathrm{i}\epsilon)/((p-k)^2-M^2+\mathrm{i}\epsilon)$. The interaction
is encoded in the kernel $V(p',p;p)$, which is typically derived from an effective field theory.

Eq.~\eqref{eq:BSE} is a genuine $d$-dimensional integral equation which can be interpreted as
an infinite series of Feynman diagrams, see the right panel of Fig.~\ref{fig:2b-unitarity}.
In general, such an equation cannot be solved analytically, but examples exists where this has been performed~\cite{Nozawa:1989pu, Lee:1991pp, vanAntwerpen:1994vh, Borasoy:2005zg, Bruns:2010sv, Ruic:2011wf, Mai:2013cka}. One simplification can be achieved by assuming that neither $\mathcal{M}$ nor the interaction kernel $V$ have singularities in $k^0$, allowing to perform the $k^0$ integral using Cauchy's theorem. This reduces the dimension of the integral, effectively putting the intermediate particles on the mass-shell $k^2=M^2$. While this destroys manifest covariance, the resulting
equation is still covariant. It is often referred to as the \emph{quasipotential} or
\emph{Gross} equation, see, e.g., Ref.~\cite{Gross:1969rv}. Still, the equation keeps its
genuine integral form. Simplifying this further, one assumes often a projection of the
interaction kernel $V$ (on-shell) onto a partial wave of interest (most of the S-wave).
Such a projected potential $\tilde V_\ell$
can only depend on the total energy squared $s$, such that the integral equation above
indeed becomes an algebraic one with the solution
\begin{align}
\mathcal{M}_\ell(s)=\tilde V_\ell(s)-\tilde V_\ell \tilde G(s) \mathcal{M}_\ell(s)=
\frac{1}{1+\tilde V_\ell(s)\tilde G(s)}\tilde V_\ell(s)\,.
\label{eq:unitarized-amp}
\end{align}
For our example of the $\phi^4$ theory with $\tilde V(s)=-\lambda$, so that
Eq.~\eqref{eq:unitarized-amp} indeed is also the solution of the BSE~\eqref{eq:BSE}.
For a general theory this solution is referred most-commonly to as
\emph{unitarized scattering amplitude}.

So far, we have seen that the Bethe-Salpeter equation~\eqref{eq:BSE} allows to restore
unitarity, keeping close connection to the Feynman diagrams. While in general this equation
is hard to solve, it can be reduced to an algebraic one projecting the interaction kernel
to a specific (on-shell) partial wave. Obviously, this still preserves unitarity since
${\rm Im}~\mathcal{M}_\ell^{-1}(s)=-q(s)/(8\pi\sqrt{s})$ as demanded by Eq.~\eqref{eq:unitarity},
bringing this ansatz close to the general K-matrix form discussed in Sec.~\ref{sec:reso}.
Specifically, matching Eq.~\eqref{eq:K-matrix} with Eq.~\eqref{eq:unitarized-amp} relates
\begin{align}
K_\ell^{-1}(s)=\tilde V_\ell^{-1}(s)+{\rm Re}~\tilde G(s)\,.
\end{align}
Over the last decades, the BSE in integral and algebraic form has become a workhorse for
many investigations of, e.g., resonant dynamics of two-hadron scattering in the
non-perturbative regime of QCD. Deriving the interaction kernel from some effective
field theory, predictions on the resonance structure on the Riemann-sheets can be made.
However, there is a price to the flexibility and simplicity of this approach as the crossing
symmetry does no longer hold. Note that approaches to the restoration of crossing symmetry in this context have
been formulated, see, e.g., Refs.~\cite{Kolomeitsev:2003ac, Kolomeitsev:2003kt}. In general the fact that crossing
symmetry is violated, and only a subset of all possible Feynman diagrams can be considered by the above unitarization technique yields that standard renormalization methodology cannot be applied. Indeed, the scattering
amplitude \cref{eq:unitarized-amp} remains explicitly dependent on the regularization scale.
This means that when fitting two different versions of a unitarized amplitude to the same data using the
same interaction kernel  leads to different parameters and predictions outside of fitting region. This means that
such an unitarization procedure inevitably induces some model-dependence.

\subsubsection{Light quark sector: Meson-meson scattering}

The application of the BSE methodology to the light (up and down) quark sector have been
plentiful both in the meson~\cite{Oller:1997ti, Oller:1998hw, Oller:1997ng} as well as baryon
sector~\cite{Doring:2005xw, Bruns:2010sv, Inoue:2001ip}, which also includes extensions to
coupled channels. Pertinent works including also
the strange and the heavy quarks will be discussed below.

Many other unitarization techniques have been developed to address the lowest mesonic resonances. The $N/D$ approach~\cite{Chew:1960iv} derives from dispersion relations and
includes in principle both the left-hand cuts (in $N$) and unitarity right-hand cut (in $D$),
see \cref{fig:2b-unitarity}. In particular, the amplitude is written as
\begin{align}
\label{eq:n/d}
\mathcal{M}_{N/D}(s)=\frac{N(s)}{D(s)}\,.
\end{align}
After this definition, an ansatz is made for the numerator $N^{I\ell}(s)$, which typically
stems from the perturbative expansion of effective field theories or more general functions
with constraints from the behavior of partial waves near threshold~\cite{Oller:1998zr}. Finally,
one can deduce the denominator from the unitarity condition, $Im~D(s)=-N(s)\theta(s-s_{\rm thr})$
rescaled appropriately, with $s_{\rm thr}$ the physical two-body threshold. This is done by means of
a dispersion relation with respect to Mandelstam $s$ including an appropriate number of
subtractions. These subtractions are then obtained from fits to data. For applications to
$\pi\pi$ and $\pi N$ scattering see, Refs~\cite{Oller:1998zr, Igi:1998gn,Dai:2014lza} and
\cite{Meissner:1999vr}, respectively. For recent work on other hadron-hadron scattering
processes using the $N/D$ method, see e.g.
Refs.~\cite{Guo:2013rpa,Entem:2016ipb,Gulmez:2016scm,Du:2018gyn}.

Another very popular approach is the so-called \emph{inverse amplitude method}
(IAM)~\cite{Truong:1988zp, Pelaez:2006nj, GomezNicola:2007qj, Pelaez:2010fj}. Also here in addition
to unitarity the left-hand cut is included at least perturbatively. This is because the amplitude matches
the chiral amplitude up to the next-to-leading order which also makes the approach an ideal tool
to access universal parameters of, e.g., the $f_0(500)$ and $\rho(770)$ resonances from
phenomenology~\cite{Hanhart:2008mx, Pelaez:2015qba, Nebreda:2010wv} and lattice
QCD~\cite{Mai:2021lwb, Guo:2018zss, Doring:2016bdr} (at unphysical pion masses). As it turns out,
the same amplitude also fulfills the general requirements on the chiral trajectory of resonances
to all chiral orders~\cite{Bruns:2017gix}.

Specifically, the partial wave amplitude projected to isospin $I$ and angular momentum $\ell$ reads
\begin{align}
\label{eq:IAM}
\mathcal{M}_{\rm IAM}^{I\ell}(s)=\frac{(\mathcal{M}_2^{I\ell}(s))^2}{\mathcal{M}_2^{I\ell}(s)-\mathcal{M}_4^{I\ell}(s)}\,,
\end{align}
where $\mathcal{M}_n^{I\ell}(s)$ denotes the $n$th order chiral amplitudes~\cite{Gasser:1983yg, Gasser:1984gg}.
The leading order chiral amplitude ($\mathcal{M}_2^{I\ell}(s)$) is a function of energy, Goldstone-boson
mass, $M^2=B(m_u+m_d)$ and pion decay constant in the chiral limit, $F$. The next-to-leading order
amplitude ($\mathcal{M}_4^{I\ell}(s)$) involves in the two-flavor case two low-energy constants (LECs)
$\bar l_1$ and $\bar l_2$. Two additional low-energy constants $\bar l_3$, $\bar l_4$ enter the NLO
chiral amplitude when replacing the above mass and decay constants by their physical values using
the well-known one-loop results~\cite{Gasser:1983yg},
\begin{align}
M_\pi^2=M^2\left(1-\frac{M^2}{32\pi^2F^2}\bar l_3\right)
\,,\qquad
F_\pi=F\left(1+\frac{M^2}{16\pi^2F^2}\bar l_4\right)\,.
\end{align}
By construction, the LECs $\bar l_i$ do not depend on the regularization scale. Because of that, they
acquire a dependence on the quark masses\footnote{We remark that such scale-independent LECs
are specific to the two-flavor case, a similar construction is not possible for $N_f \ge 3$
flavors.}.  This is of course disadvantageous when studying results of
lattice calculations performed at unphysical pion masses, but can be overcome by using the more
conventional scale-dependent but quark-mass-independent renormalized LECs $l_i^r$. The
relation between two reads
\begin{align}
l_i^r=\frac{\gamma_i}{32\pi^2}\left(\bar l_i+\log \frac{M^2}{\mu^2}\right)\,,
\end{align}
where $\gamma_1=1/3$, $\gamma_2=2/3$, $\gamma_3=-1/2$, $\gamma_4=2$. Hence, for a fixed scale $\mu$
one can determine the renormalized LECs and then make predictions for two-particle scattering at a
different pion mass.  We note that fixing $\mu$ in physical units may lead to subtleties when addressing scale-independent quantities in lattice QCD. However, in some cases methods exists to overcome this issue, see discussions in Refs.~\cite{Beane:2006kx, Miller:2020xhy, Mai:2021lwb}.

\begin{figure}[t]
\centering
\includegraphics[width=\linewidth,trim=0 0 0 0,clip]{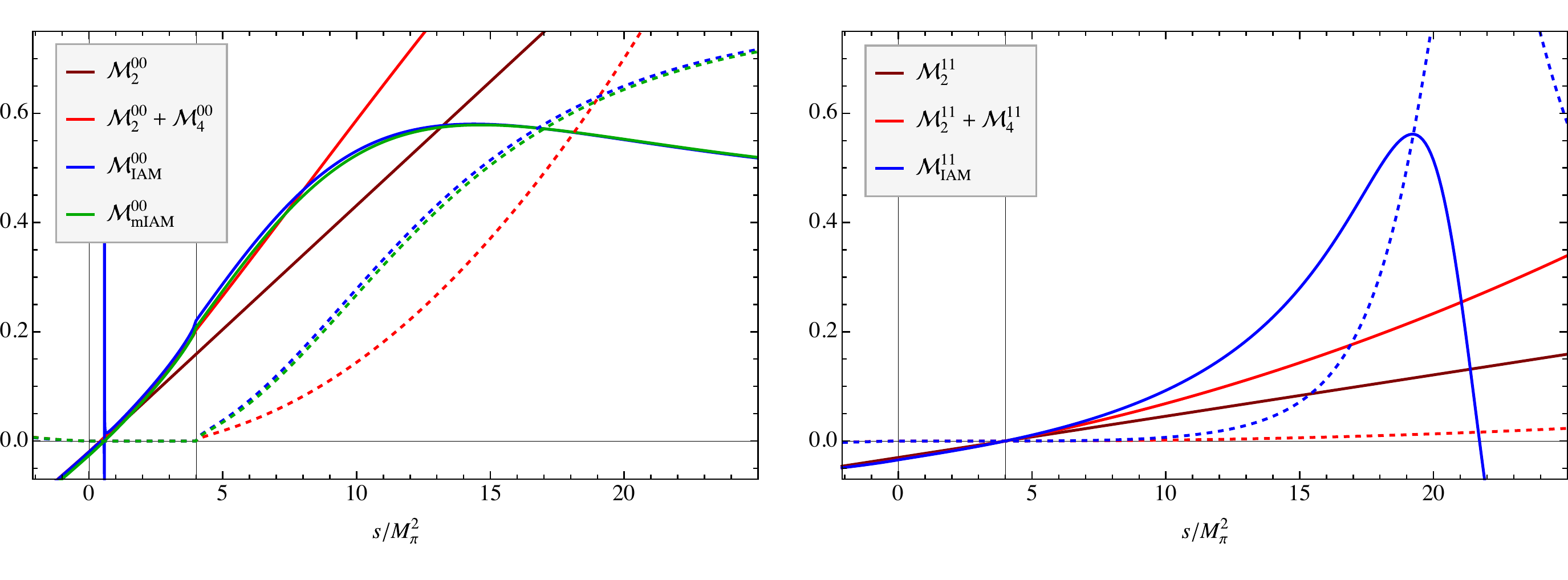}
\caption{Comparison between the perturbative chiral amplitudes with the inverse amplitude
method Eq.~\eqref{eq:IAM} for the isoscalar and isovector $\pi\pi$ channels. Full and dashed lines denote
real and imaginary (if existing) parts, respectively. In the isoscalar case the chiral amplitude is
vanishing exactly at $s\approx 1/2M_\pi^2$ (the Adler zero). This yields a pole in the IAM approach, which
is removed by following the modified inverse amplitude method Eq.~\eqref{eq:MIAM}, see text for mode details.
\label{fig:rhosigma-iam}}
\end{figure}

In the past, the IAM was used successfully to  gain deeper insights into the long-debated $f_0(500)$
isoscalar $\pi\pi$ resonance, see the dedicated review~\cite{Pelaez:2015qba}. In this context we show
a comparison of the perturbative chiral amplitude with the IAM scattering amplitude in
\cref{fig:rhosigma-iam} for isovector and isoscalar scattering.
Using LECs from a perturbative analysis~\cite{Gasser:1983yg} for demonstration only, we observe
that the unitarized amplitude $\mathcal{M}_{\rm IAM}$, indeed, is close to the next-to-leading order
chiral amplitude close to the threshold but exhibits a resonant behavior at $\sqrt{s}\approx 500~{\rm MeV}$
and $\sqrt{s}\approx 770~{\rm MeV}$, respectively, in the isoscalar and isovector channel. In the former case
we observe additionally that the perturbative amplitude vanishes exactly at $s\approx 2 M_\pi^2$,
the Adler zero~\cite{Adler:1964um}. This leads to the fact that around this energy also the denominator
of $\mathcal{M}_{\rm IAM}$ vanishes.
Notably, the same happens for the case of maximal isospin ($\mathcal{M}_2^{20}$) at $s\approx 2M_\pi^2$. Obviously, this singularity
has very small residuum, such that for studies of experimental data, this effect is  negligible.
Still from the conceptual point of view as well when embedding this two-body model into, e.g.,
multihadron setting~\cite{Mai:2017vot,Mai:2018djl} the subthreshold behavior may become important.
Fortunately, there is a convenient way to remove this spurious pole by a procedure established
in, e.g., Refs.~\cite{Boglione:1996uz,Hanhart:2008mx}.

\begin{align}
\label{eq:MIAM}
\mathcal{M}_{\rm IAM}^{I\ell}(s) \longrightarrow
\mathcal{M}_{\rm mIAM}^{I\ell}(s)&=\frac{(\mathcal{M}_2^{I\ell}(s))^2}{\mathcal{M}_2^{I\ell}(s)-\mathcal{M}_4^{I\ell}(s)+A^{I\ell}_m(s)}\,,\nonumber\\
A^{I\ell}_m(s)&=\mathcal{M}_{4}^{I\ell}(s_2)-\frac{(s_2-s_A)(s-s_2)\left(\mathcal{M}_{2}^{I\ell\prime}(s_2)-\mathcal{M}_{4}^{I\ell\prime}(s_2)\right)}{s-s_A}\,,
\end{align}
where $s_A$ and $s_2$ are the zeros of $\mathcal{M}_{2}^{I\ell}(s)-\mathcal{M}_{4}^{I\ell}(s)$ and
$\mathcal{M}_{2}^{I\ell}(s)$, respectively. In this form the inverse amplitude method has been applied
frequently~\cite{Hanhart:2008mx, Nebreda:2011di, Berengut:2013nh, Mai:2019pqr}, simultaneously describing
all three $\pi\pi$ isospin channels, and making, e.g., predictions for the quark mass dependence of the light-quark resonance parameters, see \cref{fig:pirho} and discussion thereof.

\subsubsection{Light quark sector: Meson-baryon sector}

The principle of dynamical generation of resonances has helped to understand many low-lying states
in the hadron spectrum. Including the strange quark enriches this picture even more. This is because
production thresholds are separated stronger than in the light ($u,d$) sector. Arguably the most
prominent examples in this sense is the the long-debated $\Lambda(1405)$ resonance, which indeed is
dynamically generated due to strong attraction of $\bar KN$ and interference with the $\pi\Sigma$ channels.
Several reviews~\cite{Hyodo:2020czb, Meissner:2020khl, Mai:2020ltx, Hyodo:2022xhp} have been dedicated to
the history and state of the art understanding of this enigmatic state. Thus, we will only mention facts
relevant for this review, referring the reader for more details to the quoted reviews.

The current method of choice in the studies of antikaon-nucleon scattering is based on the already
discussed Bethe-Salpeter equation. In that, the interaction kernel is taken from the chiral meson-baryon
Lagrangian at next-to-leading order, which in its full form reads
\begin{equation}
\begin{aligned}[b]
	V_{\rm off}(\slashed{q'},\slashed{q}; P)
  =
  &\Bigg(A_{WT}(\slashed{q}+\slashed{q'})
  +A_{Bs}\slashed{q'}\frac{m-\slashed{P}}{s-m^2}\slashed{q}
  +A_{Bu}\slashed{q}\frac{m-\slashed{P}+\slashed{q'}+\slashed{q}}{u-m^2}\slashed{q'}
  \Bigg)_{\rm LO}+
  \\
	&\Bigg(A_{14}(q\cdot q')
  +A_{57}[\slashed{q},\slashed{q'}]
  +A_{M}
	+A_{811}\left(\slashed{q'}\left(q\cdot P)+\slashed{q}(q'\cdot P\right)\right)\Bigg)_{\rm NLO}\,,
\label{eq:Voff}
\end{aligned}
\end{equation}
where $P=p+q=p'+q'$, and $s=P^2$, $u=(p-q')^2$ are the usual Mandelstam variables. We note that 10 combinations of ground state octet mesons and baryons have the same quantum numbers, i.e.,
$\{K^-p$, $\bar K^0 n$, $\pi^0\Lambda$, $\pi^0\Sigma^0$, $\pi^+\Sigma^-$, $\pi^-\Sigma^+$, $\eta\Lambda$, $\eta \Sigma^0$, $K^+\Xi^-$, $K^0\Xi^0\}$ meaning that the scattering amplitude $\mathcal{M}$
must describe the dynamics of all coupled-channels simultaneously. Thus, the interaction kernel $V$ is a $10\times10$ matrix build from the matrices $A_{...}$. These depend explicitly on the meson decay constants and axial couplings $D$, $F$ at leading, and on the LECs $\{b_0,b_D,b_F,b_1,...,b_{11}\}$ at
next-to-leading chiral order. This already demonstrates the rapid rise of the number of unknown LECs
as well as momentum structures compared to, e.g., light-quark $(u,d)$ meson sector (4 LECs). Also,
the higher dimension of the kernel increases the computational costs strongly. However, as discussed before,
various approximations and simplifications can be applied to the above kernel. For example off-shell effects
appear to be  rather small as studied in Ref.~\cite{Mai:2012dt}. Also, it seems safe to assume that NLO terms
are required to produce realistic scattering amplitudes.

Ultimately, various approximations and unitarization variations lead consistently to a narrow pole close to the $\bar KN$ threshold, additionally predicting a second lighter and broader pole, when using similar experimental data for fits of the free parameters. As we know now, see the recent review~\cite{Meissner:2020khl}, such unusual double pole structures are not uncommon in many other parts of the hadronic spectrum.

\subsubsection{Heavy-light quark sector: Goldstone bosons scattering off \texorpdfstring{$D$}{} mesons}
\label{sec:Dphiuni}

Unitarization methods are also successfully used in the coupled-channel scattering of Goldstone bosons
off ($\pi, K, \eta$)
$D$ (and $D^*$) mesons, which allows to dynamically generate the mysterious charmed scalar mesons,
in particular the charm-scalar mesons \cite{BaBar:2003oey,Belle:2003guh,CLEO:2003ggt} that showed that the conventional quark model was insufficient. There were considered early as hadronic molecules in a theory combining
chiral and heavy quark symmetry~\cite{Barnes:2003dj,Kolomeitsev:2003ac} or within the chiral doubling
approach~\cite{Nowak:2003ra}, see also the reviews \cite{Guo:2017jvc,Chen:2022asf,Meng:2022ozq}.
A further important observation was the two-pole structure of the $D_0^*(2300)$ \cite{Albaladejo:2016lbb}
that was first worked out explicitly in the analysis of lattice QCD data on heavy-light pseudoscalar meson
$J^P=0^+$ scattering in the strangeness-isospin $(S,I)=(0,1/2)$ sector from the Hadron Spectrum Collaboration \cite{Moir:2016srx}. This had been noted before but was not out into proper context
\cite{Kolomeitsev:2003ac,Guo:2006fu,Guo:2009ct}. Combining unitarized perturbation theory in the heavy-light
sector with data from lattice QCD and also from the LHCb collaboration on $B\to D\pi\pi$ decays
\cite{LHCb:2014ioa,LHCb:2015eqv,LHCb:2015tsv,LHCb:2015klp}.
finally led to a new paradigm for heavy-light spectroscopy \cite{Du:2017zvv} and further helped to
establish the two-pole structure as more general phenomenon in the hadron spectrum rather than being
a curiosity related to the enigmatic $\Lambda(1405)$~\cite{Meissner:2020khl} as discussed in more
detail in \cref{sec:cc}. Note that the type of three-body decays $B\to D\pi\pi$ requires
a different effective Lagrangian than the one used for $D\phi$ scattering, as will also be
discussed below.

First, let us recall the power counting for Goldstone boson scattering off matter fields, here
the triplet of $D$ (and $D^*$) mesons. To be specific, we consider the EOMS scheme. External
momenta as well as the masses of the Goldstone bosons are counted as ${\cal O}(p)$, where $p$ denotes
a small quantity. However, the nonvanishing masses of the $D$ and $D^*$ in the chiral limit introduce
new scales $M_0$ and $M_0^*$, both counted as ${\cal O}(1)$. As a result, at low
energies, the temporal components of the momenta of the $D$ and $D^*$ are
counted as ${\cal O}(1)$, while the spatial components are counted as
${\cal O}(p)$. Therefore, the virtuality $q^2-M_0^{(*)2}$ in the
propagators scales as ${\cal O}(p)$, and the propagators scale as
${\cal O}(p^{-1})$. The Goldstone boson propagators are counted as
${\cal O}(p^{-2})$ as usual. In the EOMS scheme, the power counting breaking terms are
absorbed into the redefinition of the LECs so that the resulting physical
observables obey the power counting rules. This is explicitly worked out
in Ref.~\cite{Du:2016xbh}.
Most calculation of the scattering potentials have been carried out to NLO,
but there are a few works that attempted an NNLO calculation.

The effective Lagrangian for the calculation of the unitarized $D\phi$ scattering
amplitudes up to leading one-loop order can be written as
\begin{equation}
\mathcal{L}_{\rm eff}=\sum_{i=1}^2\mathcal{L}_{\phi\phi}^{(2i)}+\sum_{j=1}^3\mathcal{L}_{D\phi}^{(j)}
+\sum_{k=1}^2\mathcal{L}_{D^\ast\phi}^{(k)}+\sum_{l=1}^3\mathcal{L}_{D^\ast D\phi}^{(l)}\,,
\end{equation}
with the superscripts specifying the chiral dimension. The terms in
the Goldstone sector are the standard ones~\cite{Gasser:1984gg} and
the terms corresponding to interactions between the $D=(D^0,D^+,D_s^+)$ mesons and
the Goldstone bosons are given by~\cite{Burdman:1992gh,Wise:1992hn,Yan:1992gz,Guo:2008gp,Yao:2015qia,Du:2016ntw}
\begin{eqnarray}
\label{eq:LDphi}
\mathcal{L}^{(1)}_{D\phi}&=& \mathcal{D}_\mu D \mathcal{D}^\mu D^\dagger-{M}_0^2D
D^\dagger\ ,\nonumber\\
\mathcal{L}^{(2)}_{D\phi} &=& D\left(-h_0\langle\chi_+\rangle-h_1{\chi}_+
+ h_2\langle u_\mu u^\mu\rangle-h_3u_\mu u^\mu\right) {D}^\dag + \mathcal{D}_\mu D\left({h_4}\langle u_\mu
u^\nu\rangle-{h_5}\{u^\mu,u^\nu\}\right)\mathcal{D}_\nu {D}^\dag\,,\\
\mathcal{L}^{(3)}_{D\phi} &=&
D\biggl[ \mathrm{i}\,{g_1}[{\chi}_-,u_\nu] +
{g_2}\left([u_\mu,[\mathcal{D}_\nu,u^\mu]] + [u_\mu,[\mathcal{D}^\mu,u_\nu]]
\right)\biggr]\mathcal{D}^\nu {D}^\dag + g_3
D\,[u_\mu,[\mathcal{D}_\nu,u_\rho]] \mathcal{D}^{\mu\nu\rho} {D}^\dag\ +h.c. \,,\nonumber
\end{eqnarray}
where $M_0$ is the $D$ meson mass in the chiral limit, the $h_i$ and $g_j$ are LECs and
the chiral building blocks are given as above
$u_\mu=\mathrm{i}[u^\dagger\partial_\mu u-u\partial_\mu u^\dagger]$, $U=u^2$ and
$\chi^\pm=u^\dagger\chi u^\dagger\pm u\chi^\dagger u$.
The covariant derivative is defined via
\begin{eqnarray}
\label{eq:cov}
\mathcal{D}_\mu H=H(\overset{\leftarrow}{\partial_\mu}+\Gamma_\mu^\dagger)\ ,
\qquad \mathcal{D}_\mu H^\dagger=(\partial_\mu+\Gamma_\mu)H^\dagger\ ,
\end{eqnarray}
and $\mathcal{D}^{\mu\nu\rho}=\{\mathcal{D}_\mu,
\{\mathcal{D}_\nu,\mathcal{D}_\rho\}\}$, where $H\in\{D,D^\ast\}$ with
$D^\ast=(D^{\ast0},D^{\ast+},D_s^{\ast+})$. The so-called chiral connection in the covariant derivatives is defined as
$\Gamma_\mu=\left(u^\dagger\partial_\mu u+u\partial_\mu
u^\dagger\right )/2.$
Similarly, the relevant terms for the interaction between the $D^\ast$ and the Goldstone bosons
 are~\cite{Burdman:1992gh,Wise:1992hn,Yan:1992gz}
\begin{eqnarray}
\mathcal{L}_{D^\ast\phi}^{(1)}= -\frac{1}{2}\mathcal{F}^{\mu\nu}\mathcal{F}_{\mu\nu}^\dagger+M_0^{\ast 2}
D^{\ast\nu} D^{\ast\dagger}_\nu\,,~~~
\mathcal{L}_{D^\ast\phi}^{(2)}=\ D_\mu^\ast\left[\tilde{h}_0\langle\chi_+\rangle+\tilde{h}_1{\chi}_+\right]
D^{\mu\ast\dagger}\ ,
\end{eqnarray}
with $\tilde{h}_{0,1}$ analogous to $h_{0,1}$ and
$\mathcal{F}_{\mu\nu}=(\mathcal{D}_\mu D^\ast_{\nu}-\mathcal{D}_\nu
D^\ast_{\mu})$. In the heavy quark limit, $\tilde{h}_{0} = h_0$ and $\tilde{h}_{1} = h_1$.
Note that due to the trace structure in the corresponding operator,
the LECs $h_0$, $h_2$ and $h_4$ are suppressed in the large-$N_c$ limit~\cite{Lutz:2007sk}.
The LEC $h_1$ can be deduced from the mass splittings in the $D$ meson triplet,
\begin{equation}
h_1 = \frac{M_{D_s}^2-M_D^2}{4(M_K^2-M_\pi^2)} = 0.427\,.
\end{equation}
The quark mass dependence of $M_D$ and $M_{D_s}$ fixes $h_0 \simeq 0.01$, which is consistent with the
expectations from large-$N_c$. The other LEC are determined from fits to lattice QCD data, see below.
Finally, the LO axial coupling has the form
\begin{eqnarray}
\mathcal{L}_{D^\ast D\phi}^{(1)}=\mathrm{i}\,{g_0}\left(D^\ast_{\mu}u^\mu D^\dagger-D\,u^\mu D^{\ast\dagger}_\mu\right)\ .
\end{eqnarray}
As pointed out in Refs.~\cite{Krebs:2009bf,Yao:2016vbz}, the resonance-exchange
contributions of $\mathcal{O}(p^2)$ and $\mathcal{O}(p^3)$ can be taken into
account by shifting the coupling in the LO resonance-exchange contribution and
the LECs in the contact terms. This also holds true for our case. Thus,
we do not need the ${\cal O}(p^2)$ and ${\cal O}(p^3)$ terms for the $D^\ast
D\phi$ coupling. The $DD^*\pi$ axial coupling constant $g_0$ can be fixed by the decay width
$\Gamma(D^{*+}\to D^0\pi^+)$. As discussed in Refs.~\cite{Yao:2015qia}, one
gets $g=(1.11\pm 0.15) ~\text{GeV}$ for the renormalized coupling $g$, which
contains the bare constants $g_0$ and one-loop chiral corrections.

\begin{table}[t!]
\begin{center}
\begin{tabular}{|l c | c c c c c |}
\hline
 $(S,I)$ & Channels & $C_{\rm LO}$ & $C_0$ & $C_1$ & $C_{24}$ & $C_{35}$
 \\
\hline
$(-1,0)$      & $D\bar{K}\to D\bar K$   & $-1$  & $M_K^2$ & $M_K^2$  & 1 & $-1$
\\
$(-1,1)$      & $D\bar{K}\to D\bar K$   & 1     & $M_K^2$ & $-M_K^2$ & 1 & 1
\\
$(2,\frac{1}{2})$ & $D_sK\to D_sK$          & 1     & $M_K^2$ & $-M_K^2$ & 1 & 1
\\
$(0,\frac{3}{2})$ & $D\pi\to D\pi$          & 1 & $M_\pi^2$ & $-M_\pi^2$ & 1 & 1
\\
$(1,1)$       & $D_s\pi\to D_s\pi$      & 0   & $M_\pi^2$ & 0        & 1 & 0
\\
              & $D K\to D K$            & 0     & $M_K^2$ & 0        & 1 & 0
\\
              & $D K\to D_s\pi$   & 1 & 0 & $-(M_K^2+M_\pi^2)/2$ & 0    & 1
\\
$(1,0)$       & $D K\to D K$            & $-2$ & $M_K^2$ & $-2M_K^2$ & 1 & 2
\\
              & $D_s\eta\to D_s\eta$ & 0 & $M_\eta^2$ & $-2M_\eta^2+2M_\pi^2/3$
             & 1 & $4/3$
\\
             & $D K\to D_s\eta$ & $-\sqrt{3}$ & 0 &
        $-\sqrt{3}(5M_K^2-3M_\pi^2)/6$ & 0 & $1/{\sqrt{3}}$
\\
$(0,\frac{1}{2})$ & $D\pi\to D\pi$       & $-2$    & $M_\pi^2$ & $-M_\pi^2$ & 1 & 1
\\
             & $D\eta\to D\eta$     & 0 & $M_\eta^2$& $-M_\pi^2/3$& 1
             & $1/3$
\\
             & $D_s\bar K\to D_s\bar K$& $-1$& $M_K^2$& $-M_K^2$& 1 & 1
\\
             & $D\eta\to D\pi$     & 0 & 0 & $-M_\pi^2$ & 0 & 1
\\
             & $D_s\bar K\to D\pi$ & $-{\sqrt{6}}/{2}$ & 0 &
             $-{\sqrt{6}}(M_K^2+M_\pi^2)/4$ & 0 & ${\sqrt{6}}/{2}$
\\
             & $D_s\bar K\to D\eta$& $-{\sqrt{6}}/{2}$ & 0 &
             ${\sqrt{6}}(5M_K^2-3M_\pi^2)/12$ & 0 & $-1/{\sqrt{6}}$
\\
\hline
\end{tabular}
\caption{\label{tab:ci}The coefficients in the scattering amplitudes $V(s,t,u)$. The
channels are labelled by strangeness ($S$) and isospin ($I$). }
\end{center}
\end{table}

Consider now the scattering process $D_1(p_1)\phi_1(p_2)\to D_2(p_3)\phi_2(p_4)$ in more detail.
There are alltogether 16 channels with different total strangeness $S$ and isospin $I$ as listed in
\cref{tab:ci}, see, e.g., Refs.~\cite{Guo:2006fu,Gamermann:2006nm,Hofmann:2003je,Guo:2008gp,Guo:2009ct,Cleven:2010aw,Wang:2012bu,Liu:2012zya,Altenbuchinger:2013vwa,Cleven:2014oka,Yao:2015qia,Guo:2015dha,Du:2016tgp,Du:2017ttu,Du:2017zvv,Guo:2018kno}.
Using the NLO chiral Lagrangian from Ref.~\cite{Guo:2008gp}, the scattering amplitudes are given by
\begin{equation}
\label{eq:v}
V(s,t,u) = \frac{1}{F_0^2} \bigg[\frac{C_{\rm LO}}{4}(s-u) - 4 C_0 h_0 +
2 C_1 h_1 - 2C_{24} H_{24}(s,t,u) + 2C_{35} H_{35}(s,t,u) \bigg]\,,
\end{equation}
where $F_0$ is the pion decay constant in the chiral limit, and the coefficients $C_i$
can be found in Tab.~\ref{tab:ci}. Further,
\begin{eqnarray}
H_{24}(s,t,u) &=& 2 h_2 p_2\cdot p_4 + h_4 (p_1\cdot p_2 p_3\cdot p_4 +
p_1\cdot p_4 p_2\cdot p_3)\,, \nonumber\\
H_{35}(s,t,u) &=& h_3 p_2\cdot p_4 + h_5
(p_1\cdot p_2 p_3\cdot p_4 + p_1\cdot p_4 p_2\cdot p_3)\,.
\end{eqnarray}
For NNLO corrections and contributions from the $D^*$ mesons, see, e.g., Ref.~\cite{Du:2017ttu}.
The appearing LECs can be determined from a fit to lattice data. In order to reduce the correlations
between the LECs, one introduces the following redefinitions of the LECs~\cite{Liu:2012zya,Yao:2015qia}
\begin{eqnarray}
h_4^\prime &=& h_4 \bar{M}_D^2\,, \quad h_5^\prime =h_5 \bar{M}_D^2\,, \quad
h_{24}=h_2+h_4^\prime \,, \quad  h_{35}=h_3+2h_5^\prime\,, \nonumber \\
g_1^\prime &=& g_1 \bar{M}_D\,,\quad g_2^\prime=g_2\bar{M}_D\,, \quad g_3^\prime=g_3\bar{M}_D^3\,,
\quad   g_{23}=g_2^\prime-2g_3^\prime\,,
\end{eqnarray}
where $\bar{M}_D$ is the average of the physical masses of the charmed mesons $D$ and $D_s$,
$\bar{M}_D=(M_D^\text{Phy}+M_{D_s}^\text{Phy})/2$. The new LECs $h_4^\prime$, $h_5^\prime$, $h_{24}$ and $h_{35}$
are dimensionless, and $g_1^\prime$, $g_3^\prime$ and $g_{23}$ have the dimension of inverse mass.

Before unitarization, a partial wave projection to a definite orbital angular
momentum $\ell$ should be performed
\begin{eqnarray}
{V}_{\ell}^{(S,I)}(s)_{D_1\phi_1\to D_2\phi_2}
= \frac{1}{2}\int_{-1}^{+1}{\rm d}\cos\theta\,P_\ell(\cos\theta)\, {V}^{(S,I)}_{D_1\phi_1\to
D_2\phi_2}(s,t(s,\cos\theta))\,,
\label{eq:pwp}
\end{eqnarray}
where $\theta$ is the scattering angle between the incoming and outgoing particles in the center-of-mass frame,
and the Mandelstam variable $t$ is expressed as
\begin{eqnarray}
t(s,\cos \theta) =
M_{D_1}^2+M_{D_2}^2-\frac{\left(s+M_{D_1}^2-M_{\phi_1}^2\right)\left(s+M_{D_2}
  ^2-M_ { \phi_2 } ^2\right) } {2s} - \frac{\cos \theta}{2s}
\sqrt{\lambda(s,M_{D_1}^2,M_{\phi_1}^2)\lambda(s,M_{D_2}^2,M_{\phi_2}^2)}\,,
\end{eqnarray}
where $\lambda(a,b,c)=a^2+b^2+c^2-2ab-2ac-2bc$ is the K{\"a}ll{\'e}n function.
In most cases, one considers S-wave scattering, and, thus, the subscript $\ell=0$ is dropped.

At this point, it is worth to discuss some problems arising in such unitarization procedures. It is well-known that any unitarization approach that relies on right-hand unitarity and the on-shell approximation has the problem of violation of unitarity
when the left-hand cut occurs in the on-shell potential. For instance, the left-hand
cut in the $K\bar K\to K\bar K$ amplitude leads a violation of unitarity for
the $\pi\pi$ scattering in the $\pi\pi$--$K\bar K$ coupled-channel
system~\cite{GomezNicola:2001as,Dai:2011bs}. Note, however, that it was shown in
Refs.~\cite{Guerrero:1998ei, GomezNicola:2001as} that
the unitarity violation is numerically small in the $\pi\pi$-$K\bar K$ case, thus,
no serious problem is generated.
The same unitarity violation happens to the $D\phi$ scattering with $(S,I)=(0,1/2)$,
which has three coupled channels: $D\pi$, $D\eta$ and $D_s\bar K$.
One of the left-hand cuts from the inelastic
channel $ D_s\bar{K}\to D\eta$ amplitude, from ($1.488$~GeV)$^2$ to ($2.318$~GeV)$^2$,
overlaps with the right-hand cut starting from the $D\pi$ threshold.
Although this left-hand cut is not numerically important, its
presence together with other left-hand cuts and right-hand cuts make the
whole real axis nonanalytic.  As a result, the
coupled-channel amplitudes do not have the
correct analytic properties even in the relevant energy region. Consequently,
a pair of pole at $(2.046\pm i 0.050)~\text{GeV}$ are found on the first Riemann
sheet  for the coupled-channel $(S,I)=(0,1/2)$ amplitude. As we know, poles on
the first Riemann sheet can only be located on the real axis below the lowest
threshold, which are associated with bound states. A pole on the first sheet
with a non vanishing imaginary part or above the lowest threshold is inconsistent
with causality. The appearance of the pole on the first sheet in the coupled-channel
$(S,I)=(0,1/2)$ is due to the existence of the coupled-channel cut. The
left-hand cuts stem from the one-loop potentials but only at NNLO.
If one considers only a single-channel such as $D\pi\to D\pi$ with  $(S,I)=(0,1/2)$, there is no such problem as it comes from the left-hand cut of the inelastic channels.

\begin{figure}[t!]
\begin{center}
\includegraphics[width=0.7\textwidth]{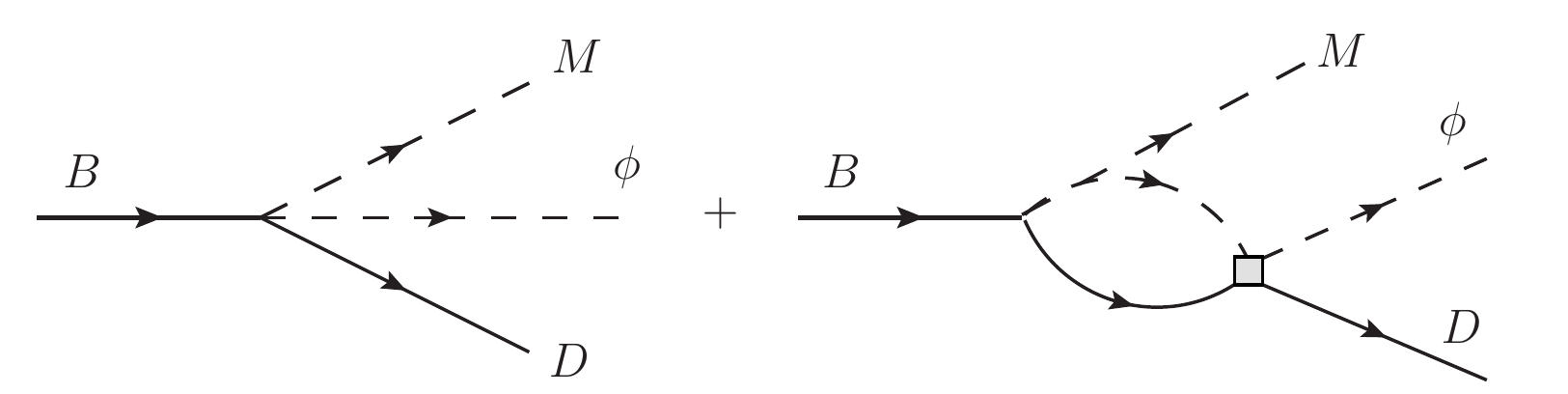}
\end{center}
\caption{The decay $B \to D\phi M$. The square denotes
  the final-state interactions between the $D$ meson and the Goldstone boson. In the loop,
  all relevant  coupled channels contribute.}\label{fig:feyndiag}
\end{figure}

Finally, we consider $D\phi$ scattering as a sub-process of the weak decays $B\to D\phi\phi$.
At low energies, for the processes with $\Delta b=1$ and $\Delta c=1$ the  interaction can be described
by the effective weak  Hamiltonian $H_\text{eff}$ which at LO has the form \cite{Gilman:1979bc}
\begin{equation}
\label{eq:heff}
H_\text{eff} = \frac{G_F}{\sqrt{2}}V_{cb}^\ast V_{ud}^{} \big( C_1^{} \mathcal{O}^d_1
+C_2^{} \mathcal{O}^d_2 \big) + (b\to s) +\text{h.c.}\,,
\end{equation}
with $G_F$ the Fermi constant, $V_{ij}$ the Cabibbo--Kobayash--Maskawa (CKM) matrix elements,
and the $C_i$ are scale-dependent Wilson  coefficients. Here, the tree-level operators read
\begin{equation}
\mathcal{O}^d_1  =  (\bar{c}_a b_b)_L(\bar{d}_bu_a)_L\,,~~~
\mathcal{O}^d_2  =  (\bar{c}_a b_a)_L(\bar{d}_bu_b)_L\,,
\end{equation}
with the subscripts $a$ and $b$ are color indices. The subscript $L$ indicates that
only the left-hand components of the quarks are involved.
Note that here the color space is irrelevant for our discussion, thus, we simply
drop the subscripts of $C_i$ and $\mathcal{O}_i$ hereafter.
One can make the effective Hamiltonian fully chirally invariant by introducing
a spurion $H_i^j$ transforming as~\cite{Bijnens:2009yr}
\begin{equation}
H_i^j \mapsto H_{i^\prime}^{j^\prime} (g_L^{})_i^{i^\prime} (g_L^\dagger)_{j^\prime}^j\,.
\end{equation}
Then the new Hamiltonian
\begin{equation}
H_\text{eff}^\prime = \frac{G_F}{\sqrt{2}}V_{cb}^\ast V_{ud}^{} H_i^j C(\bar{c} b)_L (\bar{q}_L^i q_{Lj})
\end{equation}
is chirally invariant. For Eq.~\eqref{eq:heff}, the spurion $H_i^j$
(the lower index labels rows and the upper one labels columns)
corresponds to
\begin{equation}
H = \begin{pmatrix}
0 & 0 & 0 \\
1 & 0 & 0 \\
V_{us}/V_{ud} & 0 & 0
\end{pmatrix}~ .
\end{equation}
Here, $V_{us}/V_{ud}$ is nothing but $-\sin \theta_1$,
with $\theta_1$ the Cabbibo angle. Then the component $H_2^1$ describes the Cabibbo-allowed decays and $H_{3}^1$
the Cabibbo-suppressed ones. In the matrix form, $H$ transforms under chiral symmetry as
$H \mapsto g_L^{} H g_L^\dagger$. It is more convenient to introduce a homogeneously transforming suprion as
$t=u H u^\dagger$. With those ingredients, one constructs the effective Lagrangian describing the three-body
nonleptonic decays of $B$ mesons to $D$ mesons
and two light pseudoscalars. We are interested in the region of the
invariant mass of a pair of the $D$ and one pseudoscalar not far from their
threshold, such that this light pseudoscalar can be safely treated as a soft Goldstone
boson, while the other one moves fast and can be treated as a matter field
rather than a Goldstone boson. The fast moving pseudoscalar is realized linearly in a matrix form $M$
transforming as
\begin{equation}
M \mapsto h M h^\dagger~ ,
\end{equation}
and it has the same form as $\phi$, i.e.,
\begin{equation}
M =\begin{pmatrix}
\frac{1}{\sqrt{2}}\pi^0+\frac{1}{\sqrt{6}}\eta  & \pi^+ & K^+  \\
\pi^- & -\frac{1}{\sqrt{2}}\pi^0 +\frac{1}{\sqrt{6}}\eta & K^0 \\
K^- & \bar{K}^0 & -\frac{2}{\sqrt{6}}\eta
\end{pmatrix}.
\end{equation}
Consequently, utilizing  the power counting described above, chiral symmetry implies that the  effective Lagrangian at $\mathcal{O}(p)$ has the form of~\cite{Savage:1989ub,Du:2017zvv,Du:2019oki}
\begin{align}
\label{lag:eff}
{\cal L}_\text{eff} & = \bar{B}\, \Big[ c_1(u_\mu t M+Mtu_\mu ) +c_2 (u_\mu M + M u_\mu)t
+c_3 t(u_\mu M + M u_\mu )  \nonumber \\
&\qquad\qquad + c_4 (u_\mu \langle Mt\rangle +M \langle u_\mu t\rangle
) + c_5 t \langle M u_\mu \rangle +c_6 \langle (Mu_\mu +u_\mu M )t\rangle
\Big]\, \nabla^\mu  D^\dag \nonumber \\
& + \bar{B}\,\Big[ d_1 (u_\mu tM - Mtu_\mu ) +d_2 (u_\mu M - M u_\mu ) t
+d_3 t(u_\mu M - M u_\mu ) \nonumber \\
&\qquad\qquad +d_4 (u_\mu \langle Mt \rangle - M\langle u_\mu t\rangle
)+ d_6 \langle (Mu_\mu - u_\mu M )t\rangle \Big] \, \nabla^\mu D^\dag\,,
\end{align}
where $\bar{B}= (B^-,\bar{B}^0,\bar{B}^0_s)$, the $c_i$ and $d_i$ are LECs,
and $\langle\dots\rangle$ denotes a trace
in the SU(3) flavor space. Note that the momentum operator $\nabla_\mu$ in Eq.~\eqref{lag:eff}
is chosen to act on the charmed meson field $D$. It could act on $B$ (or $M$) independently as
well. However, in our case, $M_B\gg M_D+2M_\pi$ and $M_D\gg M_\pi$ imply that
they would produce the same structures up to the LO and we can combine them by redefining the
LECs in the heavy meson limit. This effective Lagrangian considers both the chiral symmetry and
flavor SU(3) constraints (the latter  has been considered in Ref.~\cite{Savage:1989ub}). Finally, we divide the
Lagrangaian~\eqref{lag:eff} into two groups which are symmetric and antisymmetric in
the two light pseudoscalars, which correspond to the cases where the relative
orbital angular momentum of the light pseudoscalars pair is even and odd,
respectively~\cite{Savage:1989ub}.
The nonleptonic $B$-meson three-body decays $B\to D\phi M$, where $M$ denotes the fast moving light pesudoscalar
and $\phi$ denotes the soft one, provide access to the $D$-$\phi$ interaction
via the final-state interaction in the energy region where the $D\phi$ system is not far from its threshold.
Of course, there also final-state interactions between the $D$ and the hard $M$ and between
$M$ and $\phi$. These do not produce any nontrivial structure sensitive to the energy variation and thus
can encoded into an extra complex factor. The pertinent Feynman diagrams for the
decay $B\to D\phi M$ are shown in Fig.~\ref{fig:feyndiag}, where the square denotes the final-state
interactions in the  $D\phi$ subsystem.
The decay amplitude projected into the $D\phi$ channel at low energies can be decomposed into S-,
P- and D-waves, which corresponds to the orbital angular momentum of the $D\phi$ pair
$\ell=0$, $\ell=1$, and $\ell=2$, in order,
\begin{equation}
\label{eq:decayamplitudes}
\mathcal{A}(B\to D \phi M) = \mathcal{A}_0(s) +\sqrt{3}\mathcal{A}_1(s)
P_1(z) +\sqrt{5}\mathcal{A}_2(s)P_2(z)\,.
\end{equation}
Here, $\mathcal{A}_{0,1,2}(s)$ denote  the partial wave decomposed amplitudes for $D\phi$ in the S-, P-
and D-waves, respectively, and $P_\ell(z)$ are the Legendre polynomials with $z$ the cosine of the helicity
angle of the $D\phi$ system, i.e., the angle between the moving directions of the $\phi$ and the $M$
in the $D\phi$ rest frame.
For the P- and D-waves, the resonances are relatively narrow and, thus, it is reasonable to parameterize them
by Breit-Wigner amplitudes. For the S-wave, the diagrams in Fig.~\ref{fig:feyndiag} are calculated using
the effective Lagrangian~\eqref{lag:eff} and the final-state $D\phi$ interaction is determined from
the coupled channels approach described before.

Consider first the decay $B^-\to D^+\pi^-\pi^-$. Here, the relative orbital angular momenta of the two light mesons
is even. This corresponds to the first term of the Lagrangian in Eq.~\eqref{lag:eff} parameterized by the
LECs $c_i$. The corresponding production vertices for the possible
intermediate states $D^0\pi^0$, $D^+\pi^-$, $D^0\eta$ and $D_s^+K^-$ are given in Tab.~\ref{tab:DpipiS}.
\begin{table}[t!]
\begin{center}
\begin{eqnarray}
\begin{array}{|c|c|}
\hline
\text{Process} & \text{Production amplitude} \\
\hline
B^- \to D^0\pi^0 \pi^- & \dfrac{1}{F_0} (c_1+c_4)p_D\cdot p_\pi \\
B^- \to D^0\eta\pi^- & \dfrac{1}{\sqrt{3}F_0}(c_1+c_4+2c_2+2c_6)p_D\cdot p_\eta  \\
B^- \to D^+\pi^-\pi^- & \dfrac{2\sqrt{2}}{F_0} (c_1+c_4)p_D \cdot p_\pi \\
B^- \to D_s^+ K^-\pi^- & \dfrac{\sqrt{2}}{F_0} (c_1+c_4)p_{D_s} \cdot p_K  \\
\hline
\end{array}\nonumber
\end{eqnarray}
\caption{Production vertices for the possible intermediate states contributing to $B^-\to D^+\pi^-\pi^-$. The four-momenta of the charmed meson and the Goldstone boson are denoted by
$p_{D_{(s)}}$ and $p_\phi$, respectively.}
\label{tab:DpipiS}
\end{center}
\end{table}
In the heavy quark limit, we have $p_D\cdot p_\phi = M_D E_\phi$, with
$E_\phi$ the energy of $\phi$ in the rest frame of the $D\phi$ subsystem. It is convenient to
introduce two parameters $A$ and $B$ via \cite{Du:2017zvv}
\begin{eqnarray}
\label{eq:defAB}
A = \frac{\sqrt{2}}{F_0} (c_1+c_4)M_D~\,, ~~~B = \frac{2\sqrt{2}}{3F_0}  (c_2+c_6) M_D\,.
\end{eqnarray}
Consequently, the S-wave decay amplitude for $B^-\to D^+\pi^-\pi^-$ that contains the
final-state interaction can be written as
\begin{eqnarray}
\mathcal{A}_0(s) & = & 2 A E_\pi + 2 A E_\pi G_{D\pi}(s) T_{D^+\pi^- \to D^+\pi^-}(s)
+ \frac{A}{\sqrt{2}}E_\pi G_{D\pi}(s) T_{D^0\pi^0 \to D^+\pi^-} (s)\nonumber \\
&&+ \frac{A+3B}{\sqrt{6}}E_\eta G_{D\eta}(s) T_{D^0\eta\to D^+\pi^-} (s)
 + A E_K G_{D_s\bar K}(s) T_{D_s^+K^-\to D^+\pi^-}(s)\,,
\end{eqnarray}
with $s$  the center-of-mass energy squared of the $D\phi$ system, and $G_{D\phi}(s)$ is the loop function shown
in Fig.~\ref{fig:feyndiag}  coupling to  the channel $D\phi$. Unitarity allows to express this loop function
via a once-subtracted dispersion relation,
\begin{eqnarray}
\label{eq:G}
G_{D\phi}(s)&=& \frac{1}{16\pi^2}  \Bigg\{ a(\mu) +\log\frac{M_D^2}{\mu^2} + \frac{M_\phi^2-M_D^2+s}{2s}
\log \frac{M_\phi^2}{M_D^2}
 + \frac{\sigma}{2s}\bigg[ \log (s-M_D^2+M_\phi^2+\sigma) -\log (-s+M_D^2-M_\phi^2
+\sigma) \nonumber \\
&&\qquad\qquad+\log(s +M_D^2-M_\phi^2+\sigma) -\log(-s-M_D^2+M_\phi^2+\sigma )\bigg] \Bigg\}\,,
\end{eqnarray}
with $a(\mu)$ a scale-dependent subtraction constant, $\mu$ the scale of
dimensional regularization, and  $\sigma=\sqrt{\lambda(s,M_\phi^2,M_D^2)}$.
The subtraction $a$ is related to the renormalization of the interaction
vertices and varies for different processes. A change of $\mu$ can be absorbed into a corresponding change of $a$,
often one takes $\mu=1$~GeV.
The amplitudes for the final-state interactions can be expressed in the isospin basis. While
$D^+\pi^-$ can be decomposed into isospin $I=1/2$ and $3/2$ systems,  $D^+\eta$ and
$D_s^+ K^-$ can only form $I=1/2$. The relations between
the isospin basis and physical particle basis are given by~\cite{Guo:2009ct,Yao:2015qia}
\begin{eqnarray}
T_{D^0\pi^0\to D^+\pi^-} & = & -\frac{\sqrt{2}}{3}T^{3/2}
_{D\pi\to D\pi} +\frac{\sqrt{2}}{3}T_{D\pi\to D\pi}^{1/2}\,,~~
T_{D^0\eta^{~} \to D^+\pi^-}  =  \sqrt{\frac{2}{3}}
T^{1/2}_{D\eta \to D\pi}\,, \nonumber \\
T_{D^+\pi^-\to D^+\pi^-} & = & \frac{1}{3}T^{3/2}
_{D\pi \to D\pi} + \frac{2}{3}T^{1/2}_{D\pi \to D\pi}\,,~~
T_{D_s^+K^- \to D^+\pi^-}  =  \sqrt{\frac{2}{3}}
T_{D_s\bar K \to D\pi}^{1/2}\,,
\end{eqnarray}
where the superscripts indicate the total isospin $I$. The amplitudes in the isospin
basis can be found in Refs.~\cite{Guo:2009ct,Liu:2012zya,Yao:2015qia}. As a
result, we get the S-wave decay amplitude for the process $B^-\to D^+\pi^-\pi^-$~\cite{Du:2017zvv}
\begin{eqnarray}
\label{eq:swave}
\mathcal{A}_0(s) & = & A E_\pi \Big[ 2+ G_{D\pi}(s) \Big( \frac{5}{3}
T_{11}^{1/2}(s) +\frac{1}{3}T^{3/2}(s) \Big) \Big]
+\frac{1}{3}(A+3B)E_\eta G_{D\eta}(s) T_{21}^{1/2}(s)
+\sqrt{\frac{2}{3}}AE_KG_{D_s\bar K}(s) T^{1/2}_{31}(s)\,.\nonumber\\
&&
\end{eqnarray}
Here, we write the scattering amplitudes in the matrix form $T_{ij}^I(s)$ with the
total isospin $I$, where $i$, $j$ are channel indices with 1, 2 and 3 referring
to $D\pi$, $D\eta$ and $D_s\bar K$, respectively. Note that only two LECs
$A,B$, see Eq.~\eqref{eq:defAB}, appear in Eq.~\eqref{eq:swave}.
The production vertices responsible for other processes such as $B_s^0\to \bar{D}^0K^-\pi^+$,
$B^0\to\bar{D}^0\pi^-\pi^+$, $B^-\to D^+\pi^-K^-$, and $B^0\to\bar{D}^0\pi^-K^+$ can also be derived
from  Eq.~\eqref{lag:eff}. The weak production vertices needed for
those decays are listed in Tab.~\ref{tab:amps}.
\begin{table}[t!]
\begin{eqnarray}
\begin{array}{|l|c|}
\hline
\text{Reaction} & \text{Weak production vertex} \\
\hline
B_s^0\to \bar{D}^0K^-\pi^+ & E_K\Big( (c_2+c_4) + (d_2+d_4)\Big)  \\
B_s^0\to D^-\bar{K}^0 \pi^+ & E_K\Big((c_1+c_4) +(d_1+d_4)\Big) \\
B_s^0\to \bar{D}_s\eta\pi^+ & \dfrac{\sqrt{2}}{\sqrt{3}}\dfrac{M_{D_s}}{M_D}E_\eta\Big( (c_6-c_4)-d_4  \Big) \\
B_s^0\to \bar{D}_s\pi^0\pi^+ & \sqrt{2}\dfrac{M_{D_s}}{M_D}E_\pi d_6 \\
\hline
B^0\to \bar{D}^0\pi^-\pi^+ & E_\pi\Big( (c_2+c_3+c_4+2c_5)+(d_2-d_3+d_4)\Big) \\
B^0\to D^-\eta \pi^+ & \dfrac{1}{\sqrt{6}}E_\eta \Big( (c_1+2c_3+c_4+2c_6)+(d_1+d_4)\Big) \\
B^0\to D_s^- K^0\pi^+ & \dfrac{M_{D_s}}{M_D}E_K\Big( (c_3+c_4) -(d_3-d_4) \Big)\\
B^0\to  D^- \pi^0\pi^+ & -\dfrac{1}{\sqrt{2}}E_\pi\Big( (c_1+c_4) +(d_1-2d_3+d_4-2d_6)\Big) \\
\hline
B^0\to \bar{D}^0\pi^-K^+ & -\sin\theta_1 E_\pi\Big( (c_2+c_4)+(d_2+d_4)\Big)\\
B^0\to D^-\pi^0K^+ & -\sin\theta_1\dfrac{1}{\sqrt{2}}E_\pi\Big( -(c_4-c_6) -(d_4-d_6) \Big) \\
B^0\to D^-\eta K^+ & -\sin\theta_1\dfrac{1}{\sqrt{6}}E_\eta \Big( (c_4-c_6)+(d_4+3d_6)\Big) \\
B^0\to D_s^-K^0 K^+ & -\sin\theta_1\dfrac{M_{D_s}}{M_D}E_K \Big( (c_1+c_4)+(d_1+d_4) \Big) \\
\hline
B^-\to D^0\pi^0 K^- & -\sin\theta_1 \dfrac{1}{\sqrt{2}}E_\pi \Big( (c_1+c_4+c_2+c_6) -(d_1-d_2-d_4-d_6)\Big)\\
B^-\to D^0\eta K^- & -\sin\theta_1 \dfrac{1}{\sqrt{6}}E_\eta \Big( (c_1+c_4-c_2-c_6) -(d_1-3d_2-d_4-3d_6) \Big) \\
B^-\to D^+\pi^- K^- & -\sin\theta_1 E_\pi \Big( (c_1+c_4) -(d_1-d_4) \Big) \\
B^-\to D_s^+K^- K^+ & -\sin\theta_1\dfrac{2M_{D_s}}{M_D} E_K(c_1+c_4 )\\
\hline
\end{array}\nonumber
\end{eqnarray}
\caption{Weak amplitudes contributing to the decays $B_s^0\to \bar{D}^0K^-\pi^+$, $B^0\to\bar{D}^0\pi^-\pi^+$,
$B^-\to D^+\pi^-K^-$, and $B^0\to\bar{D}^0\pi^-K^+$ through coupled-channel effects. Note that an overall factor $\sqrt{2}M_D/{F_0}$ has been absorbed in the LECs $c_i,d_i$.}
\label{tab:amps}
\end{table}
For all these decays one expects that, at least in the low-energy tails of the invariant mass of the
$D\phi$ subsystems, the effects from the crossed-channel final-state interactions, that is the interactions between
the soft and hard light mesons and those between the $D$ meson and the hard pseudoscalar meson, do not
produce any nontrivial structure in the $D\phi$ distributions. This is supported by the analyses in
Refs.~\cite{Albaladejo:2016jsg,LHCb:2015eqv,LHCb:2015tsv,LHCb:2015klp}. Consequently,  effects from the
crossed-channel final-state interactions can be represented by an extra undetermined complex factor
for each partial wave, similar to what is done in isobar models. As was done above for the decay
$B^-\to D^+\pi^-\pi^-$, one obtains the S-wave amplitudes for the various other decays:
\begin{eqnarray}
\label{eq:amp:1}
\mathcal{A}_0(B_s^0 \!\!&\to&\!\! \bar{D}^0K^-\pi^+) =   (c_2+c_4+d_2+d_4)E_K +
d_6 E_\pi G_{D_s\pi}(s)T_{D_s\pi\to\bar{D}\bar{K}}^1(s) 
+ \frac{1}{2}(c_2-c_1+d_2-d_1)E_K G_{DK}(s) T_{\bar{D}\bar{K}\to \bar{D}\bar{K}}^1(s) \nonumber\\
&+& \frac{1}{2}(c_1+c_2+2c_4+d_1+d_2+2d_4)E_K G_{DK}(s) T_{\bar{D}\bar{K}\to \bar{D}\bar{K}}^0(s)
+ \sqrt{\frac13}(c_4-c_6+d_4)E_\eta G_{D_s\eta}(s)T_{\bar{D}_s\pi\to \bar{D}\bar{K}}^0(s)\,,
\end{eqnarray}
\begin{eqnarray}
\label{eq:amp:2}
\mathcal{A}_0(B^0 \!\!&\to&\!\! \bar{D}^0\pi^-\pi^+)\ =
(c_2+c_3+c_4+2c_5+d_2-d_3+d_4)E_K +\frac13E_\pi G_{D\pi}(s) T_{D\pi\to D\pi}^{1/2}(s)\nonumber\\
&\times& (c_1+2c_2+2c_3+3c_4+4c_5+d_1+2d_2-4d_3+3d_4-2d_6)\nonumber\\
&+& \frac{1}{3} (c_2-c_1+c_3+2c_5-d_1+d_2+d_3+2d_6)E_\pi G_{D\pi}(s)T_{D\pi\to D\pi}^{3/2}(s)
+ \frac13 (c_1+2c_3+c_4+2c_6+d_1+d_4)E_\eta \nonumber\\ &\times&G_{D\eta}(s)T_{D\eta\to D\pi}^{1/2}(s)
 + \sqrt{\frac23}(c_3+c_4-d_3+d_4)E_KG_{D_s\bar K}(s)T_{D_s\bar{K}\to D\pi}^{1/2}(s)\,,
\end{eqnarray}
\begin{eqnarray}
\label{eq:amp:3}
\mathcal{A}_0(B^-\!\!&\to&\!\! D^+\pi^- K^-) = -\sin\theta_1 (c_1+c_4-d_1+d_4)E_\pi -2\sqrt{\frac23}\sin\theta_1(c_1+c_4)
E_KG_{D_s\bar K}(s)T^{1/2}_{D_s\bar{K}\to D\pi}(s) \nonumber\\
&-& \sin\theta_1(3c_1+c_2+3c_4+c_6-3d_1+d_2+3d_4+d_6)E_\pi G_{D\pi}(s)T^{1/2}_{D\pi\to D\pi}(s)
- \sin\theta_1(c_1-c_2+c_4-c_6\qquad\quad\quad\nonumber\\
&-&d_1+d_4+3d_2+3d_6)E_\eta G_{D\eta}(s)T^{1/2}_{D\eta\to D\pi}(s)
+ \frac13 \sin\theta_1(c_2+c_6+d_2+d_6) E_\pi G_{D\pi}(s)T^{3/2}_{D\pi\to D\pi}(s)\,,
\end{eqnarray}
\begin{eqnarray}
\label{eq:amp:4}
\mathcal{A}_0(B^0\!\!&\to&\!\! \bar{D}^0 \pi^- K^+)  = -\sin\theta_1(c_2+c_4+d_2+d_4)E_\pi
-\frac{1}{3} \sin\theta_1(2c_2+3c_4-c_6+2d_2+3d_4-d_6)E_\pi G_{D\pi}(s)T_{D\pi\to D\pi}^{1/2}(s)\nonumber\\
&-& \frac{1}{3} \sin\theta_1(c_4-c_6+d_4+3d_6)E_\eta G_{D\eta}(s)T_{D\eta \to D\pi}^{1/2}(s)
- \sqrt{\frac{2}{3}}\sin\theta_1(c_1+c_4+d_1+d_4)E_K G_{D_s\bar K}(s)T_{D_s\bar{K}\to D\pi}^{1/2}(s) \nonumber\\
&-& \frac{1}{3} \sin\theta_1(c_2+c_6+d_2+d_6)E_\pi G_{D\pi} T^{3/2}_{D\pi\to D\pi}(s)\,.
\end{eqnarray}
It can be shown easily that the so defined amplitudes indeed fulfill two-body unitarity,
$\mathcal{A}-\mathcal{A}^\ast =-2\mathrm{i}T\rho \mathcal{A}^\ast= -2\mathrm{i} T^\dag \rho \mathcal{A}$, with
$\rho$ the two-body phase space factor. As noted before, the complex decay amplitudes for P- and
D-waves are described by an isobar model as coherent sums of intermediate resonant decays.
This is reasonable because of the relatively narrow widths of the resonances in P- and D-waves.
These are the $D^\ast$, $D^\ast_2(2460)$, $D^\ast_1(2680)$, $D_s^\ast$ and the $D_{s2}(2573)$.
Explicit expressions for their contributions are given in, e.g., Ref.~\cite{Du:2019oki}.
The S-wave amplitudes listed above contain
alltogether 11 LECs, but only  10 combinations are independent. Furthermore, to reduce the their
correlations in the fit procedure, one instead uses the following combinations of LECs,
\begin{eqnarray}
A &=&c_1+c_4\,, \quad B= \frac{3}{2} (c_2+c_6), \quad C= c_2+c_4\,, \quad D= c_3+2c_5,
\quad E = c_3+c_6\,, \nonumber \\
A^\prime &=& d_1-d_4\,, ~~\quad B^\prime = d_2+d_6\,, \quad C^\prime = d_4-d_6\,, \quad
D^\prime = d_4+d_6\,, \quad E^\prime = d_3\,.\quad ~~
\end{eqnarray}
For example, if one fits the data for the decay $B^-\to D^+\pi^-\pi^-$, there are just two free
parameters, namely the ratio $B/A$ and the subtraction constant $a(\mu)$ (the quantity $A$ alone
can not be determined as it disappears in the normalization). For the others channels, one
could use the same value of the subtraction constant to minimize the number of parameters, and, thus, one would be left with one parameter fits. The P- and D-wave resonances (with spin $1$ and $2$, respectively) can be taken as in the LHCb analyses ~\cite{LHCb:2014ioa,LHCb:2016lxy} and, thus, introduce no new parameters.

At last, we collect the formulae for the angular moments used in the LHCb analysis. These  contain
important information about the partial-wave phase variations. The angular moments $\langle P_\ell\rangle$ are
obtained by weighting the event distribution in the invariant mass by the Legendre polynomial of order $\ell$
with respect to $z$,
\begin{equation}
\langle P_\ell(s)\rangle = \int_{-1}^{+1} d z \frac{d\Gamma}{d\sqrt{s} dz}P_\ell(z)\,.
\end{equation}
The angular moments are most powerful when a resonance is present only in one invariant mass combination.
The structures show up in moments up to 2$J$, where $J$ is the spin of the contributing
resonance~\cite{LHCb:2014ioa}. Neglecting partial waves with $\ell \geq 3$, the first few moments, that are
normalized relative to each another, read
\begin{eqnarray}
\label{eq:angularm}
\langle P_0\rangle & \propto & |\mathcal{A}_0|^2+|\mathcal{A}_1|^2+|\mathcal{A}_2|^2\,,~~
\langle P_1 \rangle  \propto \dfrac{2}{\sqrt{3}}|\mathcal{A}_0||\mathcal{A}_1 | \cos (\delta_0-\delta_1 )
+\dfrac{4}{\sqrt{15}} |\mathcal{A}_1|  |\mathcal{A}_2| \cos (\delta_1 -\delta_2 )\,, \nonumber \\
\langle P_2 \rangle & \propto & \dfrac{2}{5}|\mathcal{A}_1|^2 + \dfrac{2}{7}|\mathcal{A}_2|^2
+\dfrac{2}{\sqrt{5}}|\mathcal{A}_0| |\mathcal{A}_2| \cos (\delta_0-\delta_2 )\,,~~
\langle P_3\rangle  \propto  \dfrac{6}{7}\sqrt{\dfrac{3}{5}}
|\mathcal{A}_1||\mathcal{A}_2|\cos (\delta_1-\delta_2)\,,
\end{eqnarray}
where $\delta_i$ is the phase of $\mathcal{A}_i$, i.e., $\mathcal{A}_i=|\mathcal{A}_i|e^{\mathrm{i}\delta_i}$. Instead
of $\langle P_1\rangle$ and $\langle P_3\rangle$, it is advantageous to analyze their linear combination
as proposed in
Ref.~\cite{Du:2017zvv},
\begin{equation}
\label{eq:angularm13}
\langle P_{13}\rangle = \langle P_1\rangle -\frac{14}{9}\langle P_3\rangle
\propto \frac{2}{\sqrt{3}} |\mathcal{A}_0||\mathcal{A}_1|\cos (\delta_0-\delta_1)\,,
\end{equation}
which only depends on the S-P interference up to $\ell=2$ and is particularly sensitive to the S-wave phase motion.


\section{Results: Well separated resonances}
\label{sec:well}
After discussing methods based on effective field theories and the
specifics of the unitarization methodology we now turn to the
discussion of recent theoretical determinations of resonance parameters. In this section we start with well separated resonances and with results from lattice QCD.
As discussed in Sect.~\ref{sec:LQCD} the main steps of
this approach include the calculation of finite-volume energy eigenvalues
which are then related to the physical quantities, such as
phase-shifts. Subsequently, these quantities can be
used to determine the universal parameters of the resonances. Obviously, the
last step of this strategy is not much different from the
typical approach to estimating resonance parameters from experimental data.
Here, however, we will focus on the final results given in the respective publications and compare, where applicable, to results obtained from experimental data.

\subsection{The \texorpdfstring{$\rho(770)$}{Rho(770)}-resonance}

\begin{figure}[t!]
  \centering
  \includegraphics[width=.9\textwidth]{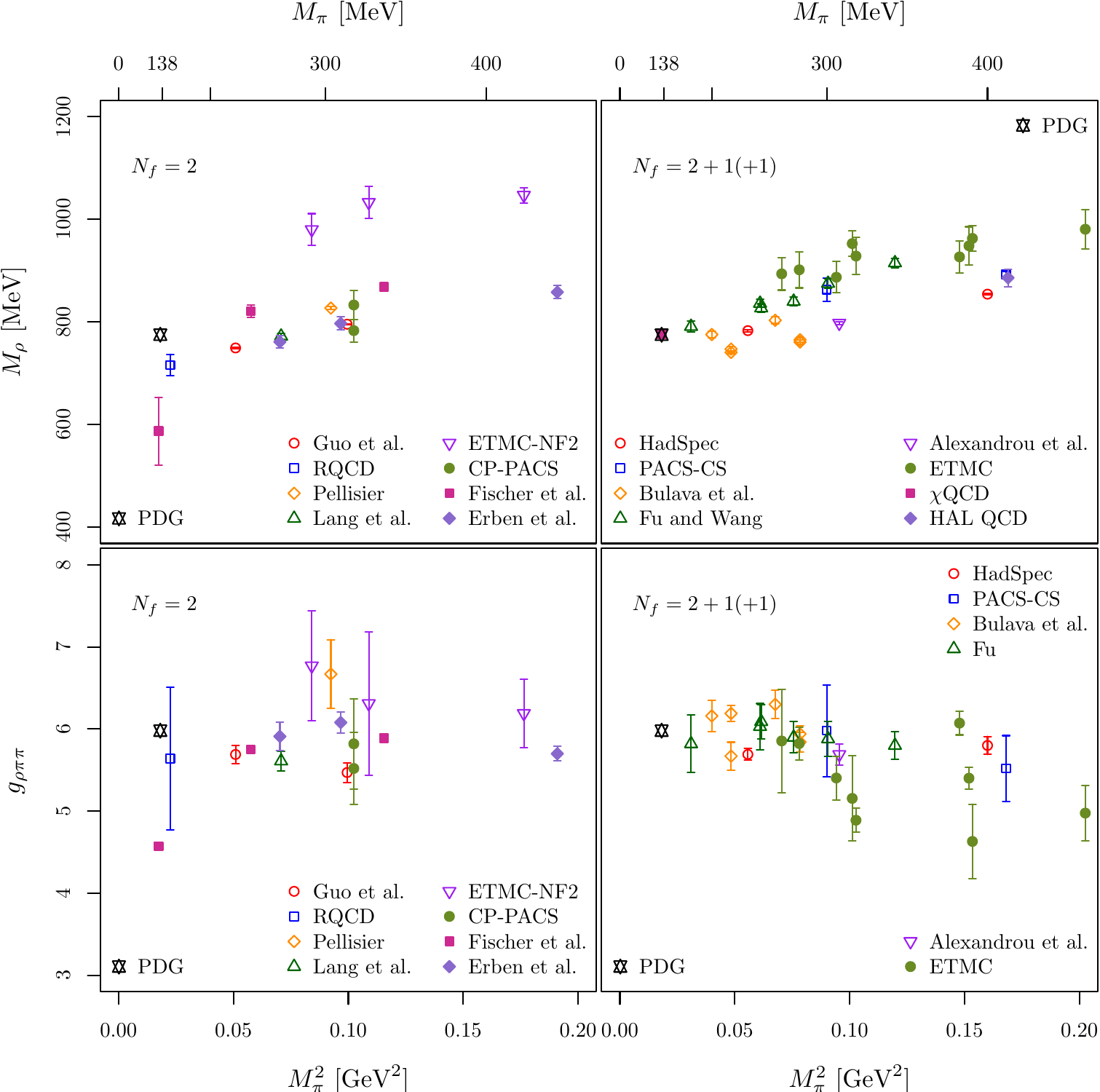}

\caption{$M_\rho$ and $g_{\rho\pi\pi}$ as functions of $M_\pi^2$ for
  lattice results available in the literature with $N_f=2$, $N_f=2+1$
  and $N_f=2+1+1$ dynamical quark flavors. For the references see the text.}
\label{fig:rhoMpisq}
\end{figure}

The lightest vector-isovector unflavored meson, the so-called $\rho$-resonance, has been
discussed throughout the current manuscript using different
approaches. Due to its relatively small width and position far from
the relevant inelastic thresholds it appears clearly as a prototypical
resonance signal in, e.g., the phase-shifts of $\pi\pi$ scattering. Since the $\rho$ also represents the second lightest resonance in QCD, it became a benchmark state for lattice QCD
calculations with the Lüscher method with first pioneering
(pre-Lüscher) studies available in
Refs.~\cite{Gottlieb:1983rh,Weingarten:1982qe}.

Compared to other states, there are many publications available focusing on the $\rho$-resonance. Thus,
we will concentrate in the following on results obtained with at least
$N_f=2$ dynamical quark flavors and leave out quenched
results. We will also not comment further on more approximate
methods like the one presented in
Refs.~\cite{McNeile:2002fh,Michael:2006hf}. Thus, we summarize  the
activity of studying the $\rho$-resonance on the lattice with the
methodology discussed in Sec.~\ref{sec:LQCD}. In order to
avoid repetitions we refer for complimentary discussions to
related reviews~\cite{Briceno:2016mjc,FlavourLatticeAveragingGroup:2019iem,Padmanath:2018zqw}.

A variety of lattice QCD calculations of the $\rho$-resonance with
$N_f=2$, $N_f=2+1$, and $N_f=2+1+1$ dynamical quark
flavors exists. As discussed before, the corresponding phase-shift is estimated at
different values of the pion mass, sometimes including ensembles with
physical or almost physical value of the pion mass. Let us first focus on the results for mass and width (or coupling) of the $\rho$ obtained at the different (not necessarily physical) pion mass values and compare. Often, a Breit-Wigner fit to the phase-shift data is used to determine the resonance parameters, which for the $\rho$ does usually not differ significantly from better parametrizations like the IAM.

The bare lattice results for $M_\rho$ and the coupling
$g_{\rho\pi\pi}$, related to $\Gamma_\rho$ via
\begin{equation}
 \Gamma_{\rho} = \frac{2}{3} \frac{g_{\rho\pi\pi}^2}{4\pi} \frac{q_\pi^3}{M_{\rho}^2}\,, \qquad q_\pi =  \sqrt{\frac{M_{\rho}^2}{4}-M_\pi^2}\,,
  \label{eq:rho_width}
\end{equation}
are plotted in Fig.~\ref{fig:rhoMpisq} as functions of
$M_\pi^2$. From the four panels, the two in the upper row depict $M_\rho$ and
in the lower one $g_{\rho\pi\pi}$, while the two in the left column are for $N_f=2$ and in the right one for $N_f=2+1(+1)$. In \cref{fig:rhoMpisq} we include $N_f=2$ results from
Guo et al.~\cite{Guo:2016zos}, CP-PACS~\cite{CP-PACS:2007wro},
RQCD~\cite{Bali:2015gji},  Lang et al.~\cite{Lang:2011mn},
ETMC-NF2~\cite{Feng:2010es}, Fischer et al.~\cite{Fischer:2020yvw},
Pelissier~\cite{Pelissier:2012pi}, and Erben et al.~\cite{Erben:2019nmx}.
For $N_f=2+1$ we
include PACS-CS~\cite{CS:2011vqf}, HadSpec~\cite{Wilson:2015dqa},
Bulava et al.~\cite{Bulava:2016mks}, Fu and Wang~\cite{Fu:2016itp},
$\chi$QCD~\cite{Sun:2015enu}, Alexandrou et
al.~\cite{Alexandrou:2017mpi}, Andersen et
al.~\cite{Andersen:2018mau},
HAL QCD~\cite{Akahoshi:2021sxc}
and for $N_f=2+1+1$ ETMC~\cite{ExtendedTwistedMass:2019omo}.

Let us mention that $\chi$QCD~\cite{Sun:2015enu} has for the first
time determined also $f_\rho$ using a partially quenched, mixed action
approach with overlap valence fermions on a staggered sea. As a value
they find $f_{\rho^-}=208.5(5.5)(0.9)\ \mathrm{MeV}$ in good agreement with the
experimental value. While their value for $M_\rho$ is spot on and,
thus, hiding behind the PDG value in \cref{fig:rhoMpisq}, they do not
determine the width or the coupling. The coupling determined by HAL
QCD in Ref.~\cite{Akahoshi:2021sxc} has a value around $g_{\rho\pi\pi}\approx12$ and is, therefore, not
visible in Fig.~\ref{fig:rhoMpisq}.

In general the picture emerging from the different panels in \cref{fig:rhoMpisq} is overall
consistent, in particular for $N_f=2+1(+1)$ results. Exact agreement
of all these results is not to be expected, because they are not
extrapolated to the continuum limit and also different scale setting
procedures are used. However, one can certainly conclude that the
coupling has little pion mass dependence towards the chiral limit.
The notable exception is the value for the coupling determined by HAL
QCD, which is double the experimental value.

From \cref{fig:rhoMpisq} it can be observed that the $N_f=2+1(+1)$ results appear to be more consistent. There is, in our opinion, no reason of principle for this. We rather assume that most of the $N_f=2+1(+1)$ results are more recent and have, thus, less uncontrolled uncertainties.

Let us next turn to extrapolations of the lattice results presented above to the physical pion mass value (assuming that the strange and charm quarks have been tuned to their physical values, if present.) The extrapolations have been performed either by the lattice practitioners themselves, or by other groups. The latter is the case for Niehus et al.~\cite{Niehus:2020gmf}, where lattice data from Ref.~\cite{Bulava:2016mks} and Ref.~\cite{Wilson:2015dqa} is being
analysed using different methods (see below) to extrapolate to the
physical point. Similarly, Mai et al.~\cite{Mai:2019pqr} use data from
Ref.~\cite{Guo:2016zos}.
In the works of $\chi$QCD~\cite{Sun:2015enu} and RQCD~\cite{Bali:2015gji} the $\rho$-resonance parameters have been even estimated
directly with physical (or nearly physical) value of the pion mass, making an extrapolation unnecessary.

\begin{table*}[thb!]
\centering
\begin{tabular*}{\textwidth}{@{\extracolsep{\fill}}llcllll}
\hline
Reference & $N_f$ & Pion mass & Chiral & Cont. & $M_\rho$\,[MeV] & $\Gamma_\rho$[MeV] \\
&& range [MeV] & extr. & extr. & \\
\hline
ETMC-NF2-a~\cite{Feng:2010es} &2& $290-480$ & VMEFT & x & $821(24)$ & $171(31)$ \\
ETMC-NF2-b~\cite{Feng:2010es} &2& $290-480$ & mVMEFT & x & $850(35)$ & $166(49)$ \\
RQCD~\cite{Bali:2015gji} &2& $150$ & -- & x & $716(30)$ &
$113(35)$\\
Fischer et al.~\cite{Fischer:2020yvw} &2& $132-340$ & IAM NLO & x & $786(20)$
& $180(6)$ \\
Guo et al.~\cite{Guo:2016zos} &2& $226-315$ & UCHPT & x & $720(15)$ &
$120(1)$ \\
Mai et al.~\cite{Mai:2019pqr} &2& $226-315$ & IAM NLO & x & $724(4)$ &
$133(2)$ \\
\hline
$\chi$QCD~\cite{Sun:2015enu} &2+1& 139.2(4) & -- & x & $776(6)$& -\\
Fu and Wang~\cite{Fu:2016itp} &2+1&176-346 & VMEFT & x &
$780(16)$&$145(17)$\\
Niehus-a~\cite{Niehus:2020gmf} & 2+1 & $200-284$ & IAM NLO&
($\checkmark$) & $761(25)$ & $151(7)$ \\
Niehus-b~\cite{Niehus:2020gmf} & 2+1 & $200-284$ & IAM NNLO&
($\checkmark$) & $750(12)$ & $129(12)$ \\
Niehus-c~\cite{Niehus:2020gmf} & 2+1 & $200-284$ & IAM NLO&
($\checkmark$) & $752(24)$ & $145(5)$ \\
Niehus-d~\cite{Niehus:2020gmf} & 2+1 & $200-284$ & IAM NNLO&
($\checkmark$) & $738(7)$ & $129(2)$ \\
\hline
ETMC~\cite{ExtendedTwistedMass:2019omo}&2+1+1& 230-500 & VMEFT & $\checkmark$& $769(19)$ &$129(7)$\\
\hline
PDG~\cite{ParticleDataGroup:2020ssz}&&&&& $775.26(23)$& $147.4(8)$\\
\hline
\end{tabular*}
\caption{Overview over results on the $\rho$-resonance at the
physical pion mass or extrapolated to the physical point based on
lattice QCD. In some cases, when calculations at nearly
physical pion mass were performed, chiral extrapolations were not
performed ("--"). Where several errors are quoted in the original
references, we combine them in quadrature. The last row shows the
PDG value~\cite{ParticleDataGroup:2020ssz}.}
\label{tab:rho-lattice}
\end{table*}

\begin{figure}[t!]
\centering
\includegraphics[width=0.8\textwidth]{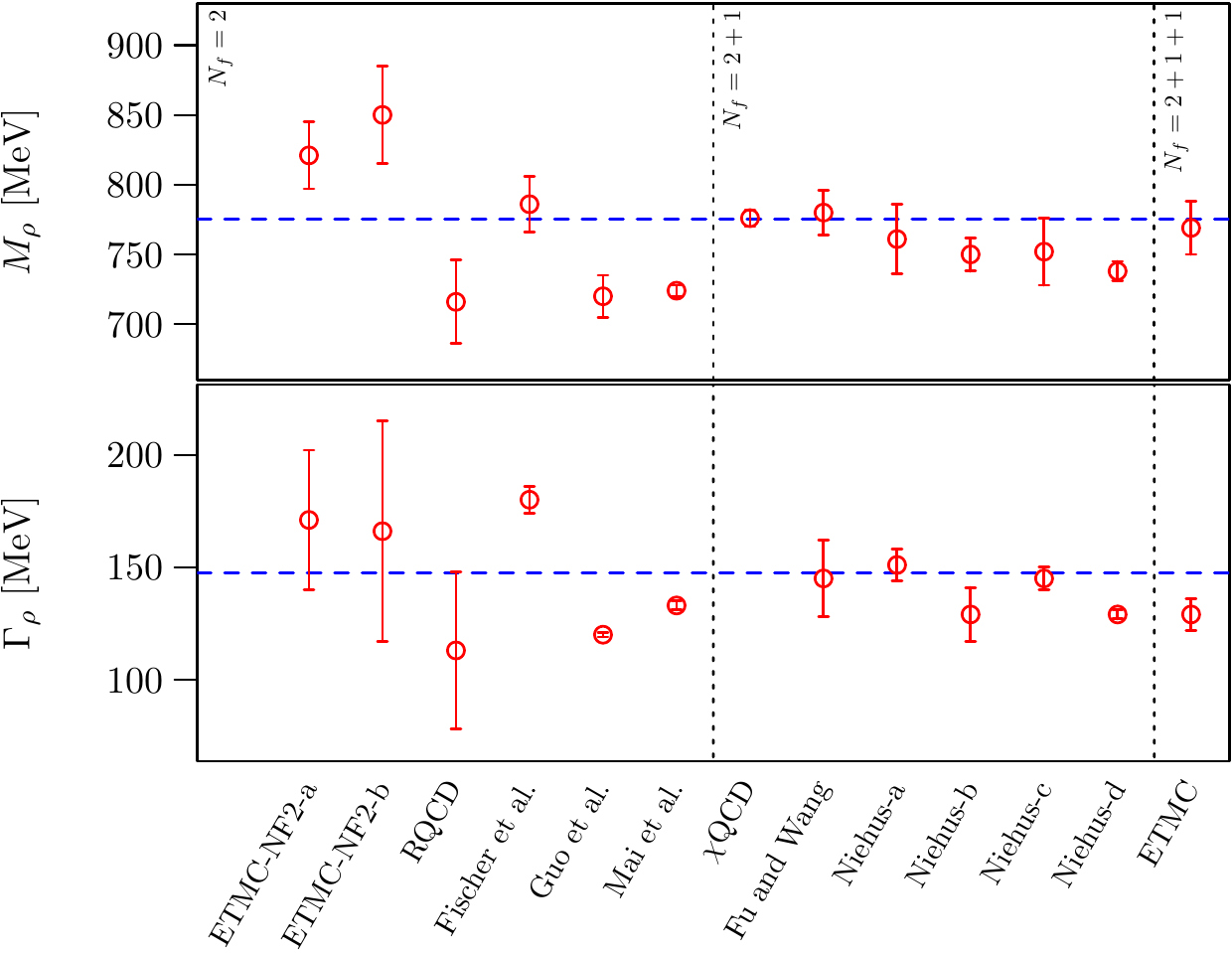}
\caption{We compare the various literature values for $M_\rho$ (upper panel)
and $\Gamma_\rho$ (lower panel) at the physical pion masses compiled in \cref{tab:rho-lattice} with the corresponding PDG values (blue, dashed line).}
\label{fig:rho-lattice}
\end{figure}

All the corresponding results are compiled in Tab.~\ref{tab:rho-lattice},
where we quote $M_\rho$ and $\Gamma_\rho$ as determined in the
corresponding references at the physical point. We also indicate the pion mass range used in the extrapolation
and whether or not chiral and continuum extrapolations have been
performed. The comparison is visualized
in \cref{fig:rho-lattice}.

For the chiral extrapolation different methods are being used:
\begin{itemize}
\item "--": the result from an ensemble directly at or close to the
physical point is quoted and, thus, no extrapolation is performed.

\item VMEFT: in Ref.~\cite{Djukanovic:2009zn} low-energy effective
theory of vector mesons is studied using complex mass
renormalization, see Sect.~\ref{sec:CMS}. The squared pion mass dependence of the complex squared pole
$Z=(M_\rho+\mathrm{i} \Gamma_\rho/2)^2$ is determined, which allows
one to extrapolate $M_\rho$ and $\Gamma_\rho$ together once these two
observables have been determined for different pion mass values.

\item mVMEFT: like EFT above, but modified according to
Ref.~\cite{Feng:2010es} where instead of using $M_\pi^2$ as proxy for
the average up/down light quark mass, the NLO chiral perturbation
theory expression is used.

\item UCHPT: unitarized chiral perturbation theory \cite{Oller:1998hw}, see \cref{sec:uni}.

\item IAM: inverse amplitude method \cite{Truong:1988zp, Dobado:1996ps, GomezNicola:2007qj, GomezNicola:2007qj}, see \cref{sec:uni}.
\end{itemize}
For the VMEFT and mVMEFT, mass and width need to be extracted for each
ensemble before the chiral extrapolation is attempted. Here, typically
a Breit-Wigner parametrization is adopted. For UCHPT and IAM one
typically fits the phase-shifts or the lattice energy levels directly.
We remind the reader that both UCHPT and IAM are representatives of a
larger class of unitary approaches, see Sec.~\ref{sec:uni}. The main
qualitative difference lies in the matching to the strictly
perturbative chiral expansion~\cite{Gasser:1983yg,Gasser:1984gg}. Assuming a Breit-Wigner form certainly
represents an approximation, which is, however, relatively well
fulfilled for the $\rho$-resonance and also used for the analysis of
experimental data.

Several comments are in order:
currently, only in Ref.~\cite{ExtendedTwistedMass:2019omo} a continuum extrapolation has been
attempted in the sense that $a^2$
effects have been included in the fit, even if the corresponding fit
parameters were found to be insignificant.
Also in Ref.~\cite{Bulava:2016mks} several lattice spacing values have
been studied. However, neither a continuum nor a chiral extrapolation
has been performed. The same data is analysed in
Ref.~\cite{Niehus:2020gmf} using the IAM at NLO or NNLO. In the latter
publication also the data at two different lattice spacing values from
Refs.~\cite{Dudek:2012xn,Wilson:2015dqa} is analysed using the same
method. In both cases, to our understanding, lattice artefacts have
not been explicitly included in the fits, though such terms appear to
be not required for describing the data.
Moreover, in most of the analyses P-wave dominance was assumed with
the notable exception of Ref.~\cite{Dudek:2012xn}. In the latter
reference it was found, however, that this assumption is well
justified.

Comparing the extrapolated lattice results in Fig.~\ref{fig:rho-lattice}
with the phenomenological information on the resonance
parameters one observes that again for the $N_f=2$ results differences to the PDG values appear larger than for $N_f=2+1(+1)$, see also the discussion above. Since the differences are positive and negative alike, we do not think that one sees the difference between $N_f=2$ QCD and nature here. We would rather conclude that these differences can be attributed to lattice artefacts or other systematic differences between the investigations.
Of course, in a $N_f=2$ flavor world the
$\rho$-resonance might have different properties as compared to
nature, but a fully controlled continuum and chiral extrapolations would be required to see such a difference.
More detailed discussions can be found in Refs.~\cite{Hu:2016shf,
Hu:2017wli, Guo:2016zos, Molina:2020qpw}.

For $N_f=2+1$ and $N_f=2+1+1$, the agreement with the PDG
value appears to be reasonable, though in some cases the deviations
represent a few $\sigma$. Notably, the results from
Ref.~\cite{Niehus:2020gmf} using IAM to NNLO deviate further from the
PDG value than the results based on IAM to NLO. The reason is likely
too sparse data to constrain all parameters.

In general, we believe that lattice QCD results of the
$\rho$-resonance at several lattice spacing values and several, close to physical pion mass values will be required to resolve the
remaining discrepancies. Unfortunately, in the only reference so far
with three lattice spacing values~\cite{ExtendedTwistedMass:2019omo}
the low pion mass region is not explored sufficiently well.

\subsection{The \texorpdfstring{$K^*(892)$}{K*(892)}-resonance}

\begin{figure}[t!]
  \centering
  \includegraphics[width=0.49\textwidth, page=1]{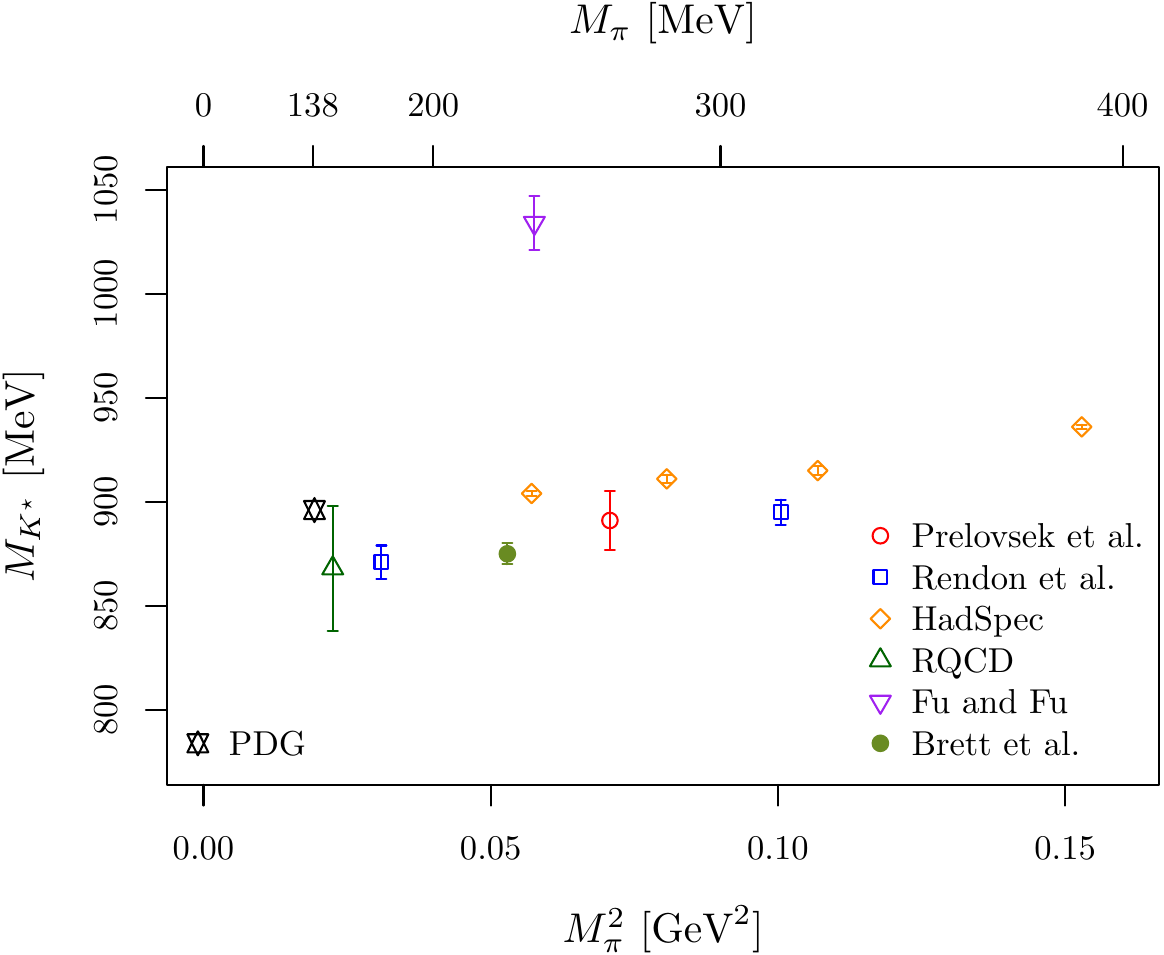}
  \includegraphics[width=0.49\textwidth, page=2]{MKstarvsMpisq}

\caption{$M_{K^*}$ and $g_{K^*-K\pi}$ as functions of $M_\pi^2$ for
  lattice results available in the literature with $N_f=2$, $N_f=2+1$
  and $N_f=2+1+1$ dynamical quark flavors. We show results by Prelovsek et al.~\cite{Prelovsek:2013ela}, Rendon et al.~\cite{Rendon:2020rtw}, HadSpec~\cite{Wilson:2019wfr}, RQCD~\cite{Bali:2015gji}, Fu and Fu~\cite{Fu:2012tj} and Brett et al.~\cite{Brett:2018jqw}.}
\label{fig:KstarMpisq}
\end{figure}

The $J^P=1^-$ $K^*$ resonance is the analogue of the $\rho$-resonance in
the $K\pi$ channel with isospin $I=1/2$ decaying in a P-wave. Experimentally, a resonance
mass of $M_{K^*} = 892\ \mathrm{MeV}$ and a coupling
$g_{K^*-K\pi} = 5.73(6)$ is quoted~\cite{ParticleDataGroup:2020ssz}.

The lattice status is certainly not yet comparable to the
$\rho$-resonance: there are fewer results available and fewer
systematic effects have been investigated and controlled.
The first lattice computation
has been performed using the Michael and McNeile method in
Ref.~\cite{McNeile:2002fh}, however, the validity of the method is
more questionable for the $K^*$ than for the $\rho$.

In Ref.~\cite{Prelovsek:2013ela} (see also Ref.~\cite{Lang:2012sv})
Prelovsek et al. then used the Lüscher method on a
single $N_f=2$ flavor lattice QCD ensemble. An additional strange
quark was added only in the valence sector. The pion mass of the
ensemble is $M_\pi = 266(4)\ \mathrm{MeV}$ and the strange quark mass
was fixed using $M_\phi$. The authors find $M_{K^*} = 891(14)$ and
$g_{K^*-K\pi} = 5.7(1.6)$ including only the P-wave, in
surprisingly good agreement to experiment. In the same reference they
also estimate the $K^*(1410)$ resonance mass including P- and D-wave mixing.

Rendon et al. study S- and P-wave scattering in the $I=1/2$ channel in Ref.~\cite{Rendon:2020rtw} with two ensembles corresponding to
pion mass values of $317$ and $176$~MeV. For the $K^*$ in the
P-wave they find $M_{K^*} = 895(6)\ \mathrm{MeV}$ and
$M_{K^*} = 871(8)\ \mathrm{MeV}$ for the heavier and the lighter
pion mass value, respectively. For the coupling they quote
$g_{K^*-K\pi} = 5.02(26)$ and $g_{K^*-K\pi} = 4.99(22)$. They
use $N_f=2+1$ flavor CLS ensembles at two different values of the
pion mass, $317$ and $175\ \mathrm{MeV}$, the lighter pion mass ensemble corresponding to the finer
lattice spacing of $a=0.088\ \mathrm{fm}$ and the heavier one to $a=0114\ \mathrm{fm}$.

The Hadron Spectrum collaboration study the $K^*$ for several
pion mass values in Ref.~\cite{Wilson:2019wfr}, after an initial study
in Ref.~\cite{Dudek:2014qha}. The pion mass values
range from $239$ to $391$ MeV and $I=1/2$ and $I=3/2$ are being
investigated.

RQCD in Ref.~\cite{Bali:2015gji} determine the $K^*$ on one
$N_f=2$ ensemble with $M_\pi=150\ \mathrm{MeV}$. They carefully investigate
systematics and arrive at $M_{K^*}=868(30)\ \mathrm{MeV}$ and
$g_{K^*-K\pi} = 4.79(49)$.

Brett et al.~\cite{Brett:2018jqw} work on a single ensemble with
$230\ \mathrm{MeV}$ pion mass and $N_f=2+1$ dynamical quark
flavors. Their result is $M_{K^*}/M_\pi = 3.808(18)$ and
$g_{K^*-K\pi} = 5.33(20)$.

Fu and Fu~\cite{Fu:2012tj} work with $N_f=2+1$ flavor MILC ensembles
with the staggered quark discretization. The ensemble they use has a
rather coarse lattice spacing of $a=0.15\ \mathrm{fm}$. They quote
$M_{K^*} = 1034(13)\ \mathrm{MeV}$ and $g_{K^*-K\pi} = 6.38(78)$.

In Fig.~\ref{fig:KstarMpisq} we summarize the current lattice status for
the $K^*$: we show $M_{K^*}$ in the left and
$g_{K^*-K\pi}$ in right panel, both as a function of $M_\pi^2$ comparing
the results available in the literature. Since no continuum limits
have been taken, the agreement is reasonable, maybe
apart from the result from Fu and Fu in Ref.~\cite{Fu:2012tj}. A possible
explanation for the discrepancy they find might be the large lattice spacing of $0.15\ \mathrm{fm}$ or the staggered fermion formulation used in Ref.~\cite{Fu:2012tj}.

\subsection{The \texorpdfstring{$\Delta(1232)$}{Delta(1232)}-resonance}

\begin{figure}[t!]
  \centering
  \includegraphics[width=0.49\textwidth, page=1]{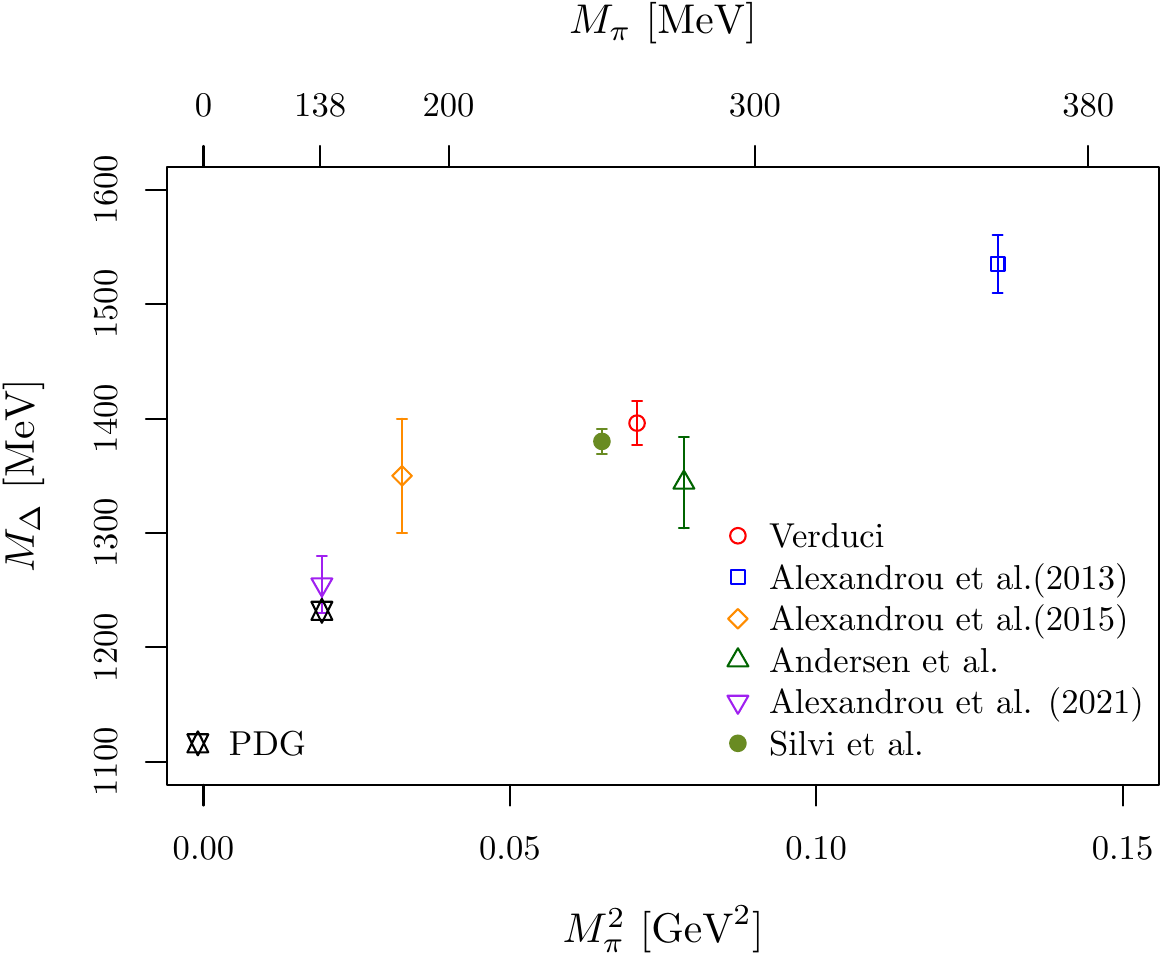}
  \includegraphics[width=0.49\textwidth, page=2]{MdeltavsMpisq}

\caption{$M_\Delta$ and $g_{\Delta-N\pi}$ as functions of $M_\pi^2$ for
  lattice results available in the literature with $N_f=2$, $N_f=2+1$
  and $N_f=2+1+1$ dynamical quark flavors. Results from Verduci~\cite{Verduci:2014btc}, Alexandrou et al.~\cite{Alexandrou:2013ata,Alexandrou:2015hxa}, Andersen et al.~\cite{Andersen:2017una}, Alexandrou et al.~\cite{Alexandrou:2021plg} and Silvi et al.~\cite{Silvi:2021uya} are shown.}
\label{fig:deltaMpisq}
\end{figure}

The $\Delta(1232)$-resonance is decaying predominantly into nucleon-pion, it
is well isolated and it represents the lowest lying spin-$3/2$ baryon
resonance. For the $\Delta$(1232)-resonance the amount of lattice
results is much sparser than for the $\rho$(770) and even the $K^\ast(892)$ discussed above. The
main reason is that the signal-to-noise ratio in the relevant
correlation functions is worse, since one is dealing with a heavier
baryonic state. An additional complication comes from the fact that
the inelastic threshold is relatively close by at $N\pi\pi$. This
means on the one hand that the calculation needs to be done at close
to physical pion mass and on the other that there are typically only
very few energy levels available in the range between $N\pi$ and
$N\pi\pi$. We remark here that the $N\pi\pi$ channel is subdominant and for this reason sometimes the elastic Lüscher formalism is applied also above the $N\pi\pi$ threshold.

Due to these limitations we include for the $\Delta(1232)$ all available
lattice results in the comparison, where the $\Delta$ is treated as a resonance. After preliminary results
for the $P_{33}$ phase-shift in pion-nucleon scattering have been
shown in Ref.~\cite{Meissner:2010ij}, results for mass and width have
been reported in Verduci's PhD thesis~\cite{Verduci:2014btc}, where
Lüscher's formalism with $N_f=2$ Wilson-clover improved dynamical
quarks was applied. The single ensemble had a pion mass value of
$266\ \mathrm{MeV}$. The $\Delta$ mass and width are determined from a Breit-Wigner fit. In the
two references by Alexandrou et al.~\cite{Alexandrou:2013ata,Alexandrou:2015hxa} the
approach developed by Michael and McNeile was used instead of the
Lüscher formalism. These calculations are based on domain wall valence
fermions on a staggered sea with two pion mass values, $180$ and
$260$~MeV. Andersen et al.~\cite{Andersen:2017una}
use one $N_f=2+1$ flavor CLS ensemble with $M_\pi =
280\ \mathrm{MeV}$. The authors apply the Lüscher
formalism and a Breit-Wigner fit to determine mass and
width. Silvi et al.~\cite{Silvi:2021uya}, again using $N_f=2+1$
dynamical quark flavors this time with a $255\ \mathrm{MeV}$ pion
mass value, apply the Lüscher method to explore for
the first time both a fit to the phase-shift data with a Breit-Wigner form and a determination of the pole in
the scattering amplitude. The difference they find between these two methods is negligible within errors. Finally, in the proceeding contribution by Alexandrou et al.~\cite{Alexandrou:2021plg} $N_f=2+1+1$ dynamical flavor
simulations are used for the first time with physical value of the
pion mass. The fit to determine the mass and width relies on a
Breit-Wigner form, but there is no value for the width quoted yet.

All these calculations are based on single lattice spacing simulations
and single pion mass values each. Therefore, the meaningfulness of
these calculations is limited. But we can learn that the lattice
calculation of the $\Delta(1232)$ is feasible and we can expect more results in the near future. Moreover, by combining all available results, one sees a more or less consistent picture emerging: in \cref{fig:deltaMpisq} we show
in the left panel $M_\Delta$ and in the right panel $g_{\Delta-N\pi}$
as functions of $M_\pi^2$. The PDG value for the mass is included in
the left figure, while in the right we take the average of the two
values from Refs.~\cite{Pascalutsa:2005vq,Hemmert:1994ky}\footnote{Note that the $\Delta$ mass and coupling values have been compiled already in
Ref.~\cite{Silvi:2021uya}, but not including the results from
Ref.~\cite{Alexandrou:2021plg} yet.}.

The coupling $g_{\Delta-N\pi}$ is defined here as in Ref.~\cite{Pascalutsa:2005vq} via the width
\begin{equation}
\Gamma_\Delta = \frac{(g_{\Delta-N\pi})^2}{48\pi m_N^2} \frac{E_N +
m_N}{E_N + E_\pi}p_\Delta^3
\end{equation}
in order to be able to compare all the results. $E_N$ and $E_\pi$ are
the energies of the nucleon and pion, respectively, with total
momentum $p_\Delta$. Note that first steps towards studying the $\Delta$ resonance by HAL
QCD can be found in Ref.~\cite{Murakami:2021tph}.

The status of the $\Delta(1232)$ from lattice QCD is certainly not yet final. But the existing, though still more exploratory investigation indicate that this baryon resonance can be studied with good precision using lattice QCD. Like for the $\rho$-resonance, it will be interesting to see the continuum extrapolation of the data, but even more so the inclusion of $N\pi\pi$ in the investigations. It is likely that eventually the continuum limit needs to be taken at the physical point, since the applicability of CHPT for this state is at least questionable.

Let us also remark that it will be also very interesting to estimate the scattering length in the $I=1/2$ and $I=3/2$ channels as precisely as possible, since these scattering length can be compared to very precise experimental measurements. These scattering lengths are also important in the analysis of the pion-nucleon $\sigma$-term~\cite{Hoferichter:2016ocj}.

\subsection{The \texorpdfstring{$f_0(500)$}{f0(500)}-resonance}
\label{sec:sigma}

The scalar isoscalar unflavored mesonic resonance $f_0(500)$ (also referred to as $\sigma$-resonance) is the lightest excited state in the spectrum of hadrons. At the same time it
is one of the most controversial as its appearance in the line-shapes
of, e.g., $\pi\pi$ scattering is hard to distinguish from the
unstructured background, see the extensive
review~\cite{Pelaez:2015qba}. Ultimately, many precise analyses led to
the currently accepted ranges for mass and
width~\cite{ParticleDataGroup:2020ssz}.

Access to this ($I=\ell=0$) channel directly from quark-gluon dynamics
using the above described methods of lattice QCD was hindered for a long
time by the presence of disconnected diagrams, despite early
pioneering works~\cite{Alford:2000mm,Prelovsek:2010kg,Fu:2012gf,Fu:2013ffa}. The first full fledged calculation was finally performed
by the Hadron Spectrum Collaboration~\cite{Briceno:2016mjc},
extracting also phase-shifts using the Lüscher
framework. These results have
then been analyzed and extrapolated to the physical point using the
mIAM in Ref.~\cite{Doring:2016bdr}. Only a little later,
a two-flavor lattice QCD study by the ETMC~\cite{Liu:2016cba}
was performed, allowing to extract and extrapolate $I=0$ $\pi\pi$
S-wave scattering length to the physical point. Note that recently
a study in chiral perturbation theory has been performed in
Ref.~\cite{Draper:2021wga}, results of which require to reassess the
uncertainties quoted in Ref.~\cite{Liu:2016cba}: the chiral
perturbation theory result suggests that these are of the order of the
result itself.

In Ref.~\cite{Fu:2017apw} the scattering lengths were determined in the
three-flavor formulation. The second calculation covering a larger
energy region was performed by the GWQCD group in the two-flavor
setup~\cite{Guo:2018zss} extrapolating the obtained resonance
parameters to the physical point using a chiral unitary approach.\\
Subsequently, the obtained energy eigenvalues were reanalyzed in a
cross-channel ($I=0,1,2$) study~\cite{Mai:2019pqr}, including chiral
extrapolations using the mIAM approach~\cite{Truong:1988zp,Dobado:1996ps,GomezNicola:2007qj,GomezNicola:2007qj}, see \cref{sec:uni}, leading to tighter constraints on the resonance parameter values at the physical point.

In Ref.~\cite{Briceno:2017qmb} coupled channel $\pi\pi$, $K\bar{K}$ and $\eta\eta$ scattering with isospin $I=0$ was studied for the first time. The authors work with one ensemble with a pion mass value of $391\ \mathrm{MeV}$ with $N_f=2+1$ dynamical quark flavors. The pion mass value of almost $400\ \mathrm{MeV}$ has implications for the thresholds. But in this setting they find a bound-state $f_0(500)$ pole on the physical Riemann sheet ${[+++]}$ at $745(5)\ \mathrm{MeV}$ mass and an $f_0$ resonance on the unphysical ${[-++]}$ Riemann sheet at
$1166(45) - \mathrm{i} 181(68)/2\ \mathrm{MeV}$. The $f_0(500)$, of course, becomes unstable once the pion mass is reduced towards the physical point. The $f_0$ resonance is connected by the authors to the $f_0(980)$ resonance, and they claim that it is dominated by a $K\bar{K}$ molecular state, because there is no pole on ${[--+]}$ Riemann sheet.

\begin{table*}[t]
\centering
\begin{tabular*}{\textwidth}{@{\extracolsep{\fill}}llcll}
\hline
Reference & $N_f$ & Pion mass & Chiral & $\sqrt{s_\sigma}$\,[MeV]\\
&& range [MeV] & extr. & \\
\hline
GWQCD~\cite{Guo:2018zss}&2& 227-315 & UCHPT & $440(52)-\mathrm{i}240(32)$  \\
Mai et al.~\cite{Mai:2019pqr}&2& 227-315 & mIAM & $443(3)\phantom{0}-\mathrm{i}221(6)$  \\
HadSpec~\cite{Briceno:2016mjc} Döring et al.~\cite{Doring:2016bdr}&2+1& 236-391 & mIAM &   $449(17)-\mathrm{i}169(24)$\\
\hline
PDG~\cite{ParticleDataGroup:2020ssz}&&&& $450(50)-\mathrm{i}275(75)$\\
\hline
\end{tabular*}
\caption{The $\sigma$-resonance pole position $\sqrt{s_\sigma}$ at the
physical pion mass value determined based on lattice QCD data. The last row shows the corresponding PDG value~\cite{ParticleDataGroup:2020ssz}. For none of these results the continuum limit has been studied.}
\label{tab:sigma-lattice}
\end{table*}

The currently available lattice QCD results for the $f_0(500)$ or $\sigma$-resonance pole position $\sqrt{s_\sigma}$ at the physical point are compiled in \cref{tab:sigma-lattice}, where we also quote the PDG value. The agreement is reasonable, keeping in mind that most of the systematic effects are still not controlled in the corresponding lattice simulations.


\section{Results: Coupled channels/thresholds}
\label{sec:cc}
The vast majority of resonant states appear in settings allowing (energetically) for multiple decay channels. Such cases are harder to access both in performing numerical lattice QCD calculations as well as in view of the analysis of the results, extracting for example universal parameters of resonances from poles on the second Riemann sheet. In the following we review the current status on such states accessed from Lattice QCD. For the $D_0^\ast(2300)$ we also discuss recent extractions of the pole positions from recent LHCb data.

\subsection{Light mesons}

The mesonic scalar states $a_0(980)$, $f_0(980)$, $f_0(500)$ and $\kappa(800)$ were studied in Ref.~\cite{Prelovsek:2008rf} using quenched simulations for pion masses 344-576~MeV. In that, an effort was made to identify the tetraquark content of these states and searching for new states. Similar goals are followed in Ref.~\cite{Prelovsek:2010kg} for isospin  $0$, $2$, $1/2$ and $3/2$ states. The question about the tetraquark content of the $a_0(980)$ and the $\kappa(800)$ was then further studied in an unquenched simulation~\cite{Alexandrou:2012rm}, finding no evidence for such an interpretation. The $a_0(980)$ was extracted in a coupled-channel meson-meson scattering in lattice QCD in Ref.~\cite{Dudek:2016cru}. In the latter reference a global fit to lattice finite-volume energy levels from $\pi\eta$ scattering and to relevant experimental data on the $\pi\eta$ event distribution in $B$ decays together with the
$\gamma\gamma\to\pi\eta$ cross section is performed. Both the leading and next-to-leading-order analyses lead to similar and successful descriptions of the finite-volume energy levels and the experimental data. However, these two different analyses yield different $\pi\eta$ scattering phase
shifts for the physical masses of the pseudoscalar mesons. Both the $a_0(980)$ and the $a_0(1450)$ poles and their properties could be extracted at NLO.

The $f_0(980)$, discussed in \cref{sec:sigma}, was studied in Ref.~\cite{Briceno:2017qmb} by investigating isoscalar $\pi\pi$, $K\bar{K}$ and $\eta\eta$ coupled channel scattering. The authors find a candidate state for the $f_0(980)$ and its properties suggest that it closest resembles a $K\bar{K}$ molecule. In this context
it is worth emphasizing that the opening of the $K\bar{K}$ threshold induces a level crossing thus
mocking up a resonance~\cite{Bernard:2010fp}. A possible resonance and a threshold can be distinguished by their different quark mass dependence.

Systems with non-vanishing strangeness quantum number ($\pi K$, $\eta K$) have been studied for a single pion mass ($\approx 400$ MeV) by the Hadron Spectrum Collaboration~\cite{Dudek:2014qha}. In that a generic parametrization of the scattering amplitude (Chew-Mandelstam) was utilized to obtain the complex pole position on the second Riemann sheet. Extracted poles were found to be comparable to the experimental values even for the broader scalar case. For the narrow tensor resonance $K^*(1430)$ also Breit-Wigner parameters have been extracted as $M_R = 933(1)~{\rm MeV}$ and $g_R = 5.93(26)$. Later the study was repeated for a range of pion masses ($200-400$ MeV) in Ref.~\cite{Wilson:2019wfr} tracing out the chiral trajectory of the resonance pole.

Also very interesting, but also more complex are light mesons decaying to multihadron states, such as the $a_1(1260)$, which are hard to access both on the lattice and in the continuum. For the former, one key challenge lies in the construction of multihadron interpolating operators. In a pioneering study~\cite{Lang:2014tia} composite operators such as $\rho\pi$, $\omega\pi$ etc. were used. Three particle operators were not considered motivated by the fact that for the considered setup ($L\approx 2~{\rm fm}$, $M_\pi=266~{\rm MeV}$) the three pion states are above the considered energy range. With these set of interpolating operators it was observed that the inclusion of composite operators in the GEVP (see \cref{sec:Elevels}) is crucial for a stable extraction of energy levels. In this work also a finite-volume analysis was performed using the two-body Lüscher formalism and the Breit-Wigner parametrization of the scattering amplitudes, both rooting in the same approximation of stable vector mesons. The resonance masses were extracted as
$M_{a_1}=1.435^{+53}_{-121}~{\rm GeV}$, $M_{b_1}=1.435^{+36}_{-90}~{\rm GeV}$ comparing roughly with the experimental values~\cite{ParticleDataGroup:2020ssz}, $1.230(40)~{\rm GeV}$ and $1.2295(32)~{\rm GeV}$, respectively.

Later, a lattice calculation of coupled $\pi\omega$, $\pi\phi$ scattering was reported by the Hadron Spectrum Collaboration in Ref.~\cite{Woss:2019hse}. While three-meson interpolators were included systematically, the finite-volume effects of the three-body channels were only included via meson-meson channels, motivated by the large pion mass $M_\pi=391~{\rm MeV}$ of the considered ensemble. A clear resonance signal for the $b_1$-resonance was observed with the $M_R\approx 1380~{\rm MeV}$.

Recently, the resonant three-pion channel for the quantum numbers of the $a_1$-resonance was calculated on the lattice by the GWQCD collaboration~\cite{Mai:2021nul}. Here, for a pion mass value of 224~MeV the interpolator basis also included one-, two- and three-meson interpolators. Similarly to the pioneering study of Lang et al.~\cite{Lang:2014tia}, it was observed that three-body operators are, indeed, indispensable for a stable extraction of energy eigenvalues. The results of this calculation were used to determine volume-independent quantities  by the means of the recently developed three-body quantization condition~\cite{Mai:2017bge,Mai:2018djl}. Subsequently, these quantities were used to obtain a pole position on the second Riemann sheet, see left panel of Fig.~\ref{fig:pirho}. Couplings of the $a_1$-resonance could be extracted via the corresponding infinite-volume three-body formalism~\cite{Sadasivan:2021emk,Sadasivan:2020syi}. Agreement of the mass of this state with the experimental value was observed, while the width  was found to be significantly smaller than its experimental counterpart. This is certainly expected having a lattice setup with heavier than physical pion mass values.

\subsection{The Roper-resonance \texorpdfstring{$N(1440)$}{N(1440)}}

The $N(1440)$ or Roper resonance~\cite{Roper:1964zza} is the first excited nucleon state with $I(J^P) = 1/2(1/2^+)$ quantum
numbers. It appears in the baryon spectrum as a considerably lighter state than the parity partner of the nucleon $I(J^P) = 1/2(1/2^-)$ $N(1535)$ which is at odds with quark model expectations~\cite{Isgur:1977ef,Isgur:1978wd,Loring:2001kx} associating the Roper with the second radial excitation of the nucleon. This comprises one of the paramount examples of the baryon spectrum puzzles, see also the discussion in Sec.~\ref{sec:EFTbaryons}.
Note that better agreement with phenomenology can be achieved by including Goldstone bosons as effective degrees of freedom to the constituent quark model~\cite{Glozman:1995fu}. Also more recent phenomenological analyses reveal an intricate analytic structure of the Roper~\cite{Arndt:2006bf,Doring:2009yv,AlvarezRuso:2010xr} including strong coupling to the three-body channels distorting its shape from the usual Breit-Wigner form. Theoretical explanations include its dynamical generation without a $qqq$ core~\cite{Krehl:1999km}, or an interplay of genuine resonance poles and dynamical effects~\cite{Suzuki:2009nj}.

In modern LQCD calculations, the Roper channel is similarly obscure despite significant efforts made to reveal its structure, see for instance
Refs.~\cite{Alexandrou:2013fsu,Roberts:2013ipa,Alexandrou:2014mka,Liu:2014jua,Engel:2013ig,Edwards:2011jj}. Among the references from above, only in Ref.~\cite{Liu:2014jua} evidence for a state compatible in mass with the $N(1440)$ was claimed, using a chiral fermion discretization and a Bayesian ansatz to estimate the energy levels. However, the calculation is partially quenched and utilizes mixed action (even though chiral valence on chiral sea) and, maybe most importantly, the Roper is to our understanding not treated as a resonance, e.g., with the Lüscher formalism.

More sophisticated LQCD studies~\cite{Lang:2016hnn, Kiratidis:2016hda} followed later, consistently ruling out the exclusive $qqq$ interpretation of the Roper-resonance. Instead the sizable coupling to the five-quark operators (meson-baryon states) is noted in Ref.~\cite{Kiratidis:2016hda}. Even more interesting, is the missing level reported in Ref.~\cite{Lang:2016hnn,Padmanath:2017oya,Leskovec:2018lxb} to the expectation from the elastic $\pi N$ scattering alone. While there are attempts based on Hamiltonian effective field theory (HEFT)~\cite{Wu:2017qve} to interpret the measured finite-volume spectrum~\cite{Lang:2016hnn}, an obvious solution to this seem to be the not included (most notably $\pi\pi N$) inelastic channels. Notably, this is also supported by phenomenology~\cite{ParticleDataGroup:2020ssz} attributing up to 50\% of the decay branching ratio to such three-body channels in the final state, as discussed in Sec.~\ref{sec:EFTbaryons}.

In the context of $\pi N$ scattering we mention that there is also an exploratory study by Lang and Verduci~\cite{Lang:2012db} in the negative parity sector, which reports a significant change of the calculated finite-volume spectrum when extending the operator basis to incorporate the meson-baryon type operators. Finally, the situation may become even more entangled as discussed in Ref.~\cite{Doring:2013glu}. There, based on a chiral unitary approach for the meson-baryon scattering~\cite{Bruns:2010sv,Mai:2012wy} a finite-volume spectrum for the
$I(J^P) = 1/2(1/2^-)$ channel was predicted using quark masses from lattice studies~\cite{Bietenholz:2011qq,Alexandrou:2009qu}. It was found that mixing of relevant two-body thresholds ($\eta N,K\Lambda,K\Sigma$) can induce spectra looking similar to an avoided level crossing. Furthermore, poles can move on hidden Riemann sheets, thus, obscuring simple level counting arguments. Future analyses have to face such challenges in both parity sectors, in addition to the complexity of the three-body channels~(see e.g.~\cite{Severt:2020jzc}) for $N^*(1440)$ channel.

\subsection{Specific Open and Closed Charm Systems}

There is significant focus on determining properties of charmonium resonances from Lattice QCD.
One reason for this is that the charmonium spectrum was long believed to be well understood
in terms of a heavy quark potential (the positronium of QCD)
but with the appearance of  quite a number of ``exotic'' states like the $X(3872)$ (nowadays called
$\chi_{c1}(3872)$) challenged this simple picture and rekindled interest.

There is a list of lattice QCD studies of charmed meson states, which are mostly of
exploratory nature. However, the studies represent important technical progress and allow one to draw
first physical conclusions. There are several groups or collaborations which successively
worked on different states based on a fixed lattice action and fixed ensembles.

The work of the group around Lang, Mohler and Prelovsek is largely based on two ensembles: one with $M_\pi=266\ \mathrm{MeV}$ and $N_f=2$ and a second one with $M_\pi=156\ \mathrm{MeV}$
and $N_f=2+1$ dynamical quark flavors. As a consequence, the charm quark is always treated partially quenched, which has a technical advantage: a charm and anti-charm cannot be created
from the vacuum, which makes certain decays to (much) lighter states impossible. For the
$N_f=2$ ensemble the strange is partially quenched as well.

The CLQCD collaboration on the other hand works with $N_f=2$ Wilson twisted mass ensembles generated by the European Twisted Mass Collaboration (ETMC)~\cite{ETM:2009ztk,ETM:2008zte} with a range of pion mass values from $300\ \mathrm{MeV}$ to $458\ \mathrm{MeV}$. They work with a relatively fine lattice spacing
value of $0.067\ \mathrm{fm}$. Thus, while they can investigate the pion mass dependence to
some extend, lattice artefacts are also not accessible, but are expected to be small.
In that, it is notable that with Wilson twisted mass fermions at maximal twist, lattice artefacts
linear in $am_c$ are absent and contributions start only at $(am_c)^2$.

Similarly, two-meson systems involving $D$ mesons have been also studied by the Hadron Spectrum
Collaboration. Their work is based on lattice QCD ensembles with $N_f=2+1$ dynamical quark flavors and anisotropy. The relevant ensembles for this review are one with pion mass value $391\ \mathrm{MeV}$ and a second one with $M_\pi= 239\ \mathrm{MeV}$. Their usage of the distillation method~\cite{HadronSpectrum:2009krc,Morningstar:2011ka}
allows them to include a large set of interpolating operators. The resulting number of energy levels
and the precision thereof allows them to perform elaborate analyses including also coupled channels.
The price for such resource intensive investigations is that typically those are only performed on one or two ensembles, which makes the understanding of systematics and, thus, the generalisation of the results harder. For a first investigation of the $D$-meson spectrum see Ref.~\cite{Cheung:2016bym},
which, however, treats all states as stable states.

\subsubsection{The \texorpdfstring{$D_0^*(2300)$}{D0*(2300)}}

\begin{figure}
    \centering
    \includegraphics[width=0.6\textwidth]{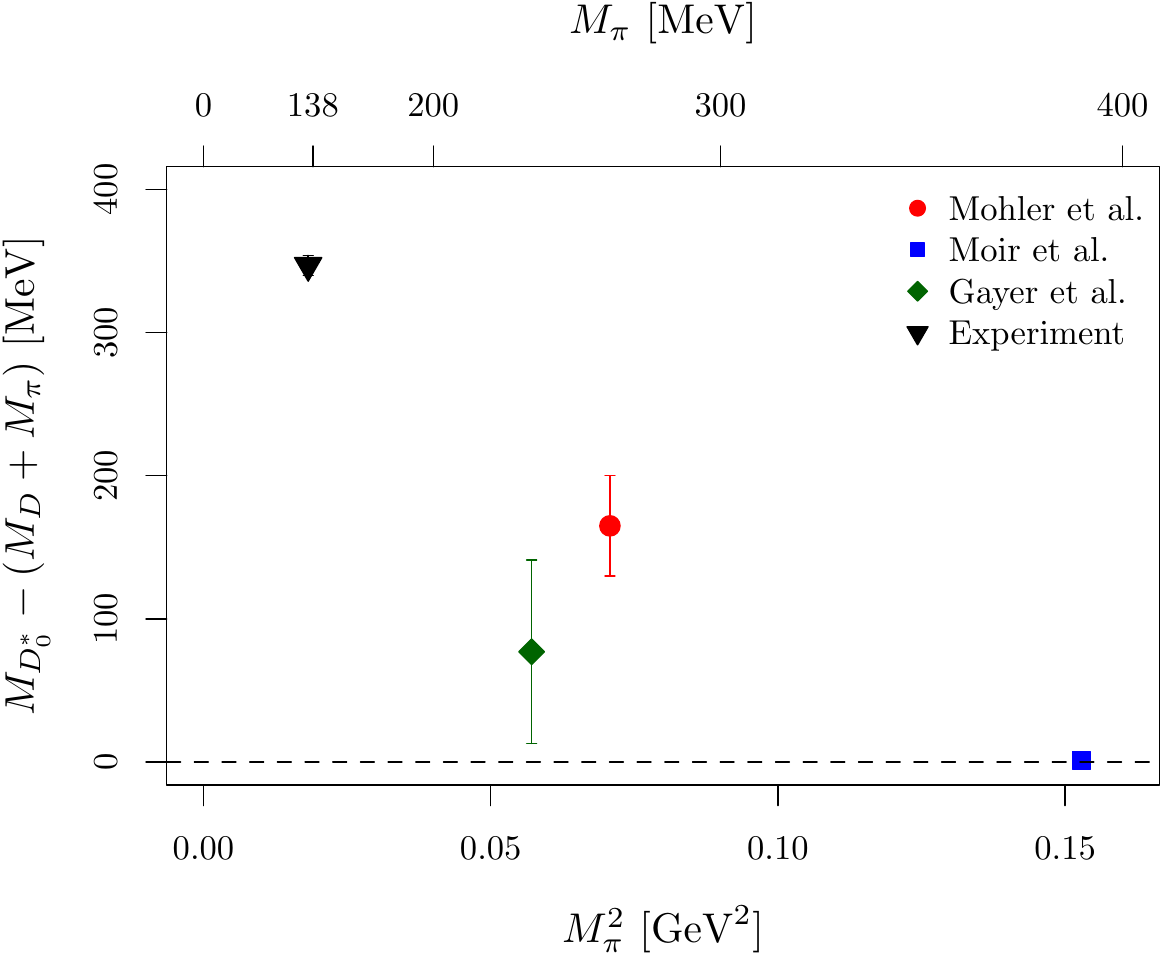}
    \caption{Comparison of $D_0^\ast(2300)$ energies from the different lattice calculations minus $M_D + M_\pi$ as a function of the squared pion mass value.}
    \label{fig:D0starDpi}
\end{figure}

The $D_0^*(2300)$ (formerly known as $D_0^\ast(2400)$) is a state with $J^P=0^+$
and isospin $I=1/2$, seen to decay into $D\pi$. The width of the state is quoted to be
around $220\ \mathrm{MeV}$ by the PDG~\cite{ParticleDataGroup:2020ssz}.

In Ref.~\cite{Mohler:2012na} Mohler and co-authors study $D$-meson pion scattering and
$D$-meson resonances. It represents an exploratory study of the scalar $D_0^*(2300)$
and the axial $D_1(2430)$ resonances on the one aforementioned $N_f=2$ ensemble with a
pion mass value of about $266\ \mathrm{MeV}$. Consequentially, the charm quarks are
treated partially quenched. Since they do not work at physical pion mass value they
quote mass differences with respect to the spin-average $(M_D + 3M_{D^\star})/4$ for the
$D_0^*(2300)$, for which they quote $351(21)\ \mathrm{MeV}$ in agreement with the
experimental value. The coupling of the $D_0^*(2300)$ to $D\pi$ is found to be
$g_{D_0^* D\pi}=2.55(21)\ \mathrm{GeV}$, only a few $\sigma$ away from the experimental
value. In addition, the authors determine energy levels for a list of states in the $D$-meson spectrum. Given the fact that they are not able to extrapolate to the continuum limit nor
to the physical pion mass value, the agreement to experiment is quite remarkable. Finally, they
also determine the S-wave scattering lengths for the $D\pi$ and the $D^*\pi$
systems with $I=1/2$, finding $a_0=0.81(14)\ \mathrm{fm}$ and $a_0=0.81(17)\ \mathrm{fm}$,
respectively. It should be noted that the spatial volume is only $16^3$.

In Ref.~\cite{Moir:2016srx} a first coupled channel analysis of $D\pi$, $D\eta$ and $D_s\bar{K}$ scattering is presented by the Hadron Spectrum Collaboration based on the aforementioned ensemble with the largest pion mass ($N_f=2+1$, $M_\pi=391\ \mathrm{MeV}$. Scattering amplitudes are determined and analytically continued to complex energies. This results in finding a shallow bound state at $2276(1)\ \mathrm{MeV}$ below the $D\pi$ threshold in the $J^P=0^+$ channel, which could be the $D_0^*(2300)$. Another -- this time deeply-bound state in the $1^-$ channel -- is found at $2009(2)\ \mathrm{MeV}$, which could be related to the $D^*(2007)$. A resonance is found in the $J^P=2^+$ channel, which is narrow and has mass and width of $2527(3)\ \mathrm{MeV}$ and $8.2(7)\ \mathrm{MeV}$, respectively. While experimentally, this could correspond to the $D_2^*(2460)$, we remark that the $D^* \pi$ is not included in this investigation.

The data of Ref.~\cite{Moir:2016srx} were reanalyzed in the framework of
UCHPT~\cite{Albaladejo:2016lbb} as discussed
in Sect.~\ref{sec:Dphiuni}. The NLO effective Lagrangian was utilized together with the
LECs determined earlier in~\cite{Liu:2012zya} to calculate the energy levels for
$J^P = 0^+ $scattering in the strangeness-isospin $(S, I) = (0, 1/2)$ in a finite volume
\cite{Doring:2011vk,MartinezTorres:2011pr}. This not only led to an amazingly precise
postdiction of the energy levels, but indeed the two pole structure of the  $D_0^*(2300)$
was revealed with the following pole positions ($\sqrt{s} =M-\mathrm{i}\Gamma/2$):
\begin{align}
\label{eq:twopoleD}
&\text{Pole~1:}~~ (2264^{+8}_{-14}-\mathrm{i}0)~{\rm MeV}\,,~~~~~~~~~
&\text{Pole~2:}~~ (2468^{+32}_{-25}-\mathrm{i}113^{+18}_{16})~{\rm MeV}~,~~
&M_\pi = 391~{\rm MeV}~, \nonumber \\
&\text{Pole~1:}~~ (2105^{+6}_{-8}-\mathrm{i}102^{+10}_{-12})~{\rm MeV}\,,~~
&\text{Pole~2:}~~ (2451^{+36}_{-26}-\mathrm{i}134^{+7}_{8})~{\rm MeV}\,,~~~ &M_\pi = 139~{\rm MeV}\,,
\end{align}
which solved the puzzle that the lowest charm-strange excitation was not heavier than the
corresponding charm-nonstrange meson. This two-pole structure can be understood easily in the
SU(3) limit. In this limit, all light and all heavy mesons take common values, see
also Refs.~\cite{Kolomeitsev:2003ac,Gamermann:2006nm}, and the heavy-light meson scattering
amplitude decomposes into irreps as $\overline{\bf 3} \otimes{\bf 8} =\overline{\bf 15} \oplus
{\bf 6} \oplus \overline{\bf 3}$ and the potential can be diagonalized accordingly. As it
turns out, at leading order only the ${\bf 6}$ and $\overline{\bf 3}$ irreps are attractive,
leading to two distinct poles. To make contact to the broken SU(3) world, linear extrapolations
in the meson masses are used (as introduced in Ref.~\cite{Jido:2003cb}), and the poles move
into the complex plane as given in Eq.~\eqref{eq:twopoleD}. In fact, the lower pole
and the $D_{s0}^\ast(2317)$ discussed below are chiral partners in such a scenario.
Using heavy-flavor symmetry, the same approach predicts a two-pole structures in the
$(0,1/2)$ sector, located at $\sqrt{s}=5537^{+9}_{-11}-\mathrm{i}116^{+14}_{-15}~{\rm MeV}$ and
$\sqrt{s}=5840^{+12}_{-13}-\mathrm{i}25^{+6}_{-5}~{\rm MeV}$. For $(S,I)=(1,0)$ the $D_0^*(2317)$ is found
at $2315^{+18}_{-28}$~MeV as in \cite{Liu:2012zya} and the corresponding state in the
$B$ meson sector is predicted at $5724^{+17}_{-24}$~MeV, i.e., it is bound by about 50~MeV. For an earlier study of these states, see~\cite{Guo:2006fu}. Using heavy quark spin
symmetry, one can make further predictions for axial $D_1$ and $B_1$ mesons. This double-pole
structure is further consolidated by the analysis of data on $B\to D\phi\phi$ decays
as discussed in Sect.~\ref{sec:BDphiphi}.

\begin{table}[t]
    \centering
    \begin{tabular}{lccc}
        \hline
        Reference & $N_f$ & $M_\pi\ [\mathrm{MeV}]$ & $M_{D_{0}^\ast}-(M_D + M_\pi)\ [\mathrm{MeV}]$        \\
        \hline
         Mohler et al.~\cite{Mohler:2012na} & 2 & 266 & 165(33) \\
         Moir et al.~\cite{Moir:2016srx} & 2+1 & 391 & -1(1) \\
         Gayer et al.~\cite{Gayer:2021xzv} & 2+1 & 239 & 77(64) \\
         \hline
         PDG~\cite{ParticleDataGroup:2020ssz} & &  & 347(7) \\
         \hline
    \end{tabular}
    \caption{The values extracted from the mentioned references for $M_{D_{0}^\ast(2300)}$ used for \cref{fig:D0starDpi}.}
    \label{tab:D0star}
\end{table}

In the $I=1/2$ channel, $D\pi$ scattering was investigated again in Ref.~\cite{Gayer:2021xzv} by the Hadron Spectrum Collaboration with one ensemble at $M_\pi=239\ \mathrm{MeV}$. A $D_0^*$ resonance pole is found in the S-wave $D\pi$ system. Mass and width of $2196(64)\ \mathrm{MeV}$ and $425(224)\ \mathrm{MeV}$, respectively, are found by analytically continuing scattering amplitudes. This study complements the earlier study~\cite{Moir:2016srx} at $M_\pi=391\ \mathrm{MeV}$ pion mass value. In contrast to the latter, the pole is now about $80\ \mathrm{MeV}$ below the $D\pi$ threshold, indicating a non-trivial pion mass dependence of this state. As discussed below, this result gives further credit
to the lower pole in Eq.~\eqref{eq:twopoleD} representing the lowest charm scalar meson,
see also Sect.~\ref{sec:BDphiphi}.

The $D_0^\ast$ energies relative to the $D\pi$ threshold are plotted in Fig.~\ref{fig:D0starDpi} as a function of $M_\pi^2$ comparing the different
available lattice estimates. Note that in Ref.~\cite{Mohler:2012na} this
energy difference is not quoted and we have determined it from the available
results in that reference. The picture, which starts to emerge, seems to indicate
that the $D_0^\ast$ is a virtual state above $M_\pi\approx400\ \mathrm{MeV}$.
Within the large uncertainties the available data lets one expect convergence towards
the experimental value once the physical pion mass is used in lattice calculations.

\subsubsection{The \texorpdfstring{$D_0^\ast(2300)$}{D0*(2300)} from Experimental Data}
\label{sec:BDphiphi}
As discussed before, certain open charm mesons do not fit into the
conventional quark model picture but rather are most probably
hadronic molecules, which explains many of their odd features that
are observed and also found on the lattice. This is further corroborated
by the analysis of the precise data from LHCb on the various decay
modes $B\to D\phi\phi$
\cite{LHCb:2016lxy,LHCb:2014ioa,LHCb:2015eqv,LHCb:2015tsv,LHCb:2015klp},
that give independent information on the thought after excited $D$ mesons in the S-wave amplitudes, that can be extracted from the so-called angular moments discussed in Sect.~\ref{sec:Dphiuni}.

In this context, it is important to stress that the chiral symmetry of
QCD requires energy-dependent pionic strong interactions at low energies.
This constraint, however, is not fulfilled by the usual Breit-Wigner (BW)
parameterization of pionic resonances, such as the open charm excitations discussed
here, leading to masses larger than the real ones~\cite{Du:2019oki}.
For an early work on this issue, see Ref.~\cite{Gardner:2001gc}. The
argument goes as follows: Neglecting for simplicity the energy dependence
of the decay width, the BW paramterization in the S-wave ($\ell = 0$)
of the $D\pi$ system reads:
\begin{equation}
\label{eq:bwori}
{\rm BW}_0(s) \propto \dfrac{1}{s-M_0^2+\mathrm{i}M_0^{}\Gamma_0^{}},
\end{equation}
with $M_0$ and $\Gamma_0$  the BW mass and width of the resonance, in order.
The peak position of this BW parameterization is given by
\begin{equation}
\dfrac{d}{ds}|{\rm BW}_0(s)|^2
\propto -\dfrac{2(s-M_0^2)}{\big[(s-M_0^2)^2+M_0^2\Gamma^2_0\big]^2}=0~.
\end{equation}
Thus,  the BW mass for a resonance corresponds to the value of the peak position.
The chiral symmetry constraints can be most simply accounted for by
modifying Eq.~\eqref{eq:bwori} with an energy-dependent prefactor:
\begin{equation}
\label{eq:bwrev}
{\rm BW}_0^\prime(s)\propto \dfrac{E_\pi}{s-M_0^2+\mathrm{i} M_0^{}\Gamma^{}_0}~, ~~
E_\pi=\dfrac{s+M_\pi^2-M_D^2}{2\sqrt{s}}\,,
\end{equation}
which is nothing but the energy of the produced soft pion in the
rest frame of the $D\pi$ system.
The peak position $s_\text{peak}$ is obtained from
\begin{equation}
\dfrac{d}{ds}|{\rm BW}_0^\prime(s)|^2 \Big\arrowvert_{s=s_\text{peak}} = 0\,,
\end{equation}
which clearly yields a shift of $s_\text{peak}$ from $M_0^2$. This shift is expected to be
small compared with $M_0$ as long as the width is small, $\Gamma_0\ll M_0$.
Setting $s_\text{peak}=(M_0+\Delta)^2$ and retaining only the linear
term  in $\Delta$ leads to
\begin{equation}
\Delta \simeq \dfrac{\Gamma_0^2(M_0^2 -M_\pi^2+M_D^2)}{2M_0^{}\big[ 2(M_0^2+M_\pi^2-M_D^2)
-\Gamma_0^2\big]} =\dfrac{\Gamma^2_0 E_D^{}}{4M_0^{}E_\pi^{}-\Gamma_0^2}\,,
\end{equation}
where $E_D$ is the energy of the produced $D$ in the rest frame of the $D\pi$ system
with total energy $M_0$. Thus, for the case $4M_0E_\pi >  \Gamma_0^2$, the shift $\Delta$ is
positive and the mass of the resonance is lower than the peak position.
Note that the modification in Eq.~\eqref{eq:bwrev} can only be applied in
a small energy region before the coupled-channel effect becomes important, and, thus, is
neither practical nor systematic. However, this little exercise clearly shows the
deficiencies of the BW parameterization for pionic resonances that are severely
constrained by chiral symmetry.

\begin{figure*}[tb!]
  \begin{center}
    \includegraphics[width=1.0\linewidth]{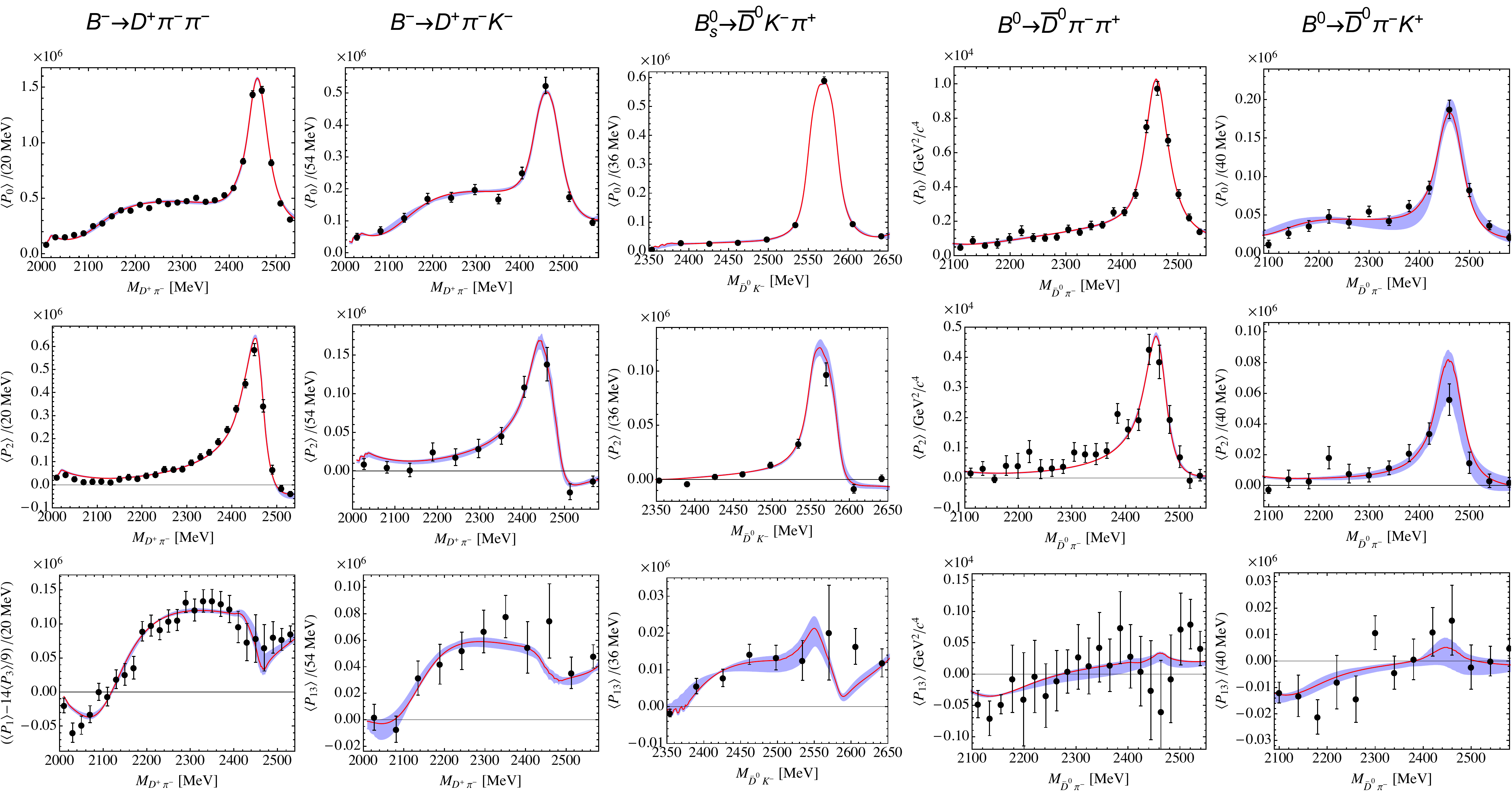}
     \end{center}
  \caption{
    Fit to the LHCb data of the angular moments $\langle P_0\rangle$, $\langle P_2\rangle$ and
    $\langle P_{13}\rangle$
    \cite{LHCb:2016lxy,LHCb:2014ioa,LHCb:2015eqv,LHCb:2015tsv,LHCb:2015klp}
    for the various $B\to D\phi\phi$ channels.
    The largest error of $\langle P_1\rangle$ and $14\langle P_3\rangle/9$ in each bin is taken as
   the error of $\langle P_{13}\rangle$. The error bands correspond to the one-sigma uncertainties
   propagated  from the input scattering amplitudes. All data sets are fit with two parameters
   (one combination of LECs and one subtraction constant) except for the fit to $B^0\to
   \bar{D}^0 \pi^- K^+$, which only requires a subtraction constant as explained in the text.
\label{fig:fitall}}
\end{figure*}

Using the framework outlined in Sect.~\ref{sec:Dphiuni}, one can now fit the LHCb data for the reactions $B^-\to D^+\pi^-\pi^-$, $B^-\to D^+\pi^- K^-$, $B_s^0\to \bar{D}^0K^-\pi^+$,
$B^0\to \bar{D}^0\pi^-\pi^+$ and $B^0\to \bar{D}^0 \pi^- K^+$. In fact, note that only
three sets of the weak production vertices in Table~\ref{tab:amps} are independent.
Thus instead of fitting the four decay amplitudes to the experimental angular moments simultaneously,
we fix the LECs in the amplitudes in Eqs.~(\ref{eq:amp:1}-\ref{eq:amp:4}) by fitting to
three of them, i.e., $B^-\to D^+\pi^- K^-$, $B_s^0\to \bar{D}^0K^-\pi^+$ and $B^0\to
\bar{D}^0\pi^-\pi^+$, and then describe the angular moments for $B^0\to \bar{D}^0 \pi^- K^+$
with the determined LECs. The data for the angular moments defined in  Eqs.~\eqref{eq:angularm}
and \eqref{eq:angularm13} are fitted up to $M_{D\pi}=2.54$~GeV as in Ref.~\cite{Du:2017zvv}
for the decays $B^-\to D^+\pi^- K^-$ and $B^0\to \bar{D}^0\pi^-\pi^+$, and up to
$M_{\bar D\bar K}=2.65$~GeV for $B_s^0\to \bar{D}^0K^-\pi^+$ as in Ref.~\cite{Du:2019oki}.
The best fit has a reasonable quality  with $\chi^2/\text{d.o.f.}=1.2$ and the comparison to the
LHCb data is shown in Fig.~\ref{fig:fitall}. The bands in this figure reflect the one-sigma errors
of the parameters in the scattering amplitudes determined in Ref.~\cite{Liu:2012zya}.
We note that in particular the linear combination of two angular moments
$\langle P_{13}\rangle = \langle P_1\rangle -{14}\langle P_3\rangle/9$ only depends on
the S-P interference as long as one restricts oneself to partial waves with $\ell \leq 2$.
Therefore, $\langle P_{13}\rangle$ is the quantity that one should focus on if one wants to better
understand the scalar charmed mesons. The LHCb angular moment data for all these decays can
be well described. The predicted $\langle P_1\rangle$ and $\langle P_3\rangle$ also
agree with the measurements. Because the final-state interactions in these fits are
taken from the unitarized chiral perturbation theory amplitudes already pinned down
in Ref.~\cite{Liu:2012zya},
this  analysis of the LHCb data implies that the poles contained in these amplitudes
can be regarded as the low-lying scalar charmed meson spectrum. Furthermore, it follows
that  such a spectrum is consistent with the LHCb data. In particular, the poles of the scalar
charm-nonstrange mesons, $\left(2105^{+6}_{-8}- \mathrm{i}\, 102^{+10}_{-11}\right)$~MeV and
$\left(2451^{+35}_{-26}-\mathrm{i}\,134^{+7}_{-8}\right)$ MeV~\cite{Du:2017zvv}, are different from the
resonance parameters of the $D_0^\ast(2400)$ listed in Review of particle Physics (RPP)~\cite{ParticleDataGroup:2018ovx},
which were extracted using a simple BW parameterization. The analysis in this work gives
a further strong support to the two-$D_0^*$ scenario as advocated in
Refs.~\cite{Albaladejo:2016lbb,Du:2017zvv}. The two-pole scenario will be at least
mentioned in the newest version of the RPP~\cite{ParticleDataGroup:2020ssz}.
This point was further strengthened by the analysis in Ref.~\cite{Du:2020pui},
the results based on unitarized CHPT were refined by using Khuri-Treiman
equations, that respect three-body unitarity. The S-wave
$D\pi$ phase-shift could be extracted and it was shown again that a BW parameterization
is not capable of describing  these data. Thus, the lightest charmed scalar meson is of similar
nature than the famous $f_0(500)$ and $K_0^\ast(700)$, namely generated by meson-meson
final-state interactions, which leads to masses at odds with quark model expectations.


\subsubsection{The \texorpdfstring{$D_{s0}^*(2317)$}{Ds0*(2317)}}

\begin{figure}[t]
    \centering
    \includegraphics[width=0.6\textwidth]{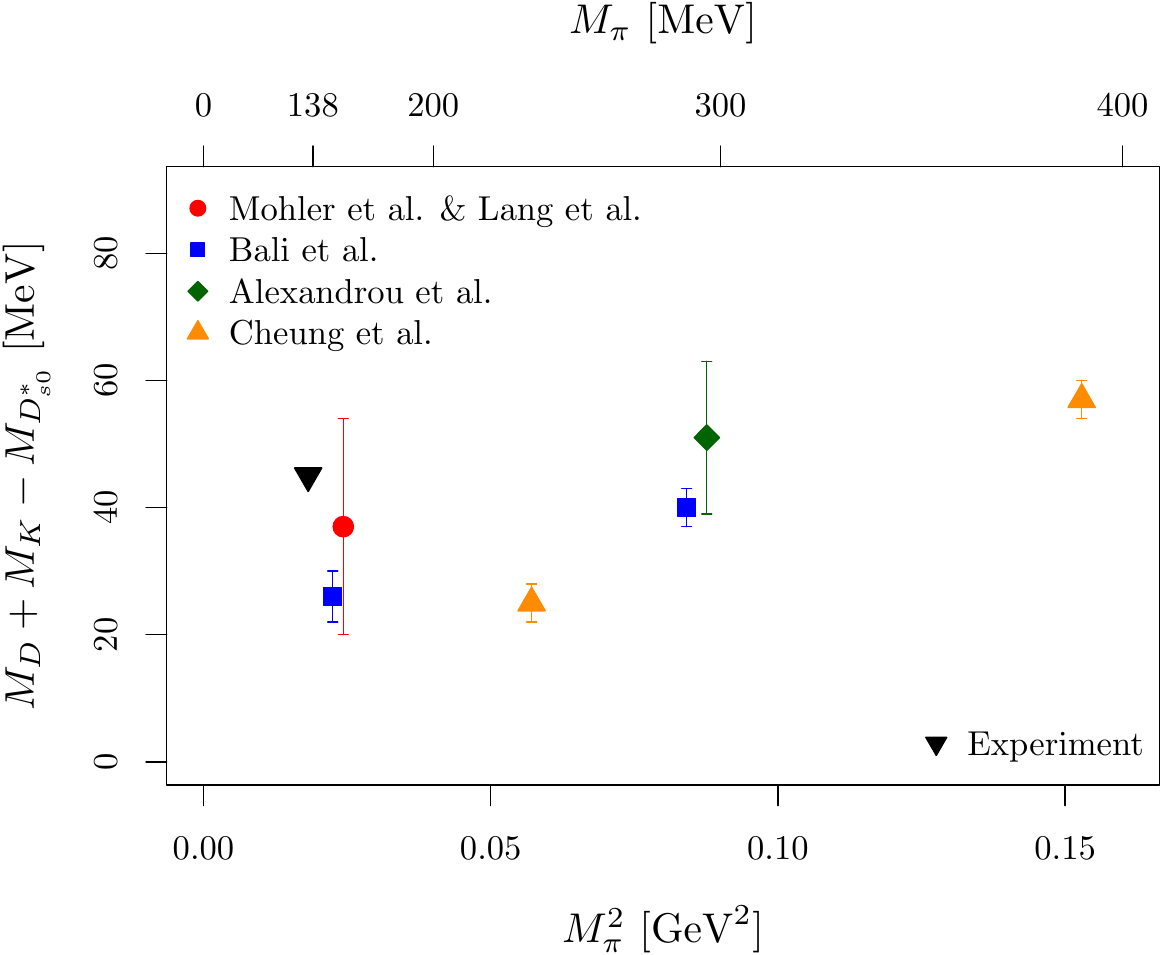}
    \caption{
    \label{fig:Ds0starDK}
    $M_D + M_K - M_{D_{s0}^\ast(2317)}$ as a function of $M_\pi^2$ comparing the available lattice determinations.}
\end{figure}

The $D_{s0}^*(2317)^\pm$ has isospin $I=0$ and appears to have quantum numbers $J^P=0^+$. It decays predominantly into $D_s^+\pi^0$ and has a width of only a few MeV. The PDG quotes  a value of $M_{D_{s0}^\ast} = 2317.8(5)\ \mathrm{MeV}$ and a width smaller than
$4\ \mathrm{MeV}$~\cite{ParticleDataGroup:2020ssz}.

The first indirect lattice calculation of the $D_{s0}^*(2317)$ was performed in
Ref.~\cite{Liu:2012zya}.
More precisely, the scattering lengths for the channels that are not affected by disconnected
diagrams, namely $I=3/2$ $D\pi$, $D_s\pi$, $D_sK$, $I=0$ $D\bar{K}$ and  $I=1$ $D\bar{K}$
(see Tab.~\ref{tab:ci}), were calculated for pion masses in the range from 300 to 600~MeV.
Using UCHPT at NLO, see Eq.~\eqref{eq:LDphi}, the LECs $h_i$ could be determined
and predictions for the channels that were not explicitly calculated on the lattice could be
made. The scattering length in the interesting channel $(S,I)=(1,0)$ was found to be
$a(DK(I=0))=-0.84^{+0.17}_{-0.22}\,$fm and the pole position of a bound state comes out as
$2315^{+18}_{-28}~{\rm MeV}$, which is very close to the PDG value of the mass of the
$D_{s0}^*(2317)$. The molecular nature of this state is supported by the Weinberg
argument, that relates the scattering length $a$ to the binding energy $\varepsilon$ and
the wave function renormalization constant $Z$, with $(1-Z)$ being the probability of
finding the molecular component in the physical state (note that $Z=0$ corresponds to a
compact multi-quark state)~\cite{Weinberg:1965zz,Baru:2003qq}. This relation reads
\begin{equation}
a = -2\left(\dfrac{1-Z}{2-Z}\right) \dfrac{1}{\sqrt{2\mu\varepsilon}} + {\cal O}(1/\beta)~,
\end{equation}
with $\mu$ the reduced mass and $1/\beta$ the range of forces. For a pure molecular state
with $Z=0$, this equation gives $a=-1.05\,$fm, which coincides with the range given above.
In fact, the trajectory of the pion mass dependence of such a molecular state is of quadratic
shape~\cite{Du:2017ttu}, consistent with the lattice data of Ref.~\cite{Bali:2017pdv}.
Finally, we note that the width of the  $D_{s0}^*(2317)$ is very tiny, as it is entirely
driven by isospin-breaking effects~\cite{Guo:2006rp,Faessler:2007gv,Lutz:2007sk,Guo:2008gp}.

Mohler et al.~\cite{Mohler:2013rwa} carried out a similar exploratory calculation for $DK$ scattering focused on the $D_{s0}^*(2317)$. They find the $D_{s0}^*$ to be located $37(17)\ \mathrm{MeV}$ below the $DK$ threshold. The calculation is based on two the ensembles of Lang, Mohler and Prelovsek mentioned above, one with $N_f=2$ dynamical quark flavors and $M_\pi=266\ \mathrm{MeV}$ and a second one with $M_\pi=156\ \mathrm{MeV}$ and $N_f=2+1$ dynamical quark flavors. Of course, given the totally different systematics on these two ensembles a combination of the results is difficult. However, their finding is a negative value of the $DK$ S-wave scattering length, which leads the authors to the conclusion stated above.

In Ref.~\cite{Lang:2014yfa} a similar set of authors work with the exact same $N_f=2$ and $N_f=2+1$ flavor ensembles as discussed in the previous paragraph. Here, they focus on $D_s$ mesons by studying $DK$ and $D^* K$ scattering in the $J^P=0^+, 1^+, 2^+$ channels. In the $J^P=0^+$ channel they find a state $37(17)\ \mathrm{MeV}$ below the corresponding threshold, which they identify to be the $D_{s0}^*(2317)$, in agreement with Ref.~\cite{Mohler:2013rwa}. In the $1^+$ channel the $D_{s1}(2460)$ is $44(10)\ \mathrm{MeV}$ below $D^* K$ threshold with a considerable four-quark component. In the same channel a narrow $D_{s1}(2536)$ state above threshold for the ensemble with lighter pion mass value. In the $J^P=2^+$ channel the $D_{s2}^*$ is found close to its experimentally expected energy. The authors find that it is important to include $DK$ and $D^* K$ interpolating operators in their analyses, which is not surprising. Torres and co-authors also investigate the $D_{s0}^*(2317)$ and the $D_{s1}^*(2460)$ in Ref.~\cite{MartinezTorres:2014kpc} reanalysing data from Refs.~\cite{Mohler:2013rwa,Lang:2014yfa} discussed above but also extending the database to all available energy levels. An existence of bound state for the $KD$ and $KD^*$ channels was confirmed, improving also the determination of the scattering length compared to the original determination.

\begin{table}[t]
    \centering
    \begin{tabular}{lccc}
    \hline
    Reference & $N_f$ & $M_\pi\ [\mathrm{MeV}]$ & $M_D + M_K - M_{D_{s0}^\ast}\ [\mathrm{MeV}]$\\
    \hline
    Mohler et al.~\cite{Mohler:2013rwa} &  2 & 156 & 37(17) \\
    Bali et al.~\cite{Bali:2017pdv} & 2 & 150 & 26(4) \\
    Bali et al.~\cite{Bali:2017pdv} & 2 & 290 & 40(3) \\
    Alexandrou et al.~\cite{Alexandrou:2019tmk} & 2+1 & 296 & 51(12) \\
    Cheung et al.~\cite{Cheung:2020mql} & 2+1 & 239 & 25(3) \\
    Cheung et al.~\cite{Cheung:2020mql} & 2+1 & 391 & 57(3) \\
    \hline
    PDG~\cite{ParticleDataGroup:2020ssz} & &  & 44.7(5) \\
    \hline
    \end{tabular}
    \caption{Results for the $M_{D_{s0}^\ast(2317)}$ extracted from the various references used in \cref{fig:Ds0starDK}.}
    \label{tab:my_label}
\end{table}

Another study of the $D_{s0}^*(2317)$ and the $D_{s1}(2460)$ can be found in Ref.~\cite{Bali:2017pdv}.
Here, two pion masses of $290\ \mathrm{MeV}$ and $150\ \mathrm{MeV}$ are investigated with different volumes at one relatively fine value of the lattice spacing with $N_f=2$. Using spin-averaged quantities, reasonable agreement with experiment is found. Not averaged energies show significant deviations from experiment, which the authors argue is due to lattice artefacts. Their Lüscher analysis includes two-quark and four-quark operators in the
$J^P=0^+$ and $1^+$ sectors. Of these, the four-quark interpolating operators have proven essential to be able to extract the exotic states of interest. The authors are actually able to perform an infinite-volume extrapolation thanks to the many volumes they have
available. This extrapolation appears to work well, and it makes a significant difference
for the final energy values. They find that the mass of the $D_{s0}^\ast$ lies $26(4)$~MeV and $40(3)$~MeV (for light and heavy pion mass ensembles, respectively) below the $DK$ threshold (the authors actually quote asymmetric errors, we use here the larger of the two). The pion mass dependence found here exactly agrees with the one based on unitarized chiral perturbation theory~\cite{Du:2017ttu}.
The work of Ref.~\cite{Bali:2017pdv}
is also the first to our knowledge which computes the decay constants of these states and compares them to those of pseudoscalar and vector $D$-mesons.

The $D_{s0}^*(2317)$ is also being investigated in Ref.~\cite{Alexandrou:2019tmk} with one $N_f=2+1$ flavor dynamical quark ensemble with $M_\pi=296\ \mathrm{MeV}$. The authors report it $51\ \mathrm{MeV}$ below the relevant $DK$ threshold. They find the largest coupling in their operator basis to quark-antiquark interpolating operators with only a small coupling to $DK$ scattering states. Tetraquark interpolators essentially contribute nothing to the analysis of the $D_{s0}^*(2317)$.

In Ref.~\cite{Cheung:2020mql} the authors of the Hadron Spectrum Collaboration present an investigation of the
$D_{s0}(2317)$ with similar techniques in to Ref.~\cite{Alexandrou:2019tmk}. The authors study the elastic scattering amplitudes for $DK$ and $D\bar{K}$ scattering with $I=0$ and $I=0,1$, respectively. In this calculation two ensembles are included, the first with $M_\pi=391\ \mathrm{MeV}$ and the second one with $M_\pi=239\ \mathrm{MeV}$. They find evidence for the bound $D_{s0}^*(2317)$ below $DK$ threshold in the $J^P=0^+$ channel.

For the $D_{s0}^\ast(2317)$ all the evidence points towards a $DK$ molecular like state.
The currently available lattice results are more or less consistent. In \cref{fig:Ds0starDK} we summarize the status by plotting $M_D + M_K - M_{D_{s0}^\ast}$,
resembling a binding energy,
as a function of $M_\pi^2$. Plotting this mass difference and neglecting the width of this state is justified, because the state is so narrow. The observed differences between the
different determinations are likely to be explained with
lattice artefacts, which are uncontrolled in all the available studies.
Therefore, the necessary next step should be an investigation of lattice
artefacts in this system.

\begin{figure}[t]
    \centering
    \includegraphics[width=0.55\textwidth]{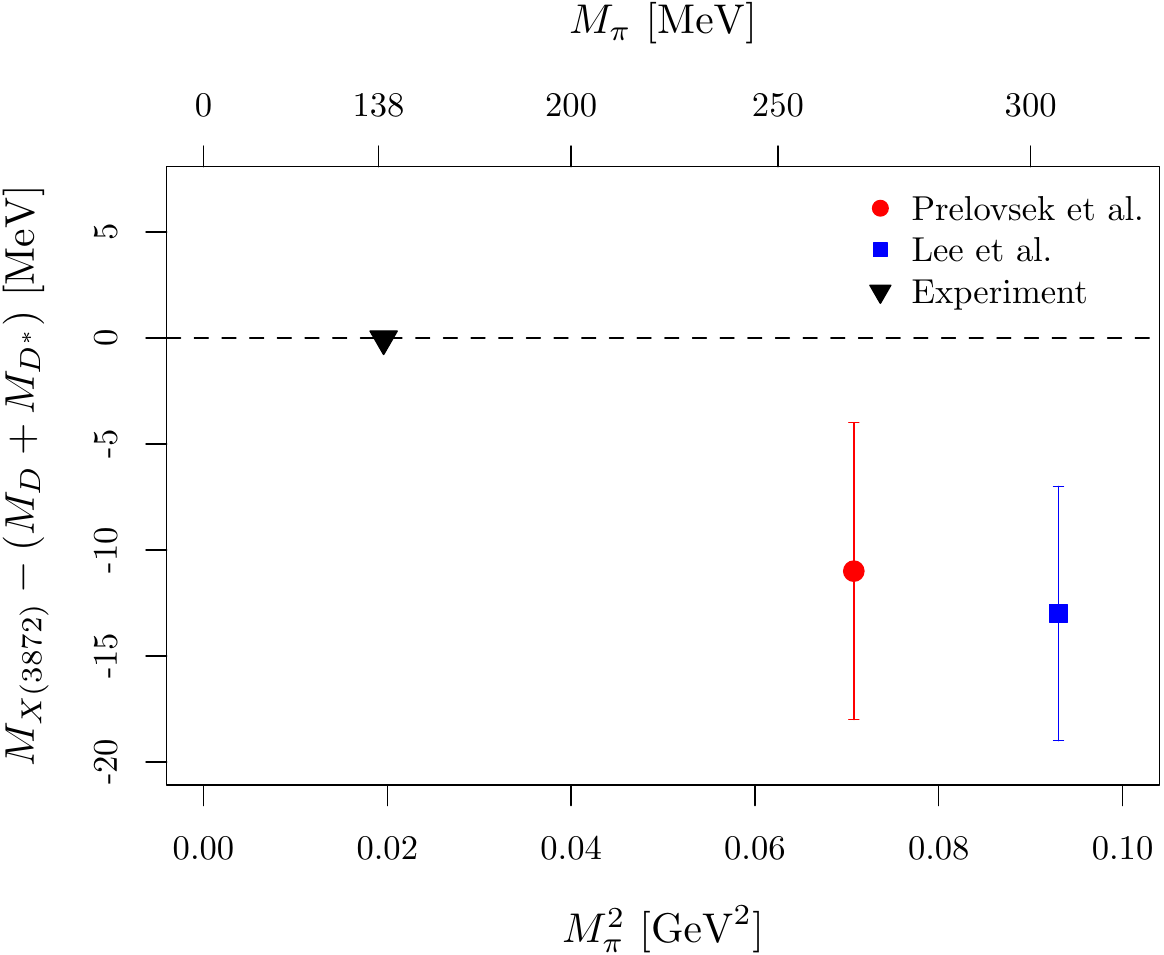}
    \caption{Comparison of the $X(3872)$ mass determined on the lattice by Prelovsek et al. in Ref.~\cite{Prelovsek:2014swa} and Lee et al. in Ref.~\cite{Lee:2014uta}.
    }
    \label{fig:X3872}
\end{figure}

\subsubsection{The \texorpdfstring{$X(3872)$}{X(3872)}}

One of the first exotic states confirmed by several experiments in different decay channels was the $X(3872)$, a narrow charmonium like state, now denoted as $\chi_{c1}(3872)$. It was rather soon hypothesized to be a
non quark-antiquark state. Its quantum numbers are $J^{PC} = 1^{++}$~\cite{LHCb:2013kgk}.
In order to understand its \emph{nature}, lattice studies aim to use different interpolating
operators to investigate the couplings of these operators to the state interpreted as the
$X(3872)$. Recently, strategies to determine the $X(3872)$ from lattice QCD more precisely were
formulated in Ref.~\cite{Garzon:2013uwa}. Finite-volume corrections to the binding energy of the $X(3872)$ are discussed in Ref.~\cite{Jansen:2015lha}.

The authors of Ref.~\cite{Chiu:2006hd} present a very first study of the $X(3872)$
exotic state. It is based on quenched lattice QCD and a chiral fermion discretization.
Even though systematic uncertainties are not well controlled in this study and, in particular, no Lüscher like analysis has been performed, they find evidence for a state in the $1^{++}$ channel with about the correct energy. In another quenched study, CLQCD investigates the $\chi_{c2}$ in Ref.~\cite{Yang:2012mya} and use their results to draw the conclusion that the $X(3872)$ has not $J^{PC}=2^{-+}$ as quantum numbers, which at the time of the publication was still not excluded.

In Ref.~\cite{Prelovsek:2013cra} Prelovsek and Leskovec study the $X(3872)$ state with $J^{PC}=1^{++}$ and zero isospin. This calculation is again exploratory in the sense that it is based on the same single $N_f=2$ ensemble with pion mass value of $266\ \mathrm{MeV}$ as used on Ref.~\cite{Mohler:2012na} discussed above. They find a candidate for the $X(3872)$ just $11(7)\ \mathrm{MeV}$ below the $D\bar{D}^*$ threshold. Importantly, they find a large and negative value for the scattering length of the $D\bar{D}^*$ system of $-1.7(4)$, which allows them to identify the $X(3827)$ in their lattice data.

In the proceeding contributions~\cite{Lee:2014uta} a candidate $X(3872)$ state is found
just $13(6)\ \mathrm{MeV}$ below the $DD^\ast$ threshold with $I=0$. This study is based
on a single ensemble with the HISQ staggered fermion discretization and $N_f=2+1+1$ dynamical quark flavors. This
represents the only study of the $X(3872)$ including also a dynamical charm quark, though the character of this investigation is quite exploratory.

The $X(3872)$ was again studied in Ref.~\cite{Padmanath:2015era} by Padmanath and co-authors together with the $Y(4140)$. The authors work with the aforementioned $N_f=2$ ensemble at $M_\pi=266\ \mathrm{MeV}$. They find that it depends on the list of operators included in the analysis whether or not a $X(3872)$ candidate in the $I=0$ $J^{PC}=1^{++}$ channel can be found: $D\bar{D}^*$ and $c\bar{c}$ like interpolators appear to be a must. In the $I=1$ channel no neutral or charged $X(3872)$ candidate is found. Likewise, they find no evidence for a $Y(4140)$ state with $\bar{c}c\bar{s}s$ quark content.

The two available results from LQCD for $M_{X(3872)} - (M_D + D_{D^\ast})$~\cite{Prelovsek:2014swa,Lee:2014uta} are compiled in \cref{fig:X3872}. Both have large uncertainty and were obtained for pion masses larger than $250\ \mathrm{MeV}$. Thus, it is probably not yet the time to compare with experiment on a quantitative level.

\subsubsection{The \texorpdfstring{$Z_c(3900)$}{Zc(3900)}}

The $Z_c(3900)$ represents a closed charm state which might be of tetraquark nature with
quark content $\bar{c}c\bar{d}u$.
It has isospin $I^G=1^+$ and $J^{PC}=1^{+-}$. With a width of around $30\ \mathrm{MeV}$ it is
relatively narrow and it is seen in different decay channels such as in $J/\psi\pi$, $\eta_c\pi\pi$,
$D\bar{D}^*$ and $DD^*$. It was observed as a decay product of the exotic $Y(4260)$ state.
We remark in passing that in Ref.~\cite{Liu:2019gmh} methodological developments for an
improved lattice investigation of the $Z_c$ were presented.

Prelovsek and Leskovec search for a candidate $Z_c(3900)$ state in Ref.~\cite{Prelovsek:2013xba} in the $J^{PC}=1^{+-}$ with $I=1$, but do not
find any evidence with $N_f=2$ dynamical quark flavors at $M_\pi=266\ \mathrm{MeV}$.
This investigation was extended in Ref.~\cite{Prelovsek:2014swa}.
Unfortunately, this study could also not reveal evidence for an extra $Z_c^+(3900)$ state in addition to all the expected two-meson states. The authors provide a detailed discussion as to why this is the case.

The CLQCD Collaboration studies the $Z_c(3900)$ in Ref.~\cite{Chen:2014afa} in the scattering of
$D\bar{D}^*$. They work with the three ensembles ($M_\pi=485, 420, 300$~MeV) with $N_f=2$ dynamical quarks discussed above.
Noteworthy is their much smaller lattice spacing value than used in the studies by Prelovsek et al.~\cite{Prelovsek:2014swa} ($a=0.067\ \mathrm{fm}$ versus $a=0.125\ \mathrm{fm}$). They find a weak repulsive interaction between
the $D$ and the $\bar{D}^*$ for all three pion mass values. Thus, their simulation results
cannot support a bound state in the corresponding $J^P=1^+$ channel, where the $Z_c^\pm(3900)$
is expected.

We mention also two proceeding contributions investigating the $Z_c(3900)$ from lattice QCD. The first one~\cite{Lee:2014uta} uses the staggered quark formulation, finding no candidate $Z_c^+(3900)$ state.
The second~\cite{Liu:2014mfy} used $N_f=2$
Wilson twisted mass fermions, also not being able to corroborate the $Z_c$ interpretation as a $D\bar{D}^\ast$ bound state.

The HAL QCD collaboration studies the $Z_c(3900)$ in Ref.~\cite{HALQCD:2016ofq,Ikeda:2017mee} in a range of pion mass values from $400\ \mathrm{MeV}$ to $700\ \mathrm{MeV}$ taking into account
coupled channels. Their data indicates that the $Z_c$ is most likely a threshold cusp.

It appears rather difficult to draw any conclusion from this list of lattice
investigations. Mostly, there is no candidate $Z_c(3900)$ state found with the operators included in the respective analyses. Only HAL QCD finds indications for a threshold cusp.

Finite-volume spectra from of Prelovsek et al.~\cite{Prelovsek:2014swa} were further reanalysed in Ref.~\cite{Albaladejo:2016jsg} using a unitarized amplitude, see Sec.~\ref{sec:uni}. In that, they find that two scenarios are compatible with the lattice energy levels: a $D^\ast\bar{D}$ resonance or a virtual state. But they cannot distinguish between these two possibilities. They conclude that several volumes should be studied to obtain a better understanding of the $Z_c(3900)$.

\subsection{Other Exotic States}

In this subsection we present mainly exploratory studies of certain hadron resonances. Thus, we refrain from a  discussion of the systematics or physics implications of the results.

\subsubsection{States involving heavy-light mesons}

The CLQCD collaboration investigated in Ref.~\cite{CLQCD:2015htz} the exotic $Z_c(4025)$ state. S-wave scattering of $D^*\bar{D}^*$ was studied in the $J^P=1^+$ channel. Also here, CLQCD finds a weak repulsive interaction between the two mesons and, again, a bound state in this channel is not supported by their study. Like in the study of the $D\bar{D}^*$ this might be due to missing interpolators in the correlator matrix. But it could also be an effect of the unphysical pion mass values.

In Ref.~\cite{Lang:2015sba} Lang and co-authors investigate vector and scalar charmonium resonances. In $D\bar{D}$ scattering they find the $\psi(3770)$ in the P-wave, as expected. However, in the scalar channel they find in S-wave $D\bar{D}$ scattering an additional state with mass slightly below $4\ \mathrm{GeV}$. This narrow resonance state was unobserved so far, while the ground state $\chi_{c0}(1P)$ is well understood. They investigate several scenarios for possible states in this channel and find that a scenario with the $\chi_{c0}(1P)$ and the additional narrow resonance mentioned above leads to a phase-shift consistent with experimental data. The authors work with the two ensembles again, one with $N_f=2$ and $M_\pi=266\ \mathrm{MeV}$ and the second one with $N_f=2+1$ and $M_\pi=156\ \mathrm{MeV}$.

In Ref.~\cite{CLQCD:2016anu}, however, CLQCD find a weakly attractive interaction in $\bar{D}_1D^*$ scattering based on the same ensembles and pion mass values. S- and P-wave channels have been studied finding attraction in both channels. Indications for bound states below threshold are reported, but further studies are needed to draw conclusions on the $Z_c(4430)$ exotic state.

A lattice QCD calculation of an $NJ/\psi$ and $N\eta_c$ system with quantum numbers overlapping with the quantum numbers of the LHCb discovered pentaquark states $P_c(4380)$ and $P_c(4450)$ was performed in Ref.~\cite{Skerbis:2018lew}. The calculation was performed on two-flavor ensembles with $M_\pi= 266~{\rm MeV}$, reaching for the first time energies of the mentioned pentaquark states. The calculation resulted in a very small (consistent with zero) attractive interaction, i.e., no significant energy shift from the non-interacting case was observed.

Positive parity $B_s$ mesons are investigated by Lang and co-authors in Ref.~\cite{Lang:2015hza}. The investigation is based on a single $M_\pi=156\ \mathrm{MeV}$ ensemble with $N_f=2+1$ dynamical quark flavors. The bottom quark is added as a valence only quark with the Fermilab method, see Sect.~\ref{sec:heavyLat}. They identify the $B_{s1}(5830)$ and the $B_{s2}(5840)$ finding good agreement with experimental results. In addition they predict a $B_{s0}$ with mass $5750(25)\ \mathrm{MeV}$ in the $J^P=0^+$ channel. Based on the same single ensemble, Ref.~\cite{Lang:2016jpk} by Lang et al. represents the first $B_s\pi^+$ scattering study and its relation to the exotic $X(5568)$ from lattice QCD. As the main result, they cannot establish the $X(5568)$ from their lattice QCD simulation in the $J^P=0^+$ channel. Since the study is based on a single ensemble at a single lattice spacing only, this result does of course not exclude that such a state with these quantum numbers exists. Still its quantum numbers are also not yet finally determined experimentally and, thus, $J^P=0^+$ is only one possibility. Note that the $X(5568)$ poses severe challenges to QCD as discussed in Refs.~\cite{Burns:2016gvy,Guo:2016nhb}, for example such a mass is neither compatible with chiral symmetry nor with heavy quark symmetry.

In Ref.~\cite{Piemonte:2019cbi} the authors study the $J^{PC}=1^{--}$ and $J^{PC}=3^{--}$
channels in $\bar{D}D$ scattering. They use lattice ensembles with $280\ \mathrm{MeV}$
and two lattice volumes with $N_f=2+1$ quark flavors at a single lattice spacing value.
In this reference the conventional $\psi(3770)$ and the $X(3842)$ are found at energies
compatible with experiment. The investigation is extended to unconventional states in
Ref.~\cite{Prelovsek:2020eiw}, where charmonium like resonances with $J^{PC}=0^{++}$
and $2^{++}$ in coupled $D\bar{D}$ and $D_s\bar{D}_s$ scattering are studied. In this study, Prelovsek et al. analytically continue the scattering matrix and determine pole singularities. They find a so far unobserved $D\bar{D}$ bound state just below threshold
and a $D\bar{D}$ resonance, the latter of which they connect to the $\chi_{c0}(3860)$. Moreover, they find a narrow $J^{PC}=0^{++}$ resonance below $D_s\bar{D}_s$ threshold. They interpret it as possibly related to the $X(3915)$ or the $\chi_{c0}(3930)$ states. They also see a resonance they connect to the $\chi_{c2}(3930)$, because it is found in the D-wave.

Only very recently also the $I=0$ and $J^P=1^{+}$ channel was studied in $DD^*$ scattering
on the ensemble from above with $280\ \mathrm{MeV}$ pion mass. Padmanath and Prelovsek~\cite{Padmanath:2022cvl}
find evidence for a virtual bound state with just $10\ \mathrm{MeV}$ binding energy. This
could be the doubly charmed tetraquark state recently discovered by LHCb~\cite{LHCb:2021vvq,LHCb:2021auc}.
It has open charm quark content $cc\bar{u}\bar{d}$ and lies just order $1\ \mathrm{MeV}$ below the
$D^0D^{\ast +}$ threshold.

Also, very recently axial-vector $D_1$ hadrons in $D^\ast \pi$ scattering were studied on a single $M_\pi=391\ \mathrm{MeV}$ ensemble for the first time in Ref.~\cite{Lang:2022elg}. The dynamically coupled $^3S_1$ and $^3D_1$ channels in $D^\ast \pi$ scattering have been looked at, and the corresponding scattering amplitudes have been computed. The $^3S_1$ is dominated by the pole right below $D^\ast \pi$ threshold. The $^3D_1$ amplitude is dominated by a single, narrow resonance.

Doubly bottom tetraquarks have been investigated in Ref.~\cite{Francis:2016hui} in an exploratory study without Lüscher analysis. Follow up investigations can be found in Refs.~\cite{Francis:2018jyb,Junnarkar:2018twb}. The first Lüscher analysis for a $\bar{b}\bar{b}ud$ tetraquark with isospin $I=0$ and $J^P=1^+$ was performed in Ref.~\cite{Leskovec:2019ioa}. The work is based on domain wall fermion ensembles with $N_f=2+1$ dynamical quark flavors including an ensemble at the physical point. Their work controls most of the systematic uncertainties, the bottom quark is treated in the framework of lattice nonrelativistic QCD. They find significant evidence for the existence of a $\bar{b}\bar{b}ud$ tetraquark stable under strong and electromagnetic interactions.
We remark that there is significant effort under way to study bottomonium, see for instance Ref.~\cite{Ryan:2020iog} by Ryan and Wilson, where only two quark operators have been used so far, or Ref.~\cite{Bicudo:2019ymo}, where a formalism is proposed and derived based on static potentials. The latter is applied in Ref.~\cite{Bicudo:2022ihz} to study among others bottomonium in different partial waves with $I=0$.

In Ref.~\cite{Meinel:2022lzo} a study of the $\bar{b}\bar{b}us$ us system with quantum numbers $J^P=1^+$ and $\bar{b}\bar{c}ud$ systems with quantum numbers $I(J^P) = 0(0^+)$ and $I(J^P) = 0(1^+)$ is presented. The authors work with $N_f=2+1$ dynamical quark flavor ensembles generated byt RBC/UKQCD with two lattice spacing values and different pion mass values including the physical one. Charm quarks are treated relativistically, the bottom quarks within NRQCD. They find evidence for $\bar{b}\bar{b}us$ tetraquark bound by $86(22)(10)\ \mathrm{MeV}$. For the systems involving charm quarks their results are inconclusive.

An interesting system with exotic quantum numbers $J^{PC}=1^{-+}$ decaying in eight multihadron final states was studied on the lattice in Ref.~\cite{Woss:2020ayi}. In that, ensembles with $\rm SU(3)$ flavor symmetry ($M_\pi\approx700~{\rm MeV}$ and near physical strange quark) with six different volumes (12-24) were used. Ultimately, the finite-volume spectrum was determined featuring 61 energy levels which were used to fix the parameters of generic parametrizations of the scattering amplitudes. Those amplitudes were also used to extract the resonance parameters (complex valued mass) and couplings to individual channels. It was found that each parametrization describing the finite-volume spectrum also leads to a pole in the complex plane in infinite volume, with well restricted position yielding overall $M_R=2144(12)~{\rm MeV}$ and $\Gamma_R=12(21)~{\rm MeV}$. However, coupling to individual channels varies strongly with the choice of a parametrisation resulting in wider ranges, also reflected in partial decay widths. A simplified approach was finally undertaken in extrapolating the latter to the physical point, suggesting a potentially broad $\pi_1$ resonance.

\subsubsection{Dibaryon States}

Proposed in 1977 by Jaffe, the $H$ dibaryon with quark content $udsuds$ attracted attention by lattice practitioners early on~\cite{Mackenzie:1985vv,Iwasaki:1987db}. More modern investigations were published in 2010 by HAL QCD in Ref.~\cite{Inoue:2010es} and by NPLQCD in Ref.~\cite{NPLQCD:2010ocs} based on different methods. HAL QCD finds a bound $H$ dibaryon in the SU$(3)$ flavor limit of QCD ($N_f=3$) with several pion mass values larger or equal $670\ \mathrm{MeV}$ and several volumes using the HAL QCD method. NPLQCD works with $M_\pi=389\ \mathrm{MeV}$ with $N_f=2+1$ dynamical quark flavors. Also NPLQCD finds evidence for a bound $H$ dibaryon at this pion mass applying the Lüscher method.

In Ref.~\cite{Luo:2011ar} the authors present a quenched lattice investigation, for which they find evidence for a bound $H$ dibaryon. This study is of interest, because they can study the continuum and chiral limits due to the usage of the quenched approximation. Their results appear to agree with the ones from HAL QCD and NPLQCD. However, the systematics due to the quenched approximation are less controlled.

More recently, Francis et al.~\cite{Francis:2018qch} work with $N_f=2$ dynamical light quark flavors and a quenched strange quark with a pion mass of almost $1\ \mathrm{GeV}$. With a Lüscher analysis they find a $H$ dibaryon bound by $19(10)\ \mathrm{MeV}$ at the flavor symmetric point. Interestingly, they claim there is no evidence for bound dineutron.

Finally, most recently the $H$ dibaryon was again studied in Ref.~\cite{Green:2021qol} at the flavor symmetric point with $M_\pi=M_K=420\ \mathrm{MeV}$ ($N_f=3$). They even take the continuum limit based on six values of the lattice spacing applying the Lüscher method. They find evidence for a weakly bound $H$ dibaryon. One rather interesting result from this publication is the finding of rather large discretization effects in the binding energy, despite $\mathcal{O}(a)$ improvement. In particular, while the binding energy is still around $30\ \mathrm{MeV}$ for $a\approx 0.1\ \mathrm{fm}$, only $B_H=4.5\ \mathrm{MeV}$ survives the continuum limit.

Some critical remarks on these results are in order. It was shown early in
Refs.~\cite{Shanahan:2011su,Haidenbauer:2011ah,Haidenbauer:2011za} that  extrapolations in the pion mass based on chiral symmetry indeed do not support the picture of a $\Lambda\Lambda$ bound state.
Indeed, as shown in Ref.~\cite{Haidenbauer:2011ah} using an EFT for baryon-baryon interactions with $S=-2$ \cite{Polinder:2007mp,Haidenbauer:2015zqb}, it was shown that SU(3) breaking effects
induced by the differences of the pertinent two-baryon thresholds ($\Lambda\Lambda$, $\Xi N$, $\Sigma\Sigma$)
have a pronounced impact that need to be incorporated properly in the lattice QCD simulations. Furthermore,
it was pointed out that if the H-dibaryon is a two-baryon bound state, its dominant component is
$\Xi N$ rather than $\Lambda\Lambda$ as a consequence of the approximate SU(3) flavor symmetry of the
two-baryon interactions.

HAL QCD has employed their method for further interesting systems. In Ref.~\cite{Gongyo:2020pyy} the $\Delta\Delta$ dibaryon state, the $d^\ast(2380)$ dibaryon with $J^P=3^+$ and isospin $I=0$, is studied. The investigation is again based on $N_f=3$ heavy pion mass ensembles. They find an short range attractive interaction and a state below the $\Delta\Delta$ threshold. Recently, there also appeared a study of the $N\phi$ system using the HAL QCD method~\cite{Lyu:2022imf}.

Much closer to physical are the quark mass values in the $N_f=2+1$ flavor study of HAL QCD of the $N\Omega$ $^5S_2$ system~\cite{HALQCD:2018qyu}: here the pion mass is estimated to be $145\ \mathrm{MeV}$ and the lattice spacing $a=0.0846\ \mathrm{fm}$. With this, the authors estimate a potential attractive at all distances. Not including the electromagnetic interaction, the binding energy is $1.54(30)\ \mathrm{MeV}$. Including electromagnetic effects for the proton-Omega $p\Omega^-$ system leads to an increase in the binding energy by $1\ \mathrm{MeV}$.
The $\Omega N$ and $\Omega\Omega$ interactions from HAL QCD have been critically discussed in Ref.~\cite{Haidenbauer:2019utu}.

Recently progress has also been achieved in addressing resonant two baryon systems from lattice QCD. Considering system with maximal charm number ($C=3$) for each of the baryons the scattering properties were investigated in a (2+1)-flavor setup in Ref.~\cite{Lyu:2021qsh}. With nearly physical light quark masses ($M_\pi\approx146~{\rm MeV}$ and $M_K\approx525~{\rm MeV}$), the charm quarks were implemented using a relativistic heavy quark action removing higher order cutoff-effects. Finally, implementing the HAL QCD method, a quite strongly bound ($B\approx 5.7~{\rm MeV}$) dibaryon was found when discarding Coulomb repulsion. However, including the latter the state changes from a deep to a shallow bound state.

In another lattice QCD study~\cite{Mathur:2022nez} the two-baryon system with maximal bottom number $B=6$ was performed. In this work ensembles generated by the MILC collaboration with $N_f=2+1+1$ dynamical quarks were used. The bottom quarks were implemented via a non-relativistic Hamiltonian. For all considered four lattice volumes, a negative energy shift to the two-$\Omega_{bbb}$ was recorded suggesting attractive force between the two baryons. These levels were further used to determine the two-baryon scattering length, assessing uncertainties from scale setting and discretization. Overall, a deeply bound dibaryon ${}^1S_0$ state ($B\approx 89~{\rm MeV}$) was found. In contrast to the $C=6$ dibaryon system studied in Ref.~\cite{Lyu:2021qsh}, the Coulomb repulsion was found not to influence the nature of the state.


\section{Summary and conclusions}
\label{sec:summary}

\begin{figure}[t]
  \centering
  \includegraphics[height=5.13cm, trim=0 0.6cm 0 0, clip]{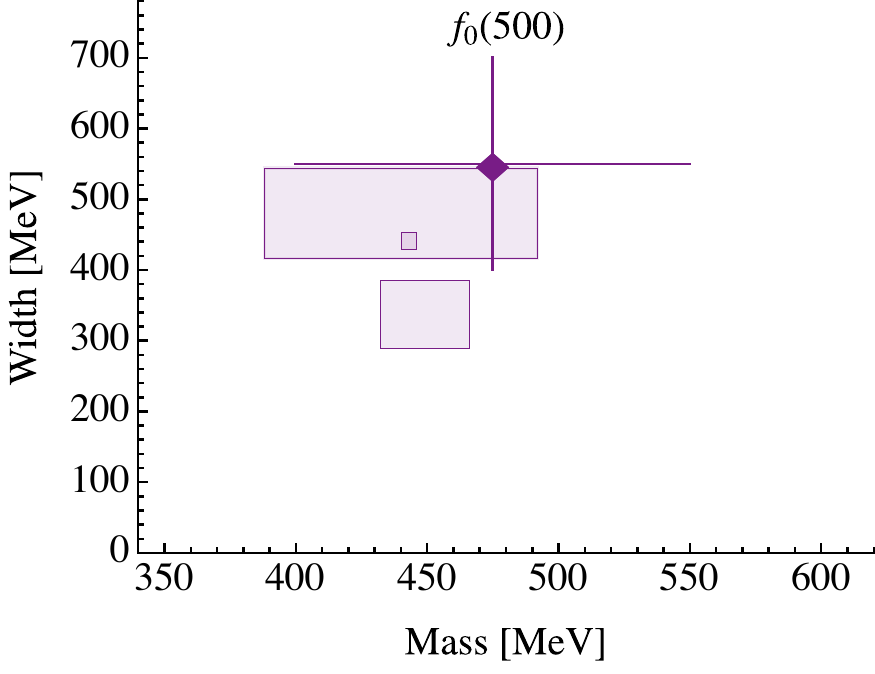}
  \includegraphics[height=5.13cm, trim=0.6cm 0.6cm 0 0, clip]{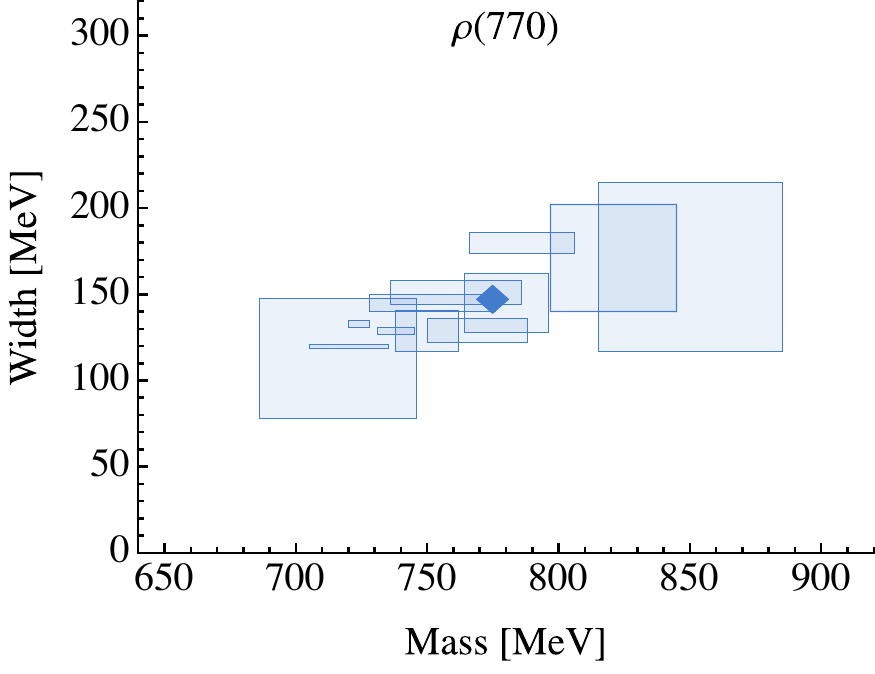}\\~
  \includegraphics[height=5.5cm, trim=0 0 0 0, clip]{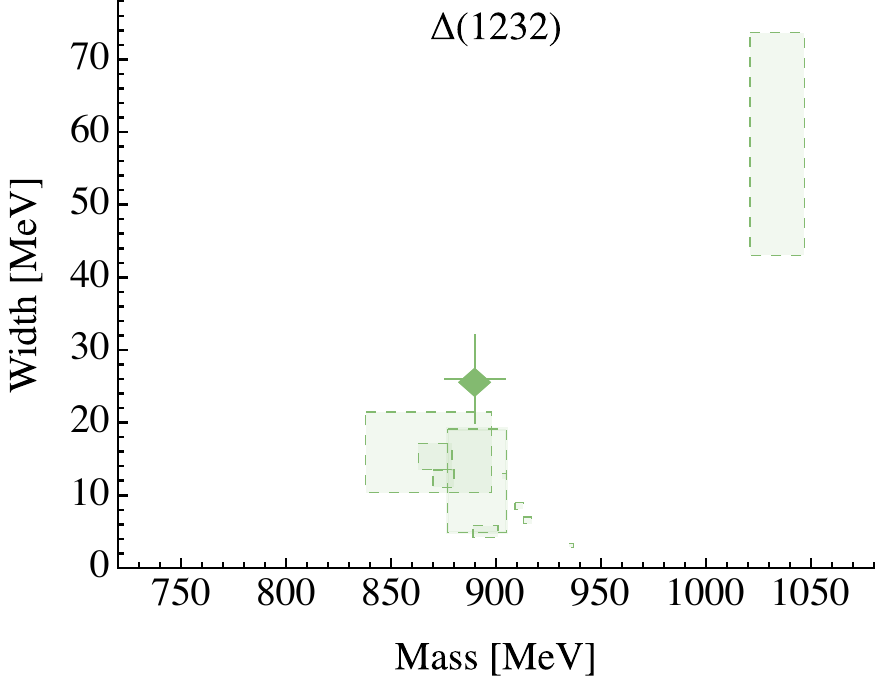}~
  \includegraphics[height=5.5cm, trim=0.6cm 0 0 0, clip]{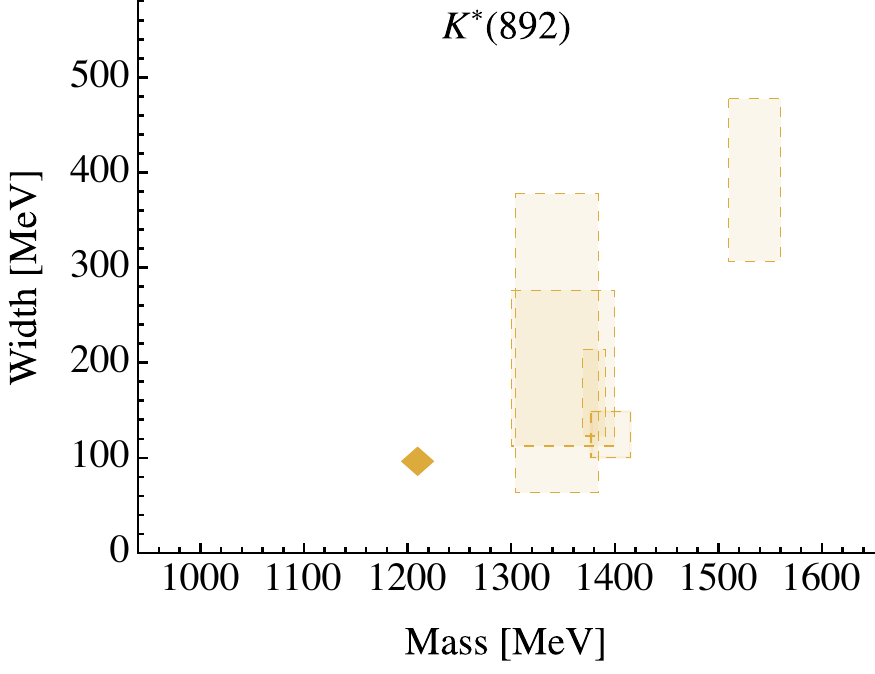}
\caption{Current state in determination of certain hadronic resonances from lattice QCD, references can be found in the main text. Dashed/full boxes denote, respectively, the lattice driven determinations at unphysical/physical and near physical pion mass as discussed in the main text. The PDG~\cite{ParticleDataGroup:2020ssz} values are denoted by the diamonds with error bars.}
\label{fig:compilation-LAT-PDG}
\end{figure}

In this review we have discussed the status of our theoretical
understanding of hadron resonances. The main tools at our disposal are
lattice QCD and effective field theories, which need to go hand in
hand to make progress. In particular, for determining phase-shifts
from lattice QCD the usage of finite-volume effective field theory
(aka the Lüscher formalism) is mandatory.
Once phase-shifts are
determined it is again effective field theory to interpret this data
and extract the resonance pole positions. HAL QCD has developed an alternative method to determine potentials from lattice QCD simulations.

There has been significant progress over the last years in lattice QCD, both methodologically and in practice for our understanding of QCD resonances. The best studied state is certainly the $\rho$-resonance, which is the only resonance for which a continuum extrapolation has been performed so far. However, the $\rho$ shows also the strongest pion mass dependence and, thus, a continuum extrapolation at physical pion mass values is likely required to finally compare to experiment. Also for the other well separated resonances $K^\ast(892)$, $\Delta(1232)$ and $f_0(500)$ there are now several lattice results available. For those the control of systematics is by far not as good as for the $\rho$-resonance. However, reasonable agreement with experiment is observed.
All available estimates for mass and width of these four states are compiled in
\cref{fig:compilation-LAT-PDG}. The figure nicely summarizes the status and the different levels of maturity of the lattice studies for the corresponding states.
It also shows that baryon resonances are particularly challenging.
We expect in the near future further results with ever more realistic simulation parameters for these four resonances.

Less well studied are resonances with strong effects from coupled channels or which are close to thresholds. This includes for instance light scalar mesons. However, the formalism for coupled channel analyses has been developed and also successfully applied, though still at unphysical parameter values. An example is the investigation by the Hadron Spectrum Collaboration of a coupled $\pi K$ and $\eta K$ system. Important lessons are learned about the interpolating operators that need to be included: in many cases multi-hadron operators are, not unexpectedly, mandatory. Also three pion resonant channels are being investigated in exploratory studies.

The Roper-resonance, despite significant effort, stays elusive: the inclusion of three-body interpolators, like for instance $\pi\pi N$, and the corresponding inelastic channels might provide a way out in the future.

Open and closed charm states represent another set of resonances relatively well covered in lattice studies. Most results are available for the $D_{s0}^\ast(2317)$ located a few MeV below the $DK$ threshold. Since it is very narrow, the energy difference to the $DK$ threshold is a meaningful quantity, and the available lattice determinations of it provide a consistent picture in agreement with experiment. For the $D_0^\ast(2300)$ fewer results are available, but it seems that it is located below the $D\pi$ threshold at large pion mass values, but moves above threshold towards physical pion mass values. Note that this state reveals a two-pole structure similar to the enigmatic
$\Lambda(1405)$. Such two-pole structures should also be seen for the axial $D$ mesons
and the corresponding $B$ mesons siblings, see, e.g., Ref.~\cite{Meissner:2020khl}.

For the $X(3872)$ the two available estimates of the mass are about $15\ \mathrm{MeV}$ below the $D\bar{D}^\ast$ threshold. These investigations are both at way too large pion mass values, such that a comparison to the experimental $X(3872)$ energy level is not yet really meaningful.
For the $Z_c(3900)$ on the other hand, a lattice study appears to be difficult: no definite candidate for this state could be identified unambiguously.

Concerning dibaryons, it is at present unclear whether such objects are really bound. In general, coupled channel analyses are required to be able to make definite conclusions, as prominently exhibited in case of the elusive $H$ dibaryon.

From all this it should have become clear that we are still far away from a detailed understanding of the hadron spectrum based on such first principles calculations. The results reported here are encouraging, as most of the required theoretical framework has been developed. As an important lesson, it has become clear that the choice of interpolating operators strongly influences the observed states. This means, on the one hand that the outcome of lattice investigations depends on the list of operators included. But on the other hand it gives hope that from the coupling of states to certain operators one can learn about the nature of the state. But, of course, it foremost means that a large list of operators need to be constructed, which was greatly simplified by methods like (stochastic) distillation, but is still very costly.
Work is ongoing to further improve the operator construction, see for instance Ref.~\cite{Knechtli:2022bji}.

Another lesson one can learn from the existing investigations is that threshold effects can be misinterpreted as resonances. Here, a careful data analysis is required.

Finally, lattice artefacts and pion mass dependence can be significant. This can mean a bound state turns into a resonance or vice versa once the limits are studied properly.

In the future we, therefore, think that it will be important to further investigate the dependence of results on the list of operators, but also to better study continuum and chiral limits. A next important step will be to tackle three-body decays:
in particular the $\omega$ meson and the Roper resonance in the baryon sector represent worthy targets.


\section*{Acknowledgements}

We thank Peter Bruns, Daniel Mohler, Chris Culver, Michael Döring, Feng-Kun Guo and Akaki Rusetsky for clarifying discussions. We thank Menglin Du for providing \cref{fig:fitall}.
This work is supported by the Deutsche
Forschungsgemeinschaft (DFG, German Research Foundation) and the NSFC through the funds provided to the Sino-German
Collaborative Research Center CRC 110 “Symmetries
and the Emergence of Structure in QCD” (DFG Project-ID 196253076 -
TRR 110, NSFC Grant No.~12070131001). Parts of this manuscript were accomplished by MM while being supported by the U.S. Department of Energy under Award Number DE-SC0016582.
The work by UGM was further supported by VolkswagenStiftung (grant No. 93562) and by the Chinese Academy of Sciences (PIFI grant 2018DM0034).

\bibliographystyle{elsarticle-num}
\bibliography{bibliography,NotInspires}

\end{document}